\documentclass[smallcondensed]{svjour3}     
\pdfoutput=1 
 \journalname{Formal Methods in System Design}
\usepackage{amsmath,amssymb,amsfonts}
\usepackage{mathtools}
\usepackage[usenames,dvipsnames]{color}
\usepackage{xcolor}
\usepackage{varioref}
\usepackage{xspace}
\usepackage{microtype}
\usepackage{regexpatch}
\usepackage{colonequals}
\usepackage{wasysym}
\usepackage{booktabs}
\usepackage{nicefrac}

\usepackage{multirow}
\usepackage{multicol}
\newcounter{oldlinenumber}

\usepackage{nicefrac}

\usepackage{paralist}

\usepackage{subfigure}
\usepackage{url}
\usepackage{graphicx}
\usepackage[pdfpagelabels=true]{hyperref}
\usepackage{algorithm}
\usepackage{listings}
\usepackage{newalg}
\usepackage{nicefrac}
\usepackage{tikz} 
\usepackage{pgfplots}
\usepackage{marvosym}
\usepackage{nicefrac}
\usepackage{upgreek}
\usepackage[numbers,sort]{natbib}

\usetikzlibrary{arrows,decorations,positioning,fit,trees,shapes,shadows,automata,calc,external} 

\tikzset{>=stealth'}

\tikzset{
  dot hidden/.style={},
  line hidden/.style={},
  dice hidden/.style={},
  dot color/.style={dot hidden/.append style={color=#1}},
  dot color/.default=black,
  line color/.style={line hidden/.append style={color=#1}},
  line color/.default=black,
  dice color/.style={dice hidden/.append style={color=#1,fill}},
  dice color/.default=white
}\def\dotsize{0.1}
\newcommand{\drawdie}[2][]{%
\begin{tikzpicture}[x=1em,y=1em,#1]
  \draw 	[thick, rounded corners=0.5,line hidden,dice hidden] (0,0) rectangle (1,1);
  \ifodd#2
    \fill[dot hidden] (0.5,0.5) circle (\dotsize);
  \fi
  \ifnum#2>1
  \fill[dot hidden] (0.25,0.25) circle (\dotsize);
  \fill[dot hidden] (0.75,0.75) circle (\dotsize);
  \ifnum#2>3
    \fill[dot hidden] (0.25,0.75) circle (\dotsize);
    \fill[dot hidden] (0.75,0.25) circle (\dotsize);
    \ifnum#2>5
      \fill[dot hidden] (0.75,0.5) circle (\dotsize);
      \fill[dot hidden] (0.25,0.5) circle (\dotsize);
    \fi
  \fi
\fi
\end{tikzpicture}%
}

\newtheorem{assumption}{Assumption}

\newcommand{\distDom}{X}

\newcommand{\distFunc}{\mu}
\newcommand{\distDomElem}{x}

\newcommand{\R}{\mathbb{R}}
\newcommand{\Q}{\mathbb{Q}}

\newcommand{\N}{\mathbb{N}}
\newcommand{\Ireal}{[0,1]\subseteq\mathbb{R}}  
\newcommand{\Irealo}{(0,1)\subseteq\mathbb{R}} 

\newcommand{\undefcustom}{\bot}

\newcommand{\pctlUntil}{\ensuremath{\mathcal{U}}}
\newcommand{\finally}{\lozenge}

\newcommand{\sinit}{s_{\mathit{I}}} 
\newcommand{\states}[1][]{S_{#1}}

\newcommand{\dtmc}{D}
\newcommand{\DtmcInit}[1][]{\ensuremath{\dtmc{#1}=(S{#1},\sinit{#1},\probdtmc{#1})}}

\newcommand{\successor}{\mathrm{succ}}
\newcommand{\predecessor}{\mathrm{pred}}
\newcommand{\sIn}{s_{1}}
\newcommand{\sOut}{s_{2}}

\newcommand{\mdp}{M}

\newcommand{\act}[1][a]{\alpha} 
\newcommand{\Act}{\mathit{Act}}        
\newcommand{\SchedSymbol}{\ensuremath{\mathit{Str}}}
\newcommand{\Sched}{\ensuremath{\SchedSymbol}}
\newcommand{\sched}{\ensuremath{\sigma}}
\newcommand{\altsched}{\ensuremath{\rho}}

\newcommand{\p}{\ensuremath{\mathbb{P}}}
\newcommand{\pr}{\ensuremath{\mathrm{Pr}}}
\newcommand{\reachPr}[2]{\ensuremath{\pr^{#1}(\finally #2)}}
\newcommand{\reachPrT}[1][]{\ensuremath{\reachPr{#1}{T}}}
\newcommand{\reachProp}[2]{\ensuremath{\p_{\leq #1}(\finally #2)}}
\newcommand{\reachProplT}{\ensuremath{\reachProp{\lambda}{T}}}
\newcommand{\reachPropneg}[2]{\ensuremath{\p_{> #1}(\finally #2)}}
\newcommand{\reachPropneglT}{\ensuremath{\reachPropneg{\lambda}{T}}}

\newcommand{\reachPrs}[3]{\ensuremath{\pr^{#1}_{#2}(\finally #3)}}
\newcommand{\rew}{\ensuremath{\mathrm{rew}}}

\newcommand{\er}{\ensuremath{\mathrm{ER}}}

\newcommand{\rewFunction}{\ensuremath{\mathrm{rew}}}
\newcommand{\e}{\ensuremath{\mathbb{E}}}

\newcommand{\mono}[1][]{\ensuremath{\mathit{Mon}[#1]}}
\newcommand{\poly}[1][]{\ensuremath{\mathbb{Q}[#1]}}
\newcommand{\ratfunc}[1][]{\ensuremath{\mathbb{Q}(#1)}}

\newcommand{\pol}{g}

\newcommand{\series}[2]{\ensuremath{#1_1,\ldots, #1_{#2}}}

\newcommand{\spOne}{\ensuremath{S_{{\text{\sl \tiny $\bigcirc$}}}}}
\newcommand{\schedOne}{\ensuremath{\sigma_{{\text{\sl \tiny $\bigcirc$}}}}}
\newcommand{\SchedOne}{\ensuremath{\Sched_{{\text{\sl \tiny $\bigcirc$}}}}}

\newcommand{\spTwo}{\ensuremath{S_{\Box}}}
\newcommand{\schedTwo}{\ensuremath{\sigma_{\Box}}}
\newcommand{\pOne}{\ensuremath{{\text{\sl \footnotesize $\bigcirc$}}}}
\newcommand{\pTwo}{\ensuremath{{\Box}}}
\newcommand{\probsg}{\probmdp}
\newcommand{\sg}{\ensuremath{G}}  

\newcommand{\probdtmc}{\probmdp}
\newcommand{\pdtmc}{\mathcal{\dtmc}}
\newcommand{\PdtmcInit}[1][]{\ensuremath{\pdtmc{#1}=(S{#1}, \Var{#1}, \sinit{#1},  \probdtmc{#1})}}
\newcommand{\probmdp}{\mathcal{P}}
\newcommand{\pmdp}{\mathcal{\mdp}}
\newcommand{\pmdpInit}[1][]{\ensuremath{\pmdp{#1}=(S{#1},\Var{#1},\sinit{#1},\Act{#1},\probmdp{#1})}}
\newcommand{\psg}{\mathcal{\sg}}
\newcommand{\psgInit}[1][]{\ensuremath{\psg{#1}=(S{#1}, \Var{#1}, \sinit{#1},\Act{#1},\probsg{#1})}}
\newcommand{\psgGeneric}{\ensuremath{\psg}}

\newcommand{\rel}[2][]{\ensuremath{\textsf{rel}_{#1}(#2)}}
\newcommand{\substAbbrev}{\ensuremath{\textsf{sub}}}
\newcommand{\subst}[2]{\ensuremath{\substAbbrev_{#2}(#1)}}
\newcommand{\substrew}[2]{\ensuremath{\substAbbrev^\rew_{#2}(#1)}}

\newcommand{\pathset}{\mathit{Paths}}
\newcommand{\paths}[1][]{\pathset^{#1}}
\newcommand{\pathsfin}[1][]{\pathset_{\mathit{fin}}^{#1}}

\renewcommand{\path}{\pi}

\newcommand{\false}{\ensuremath{\texttt{false}}\xspace}
\DeclareMathOperator{\dom}{dom}
\newcommand{\ie}{i.\,e.}
\newcommand{\eg}{e.g.\xspace}
\newcommand{\wrt}{w.\,r.\,t.\xspace}

\newcommand{\last}{\mathit{last}}
\newcommand{\Var}{\ensuremath{V}}                
\newcommand{\tool}[1]{\texttt{#1}\xspace}
\newcommand{\prophesy}{\tool{PROPhESY}}
\newcommand{\param}{\tool{PARAM}}
\newcommand{\prism}{\tool{PRISM}}
\newcommand{\prismpsy}{\tool{PRISM-PSY}}
\newcommand{\storm}{\tool{Storm}}
\newcommand{\modest}{\tool{Modest}}
\newcommand{\python}{\tool{python}}
\newcommand{\sylvan}{\tool{sylvan}}
\newcommand{\carl}{\tool{carl}}
\newcommand{\gmp}{\tool{gmp}}
\newcommand{\cln}{\tool{cln}}
\newcommand{\cocoa}{\tool{cocoa}}
\newcommand{\ginac}{\tool{ginac}}

\newcommand{\model}[1]{\textsf{#1}\xspace}

\DeclareRobustCommand{\cpp}
{\valign{\vfil\hbox{##}\vfil\cr
   \textsf{C\kern-.1em}\cr
   $\hbox{\fontsize{\sf@size}{0}\textbf{+\kern-0.05em+}}$\cr}%
}
\newcommand{\timeout}{\mbox{TO}}

\selectcolormodel{cmy}

\newcommand{\commentside}[2]{\marginpar{\color{#1}#2}}

\newif\ifdebug
\debugfalse

\ifdebug
 \newcommand{\nils}[1]{\color{red}#1\color{black}\xspace}
 
 \newcommand{\nj}[1]{\commentside{teal}{\tiny{N: #1} }}
 \newcommand{\sj}[1]{\commentside{red}{\tiny{SJ: #1} }}
 \newcommand{\cd}[1]{\commentside{blue}{\tiny{CD: #1} }}
 
 \newcommand{\mv}[1]{\commentside{green}{\tiny{MV: #1} }}

\else
 \newcommand{\nils}[1]{\color{black}#1\color{black}\xspace}
\newcommand{\nj}[1]{}
\newcommand{\sj}[1]{}
\newcommand{\fc}[1]{}
\newcommand{\cd}[1]{}
\newcommand{\mv}[1]{}
\fi

\tikzset{cross/.style={cross out, draw=black, minimum size=2*(#1-\pgflinewidth), inner sep=0pt, outer sep=0pt},
cross/.default={1pt}}

\usetikzlibrary{patterns}

\newcommand{\picscale}{1}
\newcommand{\distsonesthree}{1.5cm}
\newcommand{\distforsone}{1.5cm}

\newcommand{\hcent}[1]{\multicolumn{1}{c}{#1}}
\newcommand{\hcentl}[1]{\multicolumn{1}{c|}{#1}}

\smartqed  

\begin{document}

\title{Parameter Synthesis for Markov Models}
\subtitle{Covering the Parameter Space}



\author{Sebastian Junges \and Erika \'Abrah\'am \and Christian~Hensel \and Nils Jansen \and Joost-Pieter~Katoen \and Tim Quatmann \and Matthias Volk
\thanks{
The work has been partially supported by the DFG RTG 2236 UnRAVeL, the  European Union’s Horizon 2020 research and innovation programme under the Marie Skłodowska-Curie grant agreement No. 101008233 (Mission), and the ERC Starting Grant 101077178 (DEUCE).
}
}

\institute{
S.~Junges \at Radboud University, Nijmegen, The Netherlands
\and
N.~Jansen \at Ruhr-University, Bochum, Germany and Radboud University, Nijmegen, The Netherlands 
\and
E.~\'Abrah\'am, C.~Hensel, J.-P.~Katoen, T.~Quatmann \at Department of Computer Science, RWTH Aachen University, Aachen, Germany
\and
M.~Volk \at Eindhoven University of Technology, Eindhoven, The Netherlands
}



\maketitle
\begin{abstract}
Markov chain analysis is a key technique in formal verification. A
practical obstacle is that all probabilities in Markov models need to be
known. However, system quantities such as failure rates or packet loss ratios, etc. are often not---or only partially---known.
This motivates considering parametric models with transitions
labeled with functions over parameters.
Whereas traditional Markov chain analysis relies on a single, fixed set of probabilities, analysing parametric Markov models focuses on synthesising parameter values that establish a given safety or performance
specification $\varphi$. Examples are: what
component failure rates ensure the probability of a system breakdown to
be below 0.00000001?, or which failure rates maximise the performance, for instance the throughput, of the system?
This paper presents various analysis algorithms for parametric discrete-time Markov
chains and Markov decision processes. We focus on three problems: (a) do
all parameter values within a given region satisfy $\varphi$?, (b) which
regions satisfy $\varphi$ and which ones do not?, and (c) an approximate
version of (b) focusing on covering a large fraction of all possible
parameter values.
We give a detailed account of the various algorithms, present a software
tool realising these techniques, and report on an extensive experimental
evaluation on benchmarks that span a wide range of applications.
\keywords{Formal Methods \and Verification \and Model Checking \and Probabilistic Systems \and Parameter Synthesis \and Markov Chains}
\end{abstract}

\section{Introduction}

\paragraph{Uncertainty.}

Probabilistic model checking subsumes a multitude of formal verification techniques for systems that exhibit uncertainties~\cite{DBLP:conf/focs/CourcoubetisY88,katoen2016probabilistic,DBLP:reference/mc/BaierAFK18}.
Such systems are typically modeled by Markov chains or Markov decision processes~\cite{Put94}. 
Applications range from reliability, dependability and performance analysis to systems biology, take for instance reliability measures such as the mean time between failures in fault trees~\cite{RS15,Bozzano2010} and the probability of a system breakdown within a time limit.

The results of probabilistic model checking algorithms are rigorous, their quality depends solely on the system models.
Yet, there is one major practical obstacle: All probabilities (or rates) in the Markov model are precisely known \emph{a priori}.
In many cases, this assumption is too severe.
System quantities such as component fault rates, molecule reaction rates, packet loss ratios, etc.\ are often not, or at best partially, known. 
Let us give a few examples.
The quality of service of a (wireless) communication channel may be modelled by e.g., the popular Gilbert-Elliott model, a two-state Markov chain in which packet loss has an unknown probability depending on the channel's state~\cite{DBLP:journals/tit/MushkinB89}.
Other examples include the back-off probability in CSMA/CA protocols determining a node's delay before attempting a transmission~\cite{ieee802-11}, the bias of used coins in self-stabilising protocols~\cite{DBLP:journals/ipl/Herman90,DBLP:journals/fac/KwiatkowskaNP12}, and the randomised choice of selecting the type of time-slots (sleeping, transmit, or idle) in the birthday protocol, a key mechanism used for neighbour discovery in wireless sensor networks~\cite{DBLP:conf/mobihoc/McGlynnB01} to lower power consumption. In particular, in early stages of reliable system design, the concrete failure rate of components~\cite{Cousineau2009} is left unspecified.
Optimally, analyses in this stage may even guide the choice of a concrete component from a particular manufacturer.

The probabilities in all these systems are deliberately left unspecified.
They can later be determined in order to optimise some performance or dependability measure. Dually, some systems should be robust for all (reasonable)  failure rates. For example, a network protocol should ensure a reasonable quality of service for each reasonable channel quality.

\paragraph{Parametric probabilistic models.}
What do these examples have in common?
The random variables for packet loss, failure rate etc.\ are not fully defined, but are \emph{parametric}.
Whether a parametric system satisfies a given property or not---``is the probability that the system goes down within $k$ steps below $10^{{-}8}$''---depends on these parameters.
Relevant questions are then: for which concrete parameter values is such a property satisfied---the \emph{(parameter) synthesis problem}---and, in case of decision-making models, which parameter values yield optimal designs?
That is, for which fixed probabilities do such protocols work in an optimal way, i.e., lead to maximal reliability, maximise the probability for nodes to be discovered, or minimise the time until stabilisation, and so on.
These questions are intrinsically hard as parameters can take infinitely many different values that, in addition, can depend on each other.

\emph{This paper} faces these challenges and \emph{presents various algorithmic techniques to treat different variations of the (optimal) parameter synthesis problem}.
To deal with uncertainties in randomness, \emph{parametric} probabilistic models are adequate.
These models are just like Markov models except that the transition probabilities are specified by arithmetic expressions over real-valued parameters. 
Transition probabilities are thus functions over a set of parameters.
A simple instance is to use intervals over system parameters imposing constant lower and upper bounds on every parameter~\cite{DBLP:journals/rc/KozineU02,DBLP:journals/ai/GivanLD00}.
The general setting as considered here is more liberal as it e.g., includes the possibility to express complex parameter dependencies.
We address the analysis of \emph{parametric} Markov models where probability distributions are functions over system parameters, specifically, parametric discrete-time Markov \emph{chains} (pMCs) and parametric discrete-time Markov \emph{decision processes} (pMDPs).
\begin{figure}[t]
\centering
   \subfigure[Unfair coins]{
\scalebox{1}{
%
%
%
%
\begin{tikzpicture}[scale=1, die/.style={inner sep=0,outer sep=0},nodestyle/.style={draw,circle},baseline=(s0)]
    
    \node [nodestyle,fill=gray!50] (s0) at (0,0) {$s_0$};
    \node [] (leftdummy)  [on grid, left=1.2cm of s0] {};
    \node [] (rightdummy) [on grid, right=1.2cm of s0] {};
    \node [nodestyle] (s1) [on grid, below=1.1cm of leftdummy] {$s_1$};
    \node [nodestyle] (s2) [on grid, below=1.1cm of rightdummy] {$s_2$};
    \node [nodestyle,fill=gray!50] (s3) [on grid, below=1.3cm of s1, xshift=-0.5cm] {$s_3$};
        \node [nodestyle,fill=gray!50] (s4) [on grid, below=1.3cm of s1, xshift=0.5cm] {$s_4$};
    \node [nodestyle,fill=gray!50] (s5) [on grid, below=1.3cm of s2, xshift=-0.5cm] {$s_5$};
    
    \node [nodestyle,fill=gray!50] (s6) [on grid, below=1.3cm of s2, xshift=0.5cm] {$s_6$};

    \node[die, scale=1.5, below=0.6cm of s3, xshift=-0.3cm] (X1) {\drawdie{1}};
    \node[die, scale=1.5, right=0.31cm of X1, inner sep=0pt] (X2) {\drawdie{2}};
    \node[die, scale=1.5, right=0.31cm of X2] (X3) {\drawdie{3}};
    \node[die, scale=1.5, right=0.31cm of X3] (X4) {\drawdie{4}};
    \node[die, scale=1.5, right=0.31cm of X4] (X5) {\drawdie{5}};
    \node[die, scale=1.5, right=0.31cm of X5] (X6) {\drawdie{6}};
    
    \draw ($(s0)-(0.7,0)$) edge[->] (s0);
    \draw (s0) edge[->] node[right] {\scriptsize$\nicefrac{2}{5}$} (s1);
    \draw (s0) edge[->] node[right] {\scriptsize$\nicefrac{3}{5}$} (s2);
    \draw (s1) edge[bend left, ->] node[left] {\scriptsize$\nicefrac{7}{10}$} (s3);
    \draw (s1) edge[->] node[right] {\scriptsize$\nicefrac{3}{10}$} (s4);
    \draw (s3) edge[bend left, ->] node[left] {\scriptsize$\nicefrac{2}{5}$} (s1);
    \draw (s3) edge[->] node[left] {\scriptsize$\nicefrac{3}{5}$} (X1);
    \draw (s4) edge[->] node[left] {\scriptsize$\nicefrac{3}{5}$} (X2);
    \draw (s4) edge[->] node[right] {\scriptsize$\nicefrac{2}{5}$} (X3);
    \draw (s2) edge[bend left, ->] node[left] {\scriptsize$\nicefrac{7}{10}$} (s5);
    \draw (s2) edge[->] node[right] {\scriptsize$\nicefrac{3}{10}$} (s6);
    \draw (s5) edge[bend left, ->] node[left] {\scriptsize$\nicefrac{2}{5}$} (s2);
    \draw (s5) edge[->] node[left] {\scriptsize$\nicefrac{3}{5}$} (X4);
    \draw (s6) edge[->] node[left] {\scriptsize$\nicefrac{3}{5}$} (X5);
    \draw (s6) edge[->] node[right] {\scriptsize$\nicefrac{2}{5}$} (X6);
    
    \node[draw=white, rectangle, fit=(current bounding box)] {};
\end{tikzpicture}

	\label{fig:pkydiei}
		}
	} 
  \subfigure[Parametric probabilities]{
		\scalebox{1}{
%
%
%

\begin{tikzpicture}[scale=1, die/.style={inner sep=0,outer sep=0},nodestyle/.style={draw,circle},baseline=(s0)]
    
    \node [nodestyle,fill=gray!50] (s0) at (0,0) {$s_0$};
    \node [] (leftdummy)  [on grid, left=1.2cm of s0] {};
    \node [] (rightdummy) [on grid, right=1.2cm of s0] {};
    \node [nodestyle] (s1) [on grid, below=1.1cm of leftdummy] {$s_1$};
    \node [nodestyle] (s2) [on grid, below=1.1cm of rightdummy] {$s_2$};
    \node [nodestyle,fill=gray!50] (s3) [on grid, below=1.3cm of s1, xshift=-0.5cm] {$s_3$};
        \node [nodestyle,fill=gray!50] (s4) [on grid, below=1.3cm of s1, xshift=0.5cm] {$s_4$};
    \node [nodestyle,fill=gray!50] (s5) [on grid, below=1.3cm of s2, xshift=-0.5cm] {$s_5$};
    .
     \node [nodestyle,fill=gray!50] (s6) [on grid, below=1.3cm of s2, xshift=0.5cm] {$s_6$};

    \node[die, scale=1.5, below=0.6cm of s3, xshift=-0.3cm] (X1) {\drawdie{1}};
    \node[die, scale=1.5, right=0.31cm of X1, inner sep=0pt] (X2) {\drawdie{2}};
    \node[die, scale=1.5, right=0.31cm of X2] (X3) {\drawdie{3}};
    \node[die, scale=1.5, right=0.31cm of X3] (X4) {\drawdie{4}};
    \node[die, scale=1.5, right=0.31cm of X4] (X5) {\drawdie{5}};
    \node[die, scale=1.5, right=0.31cm of X5] (X6) {\drawdie{6}};
    
    \draw ($(s0)-(0.7,0)$) edge[->] (s0);
    \draw (s0) edge[->] node[right] {\scriptsize$p$} (s1);
    \draw (s0) edge[->] node[right] {\scriptsize$1{-}p$} (s2);
    \draw (s1) edge[bend left, ->] node[left] {\scriptsize$q$} (s3);
    \draw (s1) edge[->] node[right] {\scriptsize$1{-}q$} (s4);
    \draw (s3) edge[bend left, ->] node[left] {\scriptsize$p$} (s1);
    \draw (s3) edge[->] node[left] {\scriptsize$1{-}p$} (X1);
    \draw (s4) edge[->] node[left] {\scriptsize$1{-}p$} (X2);
    \draw (s4) edge[->] node[right] {\scriptsize$p$} (X3);
    \draw (s2) edge[bend left, ->] node[left] {\scriptsize$q$} (s5);
    \draw (s2) edge[->] node[right] {\scriptsize$1{-}q$} (s6);
    \draw (s5) edge[bend left, ->] node[left] {\scriptsize$p$} (s2);
    \draw (s5) edge[->] node[left] {\scriptsize$1{-}p$} (X4);
    \draw (s6) edge[->] node[left] {\scriptsize$1{-}p$} (X5);
    \draw (s6) edge[->] node[right] {\scriptsize$p$} (X6);
    
    \node[draw=white, rectangle, fit=(current bounding box)] {};
    
\end{tikzpicture}
	\label{fig:pkydie}
		}
	}
\caption{A (a) biased and (b) parametric variant of Knuth-Yao's algorithm. In gray states an unfair coin is flipped with probability $\nicefrac{2}{5}$ for `heads'; for the unfair coin in the white states this probability equals $\nicefrac{7}{10}$.
On the right, the two biased coins have parametric probabilities.}
\end{figure}
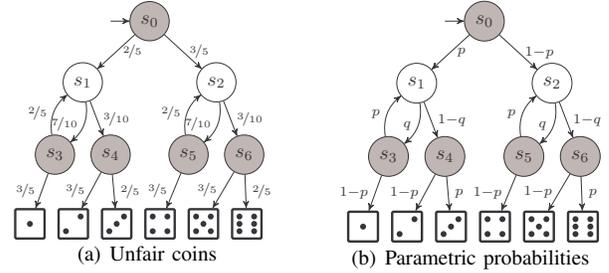
\begin{example}\label{ex:kydie}
The Knuth-Yao randomised algorithm~\cite{KY76} uses repeated coin flips to model a six-sided die.
It uses a fair coin to obtain each possible outcome (`one', `two', ..., `six') with probability $\nicefrac{1}{6}$.
Figure~\vref{fig:pkydiei} depicts a Markov chain (MC) of a variant in which two \emph{unfair} coins are flipped in an alternating fashion.
Flipping the unfair coins yields \emph{heads} with probability $\nicefrac{2}{5}$ (gray states) or $\nicefrac{7}{10}$ (white states), respectively.
Accordingly, the probability of \emph{tails} is $\nicefrac{3}{5}$ and $\nicefrac{3}{10}$, respectively.
The event of throwing a `two' corresponds to reaching the state~\drawdie[scale=0.7]{2} in the MC.
Assume now a \emph{specification} that requires the probability to obtain `two' to be larger than $\nicefrac{3}{20}$. 
Knuth-Yao 's original algorithm accepts this specification as using a fair coin results in $\nicefrac{1}{6}$ as probability to end up in \drawdie[scale=0.7]{2}.
The biased model, however, does not satisfy the specification; in fact, a `two' is reached with probability $\nicefrac{1}{10}$. 
\end{example}

\paragraph{Probabilistic model checking.}
The analysis algorithms presented in this paper are strongly related to (and presented as) techniques from probabilistic model checking.
\emph{Model checking}~\cite{BK08,clarke1999model} is a popular approach to verify the correctness of a system by systematically evaluating all possible system runs. 
It either certifies the absence of undesirable (dangerous) behaviour or delivers a system run witnessing a violating system behaviour.
Traditional model checking typically takes two inputs: a finite transition system modelling the system at hand and a temporal logic formula specifying a system requirement.
Model checking then amounts to checking whether the transition system satisfies the logical specification, which in its simplest form describes that a particular state can (not) be reached. 
Model checking is nowadays a successful analysis technique adopted by mainstream hardware and software industry~\cite{DBLP:conf/cav/Cook18,DBLP:reference/mc/Kurshan18}.

To cope with real-world systems exhibiting random behaviour, model checking has been extended to deal with probabilistic, typically Markov, models.
\emph{Probabilistic model checking}~\cite{BK08,katoen2016probabilistic,DBLP:reference/mc/BaierAFK18} takes as input a Markov model of the system at hand together with a quantitative specification specified in some probabilistic extension of LTL or CTL.
Example specifications are e.g., ``is the probability to reach some bad (or degraded) state below a safety threshold $\lambda$?'' or
``is the expected time until the system recovers from a fault bounded by some threshold $\kappa$''. 
Efficient probabilistic model-checking techniques do exist for models such as discrete-time Markov chains (MCs), Markov decision processes (MDPs), and their continuous-time counterparts~\cite{katoen2016probabilistic}. 
Probabilistic model checking extends and complements long-standing analysis techniques for Markov models. 

It has been adopted in the field of performance analysis to analyse stochastic Petri nets~\cite{DBLP:conf/qest/CerottiDHS06,DBLP:conf/apn/AmparoreBD14}, in dependability analysis for analysing architectural system descriptions~\cite{DBLP:journals/ress/BozzanoCKKMNNPR14}, in reliability engineering for fault tree analysis~\cite{DBLP:journals/tdsc/BoudaliCS10,DBLP:journals/tii/VolkJK18}, as well as in security~\cite{DBLP:journals/jcs/NormanS06}, distributed computing~\cite{DBLP:journals/fac/KwiatkowskaNP12}, and systems biology~\cite{DBLP:journals/sigmetrics/KwiatkowskaNP08}.
Unremitting algorithmic improvements employing the use of symbolic techniques to deal with large state spaces have led to powerful and popular software tools realising probabilistic model checking techniques such as \prism~\cite{KNP11} and \storm~\cite{DBLP:conf/cav/DehnertJK017}.

\subsection{Problem statements}
We now give a more detailed description of the parameter synthesis problems considered in this paper.
We start off by establishing the connection between parametric Markov models and concrete ones, i.e., ones in which the probabilities are fixed such as MCs and MDPs.
Each parameter in a pMC or pMDP (where p stands for parametric) has a given parameter range. 
The \emph{parameter space} of the parametric model is the Cartesian product of these parameter ranges.
Instantiating the parameters with a concrete value in the parameter space to the parametric model results in an \emph{instantiated} model.
The parameter space defines all possible parameter instantiations, or equivalently, the instantiated models.
A parameter instantiation that yields a Markov model, e.g., results in probability distributions, is called \emph{well-defined}.
In general, a parametric Markov model defines an uncountably infinite \emph{family of Markov models}, where each family member is obtained by a well-defined instantiation. 
A \emph{region} $R$ is a fragment of the parameter space; it is well-defined if all instantiations in $R$ are well-defined.
\begin{example}[pMC]\label{ex:kydie_pMC}
Figure~\vref{fig:pkydie} depicts a parametric version of the biased Knuth-Yao die from Example~\vref{ex:kydie}.
It has parameters $\Var = \{p,q\}$, where $p$ is the probability of outcome \emph{heads} in gray states and $q$ the same for white states. 
The parameter space is $\{ (p,q) \mid 0 < p,q < 1 \}$.
The probability for \emph{tails} is $1{-}p$ and $1{-}q$, respectively.
The sample instantiation $u$ with $u(p) = \nicefrac{2}{5}$ and $u(q) = \nicefrac{7}{10}$ is well-defined and results in the MC in Figure~\vref{fig:pkydiei}.
The region \[R= \{ u \colon \Var \to \R \mid \nicefrac{1}{10} \leq u(p) \leq \nicefrac{9}{10} \text{ and } \nicefrac{3}{4} \leq u(q)\leq \nicefrac{5}{6} \}\] is well-defined.
Contrarily, region 
\[R' = \{ u \mid \nicefrac{1}{5} \leq u(p) \leq \nicefrac{6}{5} \text { and } \nicefrac{2}{5} \leq u(q) \leq \nicefrac{7}{10} \}\] is not well-defined, as it contains the instantiation $u'$ with $u'(p) = \nicefrac{6}{5}$ which does not yield an MC. For pMCs whose transition probabilities are high-degree polynomials, it is not always obvious whether a region is well-defined.
\end{example}
We are now in a position to describe the three  problems considered in this paper.
\paragraph{The verification problem is defined as follows:}

\begin{center}
\smallskip\noindent\fcolorbox{black}{black!10}{%
    \parbox{0.95\columnwidth}
    {\textbf{The verification problem.}\label{informal_verification}
    Given a parametric Markov model $\pdtmc$, a well-defined region $R$, and a specification $\varphi$, the \emph{verification problem} is to check whether \emph{all} instantiations of $\pdtmc$ within $R$ satisfy $\varphi$.
}}
\end{center}
\noindent 
Consider the following possible outcomes:
\begin{itemize}
\item 
If $R$ only contains instantiations of $\pdtmc$ satisfying $\varphi$, then the verification problem evaluates to \texttt{true} and the Markov model $\pdtmc$ on region $R$ \emph{accepts} specification $\varphi$. 
Whenever $\pdtmc$ and $\varphi$ are clear from the context, we call $R$ \emph{accepting}.
\item 
If $R$ contains an instantiation of $\pdtmc$ refuting $\varphi$, then the problem evaluates to \texttt{false}.
If $R$ contains only instantiations of $\pdtmc$ refuting $\varphi$, then $\pdtmc$ on $R$ \emph{rejects} $\varphi$. 
Whenever $\pdtmc$ and $\varphi$ are clear from the context, we call $R$ \emph{rejecting}.
\item 
If $R$ contains instantiations satisfying $\varphi$ as well as instantiations satisfying $\neg \varphi$, then $\pdtmc$ on $R$ is inconclusive \wrt\ $\varphi$.
In this case, we call $R$ \emph{inconsistent}.
\end{itemize}
In case the verification problem yields \false for $\varphi$, one can only infer that the region $R$ is not accepting, but not conclude whether $R$ is inconsistent or rejecting. 
To determine whether $R$ is rejecting, we need to consider the verification problem for the negated specification $\neg\varphi$.
Inconsistent regions for $\varphi$ are also inconsistent for~$\neg\varphi$.
\begin{example}[Verification problem]\label{ex:verif}
Consider the pMC $\pdtmc$, the well-defined region $R$ from Example~\vref{ex:kydie_pMC}, and the specification $\varphi' := \neg \varphi$ that constrains the probability to reach \drawdie[scale=0.7]{2} to be at most $\nicefrac{3}{20}$.
The verification problem is to determine whether all instantiations of $\pdtmc$ in $R$ satisfy $\varphi'$. 
As there is no instantiation within $R$ for which the probability to reach \drawdie[scale=0.7]{2} is above $\nicefrac{3}{20}$, the verification problem evaluates to \texttt{true}.
Thus, $R$ \emph{accepts}~$\varphi'$.
\end{example}
Typical structurally simple regions are described by hyperrectangles or given by linear constraints, rather than non-linear constraints; we refer to such regions as \emph{simple}.
A simple region comprising a large range of parameter values may likely be inconsistent, as it contains both instantiations satisfying $\varphi$, and some satisfying $\neg\varphi$.
Thus, we generalise the problem to synthesise a partition of the parameter space.

\paragraph{The exact synthesis problem is described as follows:}
\begin{center}
\smallskip\noindent\fcolorbox{black}{black!10}{%
    \parbox{0.95\columnwidth}
    {\textbf{The synthesis problem.}
Given a parametric Markov model $\pdtmc$ and a specification $\varphi$, the \emph{(parameter) synthesis problem} is to partition the parameter space of $\pdtmc$ into an accepting region $R_a$ and a rejecting region $R_r$ for $\varphi$.
}}
\end{center}
The aim is to obtain such a partition in an automated manner.
A complete sub-division of the parameter space into accepting and rejecting regions provides deep insight into the effect of parameter values on the system's behaviour.
The exact division typically is described by non-linear functions over the parameters, referred to as \emph{solution functions}.
%
\begin{example}\label{ex:solution_function_synth}
Consider the pMC $\pdtmc$, the region $R$, and the specification $\varphi$ as in Example~\vref{ex:verif}. 
The solution function:
\[ f_{\varphi}(p, q) = \frac{p \cdot (1-q) \cdot (1-p)}{1-p\cdot q} \]
describes the probability to eventually reach \drawdie[scale=0.7]{2}.
Given that $\varphi$ imposes a lower bound of $\nicefrac{3}{20}$, we obtain \[ R_a = \{ u \mid f(u(p), u(q)) \geq \nicefrac{3}{20} \}\text{ and }R_r = R \setminus R_a.\]
\end{example}
The example illustrates that exact symbolic representations of the accepting and rejecting regions may be complex and hard to compute algorithmically.
The primary reason is that the boundaries are described by non-linear functions.
A viable alternative therefore is to consider an approximative version of the synthesis problem.

\begin{figure}[t]
\centering
\begin{center}
\includegraphics[scale=0.5]{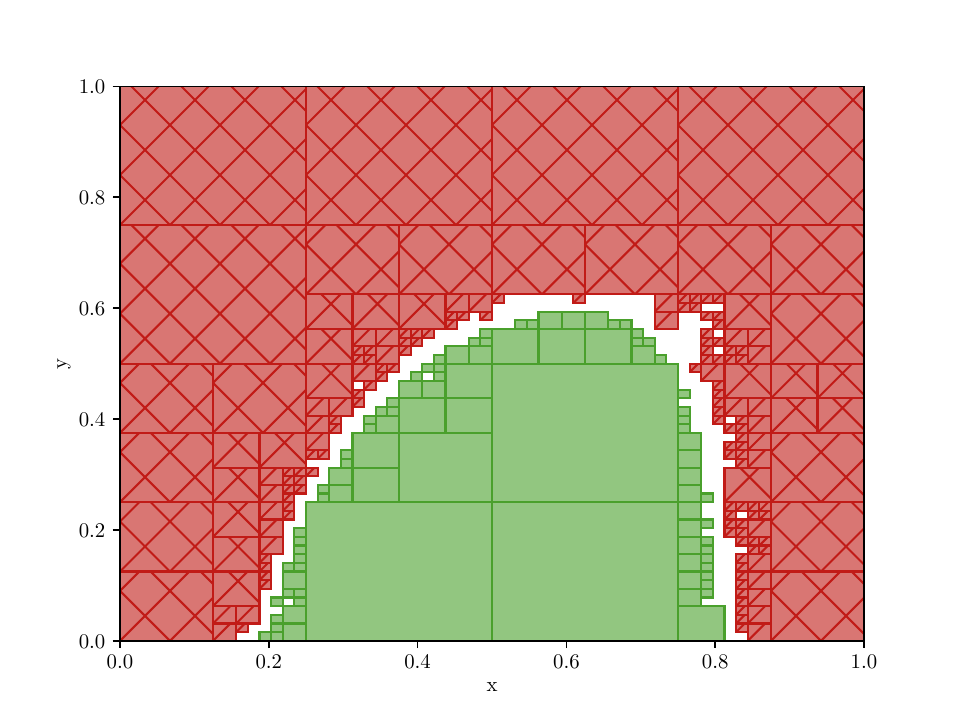}
	\end{center}
 \caption{Parameter space partitioning into accepting (green), rejecting (red), and unknown (white) regions.}
 \label{fig:regions_die}
\end{figure}

\paragraph{The approximate synthesis problem:}
As argued before, the regions obtained via exact synthesis are typically not simple.
The aim of the approximate synthesis problem is to use simpler and more tractable representations of regions. 
As such shapes ultimately approximate the exact solution function, simple regions become infinitesimally small when getting close to the border between accepting and rejecting areas.
For computational tractability, we are thus interested in \emph{approximating} a partition of the parameter space in accepting and rejecting regions, where we allow also for a (typically small) part to be covered by possibly inconsistent regions.
Practically this means that $c\,\%$ of the entire parameter space is covered by simple regions that are either accepting or rejecting, for some adequate value of $c$.
Altogether this results in the following problem description:
\begin{center}
\smallskip\noindent\fcolorbox{black}{black!10}{%
    \parbox{0.95\columnwidth}
    {\textbf{The approximate synthesis problem.}
Given a parametric Markov model, a specification $\varphi$, and a percentage $c$, the \emph{approximate (parameter) synthesis problem} is to partition the parameter space of $\pdtmc$ into a simple accepting region $R_a$ and a simple rejecting region $R_r$ for $\varphi$ such that $R_a \cup R_r$ cover at least $c$\% of the entire parameter space.
}}
\end{center}
\begin{example}
\label{ex:psp}
Consider the pMC $\pdtmc$, the region $R$, and the specification $\varphi$ as in Example~\vref{ex:verif}. 
The parameter space in Figure~\vref{fig:regions_die} is partitioned into simple regions (rectangles).
The green (dotted) area---the union of a number of smaller rectangular accepting regions---indicates the parameter values for which $\varphi$ is satisfied, whereas the red (hatched) area indicates the set of rejecting regions for $\varphi$.
The white area indicates the unknown regions.
The indicated partition covers 95\% of the parameter space.
The sub-division into accepting and rejecting (simple) regions approximates the solution function $f_{\varphi}(p,q)$ given before.
\end{example}

%

%
\subsection{Solution approaches}
We now outline our approaches to solve the verification problem and the two synthesis problems.
For the sake of convenience, we start with the synthesis problem.

\paragraph{Synthesis.}
The most straightforward description of the sets $R_a$ and $R_r$ is of the form: 
\begin{align*}
	R_a &= \{ u \mid \pdtmc[u] \text{ satisfies } \varphi \} \quad \mbox{ and} \\
	R_r &= \{ u \mid \pdtmc[u] \text{ satisfies } \neg\varphi \}.
\end{align*}
The satisfaction relation (denoted $\models$) can be concisely described by a set of linear equations over the transition probabilities~\cite{BK08}.
As in the parametric setting the transition probabilities are no longer fixed, but rather defined over a set of parameters, the equations become non-linear.
\begin{example}[Non-linear equations for reachability]\label{ex:nonlinear}
Take the MC from Figure~\vref{fig:pkydiei}.
To compute the probability of eventually reaching, e.g., state~\drawdie[scale=0.7]{2}, one introduces a variable $p_s$ for each transient state $s$ encoding that probability for $s$. 
For state $s_0$ and variable $p_{s_0}$, the corresponding linear equation reads:
	\begin{align*}
		p_{s_0} = \nicefrac{2}{5}\cdot p_{s_1} + \nicefrac{3}{5}\cdot p_{s_2},
	\end{align*}
	where $p_{s_1}$ and $p_{s_2}$ are the variables for $s_1$ and $s_2$, respectively.
	
The corresponding equation for the pMC from Figure~\vref{fig:pkydie} reads:
	\begin{align*}
		p_{s_0} = p\cdot p_{s_1} + (1-p)\cdot p_{s_2}.
	\end{align*}
	The multiplication of parameters in the model and equation variables leads to a \emph{non-linear} equation system.
\end{example}
Thus, we can describe the sets $R_a$ and $R_r$ colloquially as: 
\begin{align*}
	R_a, R_r &= \{ u \mid u \text{ satisfies a set of non-linear constraints} \}. 
\end{align*}
We provide further details on these constraint systems in Section~\ref{sec:exact_mc}.

A practical drawback of the resulting equation system is the substantial number of auxiliary variables $p_s$, one for each state in the pMC.
A viable possibility for pMCs is to simplify the equations by (variants of) \emph{state elimination}~\cite{Daws04}.
This procedure successively removes states from the pMC until only a start and final state (representing the reachability objective) remain that are connected by a transition whose label is (a mild variant of) the solution function $f_\varphi$ that \emph{exactly} describes the probability to reach a target state:
\[
	R_a = \{ u \mid f_\varphi(u) > 0 \} 
	\quad  \mbox{and}\quad
	R_r = \{ u \mid f_{\neg\varphi}(u) > 0 \}. 
\]
We recapitulate  state elimination and present several alternatives in Section~\ref{sec:solution_fct}.
\paragraph{Verification.}
The basic approach to the verification problem is depicted in Figure~\ref{fig:verifoutline}.
We use a description of the accepting region as computed via the synthesis procedure above. 
Then, we combine the description of the accepting region with the region $R$ to be verified, as follows:
\begin{figure}
\centering
\scalebox{1}{
\begin{tikzpicture}
	\node[draw,rectangle, rounded corners, inner sep=5pt, text width=4cm] (synt) {\textbf{synthesise} description of:\\ accepting region $R_a$, and\\rejecting region $R_r$};
   		        
	\node[above=0.7cm of synt, shape=ellipse, draw, text width=3.5cm, text centered] (model) {parametric MDP $\pmdp$,\\ parameter space $R$,\\ specification $\varphi$};
	
	\node[rectangle, draw, right=of synt, rounded corners, inner sep=5pt, text centered, text width=4cm] (join) {\textbf{check} $R \land R_a$  unsatisfiable,\\
	\textbf{check} $R \land R_r$  unsatisfiable };
	\node[below=0.7cm of join,shape=ellipse,draw,text centered,text width=3.5cm] (outcome) {yes for $R_a$ $\rightarrow$ reject,\\yes for $R_r$ $\rightarrow$ accept,\\otherwise $\rightarrow$ unknown};
	\draw[->] (model) -- (synt);
	\draw[->] (model) -- (join);
	
	\draw[->] (synt) -- (join);
	\draw[->] (join) -- (outcome);
\end{tikzpicture}
}
\caption{Verification via exact synthesis}	
\label{fig:verifoutline}
\end{figure}
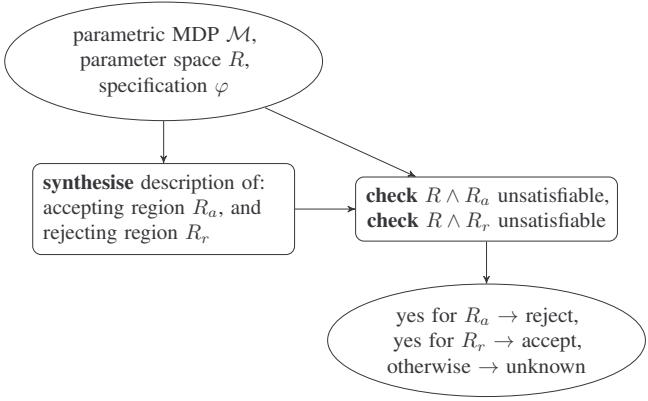
A region $R$ accepts a specification, if $R \cap R_a = R$, or equivalently, if $R \cap R_r = \emptyset$.
The existence of a rejecting instance in $R$ is thus of relevance; if such a point does not exist, the region is accepting.
Using $R_a$ and $R_r$ as obtained above, the query ``is $R \cap R_r = \emptyset$?'' can be solved via \emph{satisfiability modulo theories} (SMT) over non-linear arithmetic, checking the conjunction over the corresponding constraints for unsatisfiability.
With the help of SMT solvers over this theory like \tool{Z3}~\cite{demoura_nlsat}, \tool{MathSAT}~\cite{mathsat4}, or \tool{SMT-RAT}~\cite{DBLP:conf/sat/CorziliusKJSA15}, this can be solved in a fully automated manner.
This procedure is complete, and is computationally involved.
Details of the procedure are discussed in Section~\ref{sec:exact_mc}.


\emph{Parameter lifting}~\cite{QDJJK16} is an alternative, \emph{approximative} solution to the verification problem.
Intuitively, this approach over-approximates $R_r$ for a given $R$, by ignoring parameter dependencies. 
Region $R$ is accepted if the intersection with the over-approximation of $R_r$ is empty. 
This procedure is sound but may yield false negatives as a rejecting point may lie in the over-approximation but not in $R_r$.
Tightening the over-approximation makes the approach complete.
A major benefit of parameter lifting (details in Section~\ref{sec:approx_mc} and Section~\ref{sec:nondet}) is that the intersection with the over-approximation of $R_r$ can be investigated by standard probabilistic model-checking procedures.
This applicability of mature tools results---as will be shown in Section~\ref{sec:eval}---in a practically efficient procedure.

\paragraph{Approximate synthesis.} 
We solve the approximate synthesis problem with an iterative synthesis loop.
Here, the central issue is to obtain representations of $R_a$ and $R_r$ by simple regions.
Our approach for this \emph{parameter space partitioning} therefore iteratively obtains partial partitions of the parameter space.
The main idea is to compute a sequence $\left( R^i_a \right)_i$ of simple accepting regions that successively extend each other.
Similarly, an increasing sequence $\left( R^i_r \right)_i$ of simple rejecting regions is computed.
The typical approach is to let $R^{i+1}_a$ be the union of $R^i_a$, the approximations in the previous iteration, together with some accepting region with a simple representation. 
Rejecting regions are handled analogously.
At the $i$-th iteration, $R^i_a \cup R^i_r$ is the covered fragment of the parameter space.
The iterative approach halts when this fragment forms at least $c\,\%$ of the entire parameter space.
Termination is guaranteed. In the limit, the accepting and rejecting regions converge to the exact solution,  $\lim_{i \rightarrow \infty} R_a^i = R_a$ and $\lim_{i \rightarrow \infty} R_r^i = R_r$, under some mild constraints on the ordering of the regions $R^i$.

\begin{figure}[t]
\centering
\scalebox{1}{\begin{tikzpicture}
			\tikzstyle{outer}= [draw, text centered, shape=ellipse, minimum height=0.4cm, text width=3.5cm]
    			\tikzstyle{inner}=[draw, text centered, shape=rectangle, rounded corners, text width=3.5cm, minimum height=1.1cm, inner sep=5pt]

		    	\node[inner] (guess) at (0,0) {\textbf{refine} undecided region: \\ \textbf{guess} candidate};
		    	\node[inner, right=1cm of guess] (verification) {\textbf{verification}: either \\$\pmdp, R' \models \varphi$ (accept) or $\pmdp, R' \models \neg\varphi$ (reject)};
		    	
   		        \node[outer, above=1.9cm of guess] (model) {parametric MDP $\pmdp$,\\ parameter space $R$,\\ specification $\varphi$};

   		        \node[outer, below=1.6cm of verification] (partition) {accepting/rejecting regions};

                \node (bb) [use as bounding box, inner sep=0.2cm] {};

                \path[->,thick]
                    (model) edge[] (guess)
                    (guess) edge[bend left] node[above, text centered, text width=3.5cm] (vtext) {region $R'$\\ and hypothesis $\varphi$ or $\neg\varphi$} (verification)
		    (verification) edge[bend left] node[auto] (gtext) {not \emph{accepted}/\emph{rejected}} (guess)
                    (verification) edge[] node[auto] (ptext) {\emph{accepted}/\emph{rejected}} (partition)
                ;

                \node (box) [draw, dashed, fit = (guess) (verification) (gtext) (ptext) (vtext), inner sep = 0.2cm] {};

\end{tikzpicture}}
\caption{Approximate synthesis process using verification as black box.}	
\label{fig:problemrelation}
\end{figure}
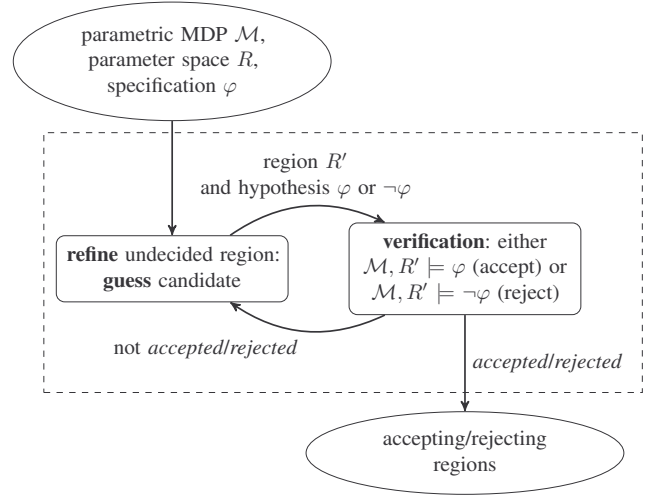

Figure~\ref{fig:problemrelation} outlines a procedure to address the approximate synthesis problem.
As part of our synthesis method, we algorithmically \emph{guess} a (candidate) region $R$ and guess whether it is accepting or rejecting. 
We then exploit one of our verification methods to verify whether $R$ is indeed accepting (or rejecting).
If it is not accepting (rejecting), we exploit this information together with any additional information obtained during verification to refine the candidate region. 
This process is repeated until an accepting or rejecting region results.
We discuss the method and essential improvements in Section~\ref{sec:psp}.
\begin{example}
Consider the pMC $\pdtmc$ and the specification $\varphi$ as in Example~\vref{ex:kydie_pMC}. 
The parameter space in Figure~\vref{fig:regions_die} is partitioned into regions.
The green (dotted) area---the union of a number of smaller rectangular accepting regions---indicates the parameter values for which $\varphi$ is satisfied, whereas the red (hatched) area indicates the set of rejecting regions for $\varphi$.
Checking whether a region is accepting, rejecting, or inconsistent is done by verification.
The small white area consists of regions that are \emph{unknown} (i.e., not yet considered) or inconsistent.
\end{example}

\subsection{Overview of the paper}

Section~\ref{sec:preliminaries} introduces the required formalisms and concepts. 
Section~\ref{sec:regionsMC} defines the notion of a region and formalises the three problems: the verification problem and the two synthesis problems. 
It ends with a bird's eye view of the verification approaches that are later discussed in detail.
Section~\ref{sec:regions} details specific region structures and procedures to check elementary region properties such as well-definedness and graph-preservedness, two prerequisites for the verification procedures. 
Section~\ref{sec:solution_fct} shows how to do exact synthesis by computing the solution function.
Sections~\ref{sec:exact_mc}--\ref{sec:nondet} present algorithms for the verification problem. 
Section~\ref{sec:psp} details the approach to reduce the synthesis problem to a series of verification problems.
Sections~\ref{sec:impl} and \ref{sec:eval} contain information about the implementation of the approaches, as well 
as an extensive experimental evaluation.
Section~\ref{sec:discussion} contains a discussion of the approaches and related work. 
Section~\ref{sec:conclusion} concludes with an outlook. 
\subsection{Contributions of this paper}
The paper is loosely based on the conference papers~\cite{DJJ+15} and~\cite{QDJJK16} and extends these works in the following ways.
It gives a uniform treatment of the solution techniques to the synthesis problem, and treats all techniques uniformly for all different objectives---bounded and unbounded reachability as well as expected reward specifications. 
The material on SMT-based region verification has been extended in the following way:
The paper gives the complete characterisations of the SMT encoding with or without solution function. Furthermore, it is the first to extend this encoding to MDPs under angelic and demonic non-determinism and includes an explicit and in-depth discussion on exact region checking via SMT checkers. 
It presents a uniform treatment of the linear equation system for Markov chains and its relation to state elimination and Gaussian elimination. 
It presents a novel and simplified description of state elimination for expected rewards, and a version of state elimination that is targeted towards MTBDDs.
The paper contains a correctness proof of approximate verification for a wider range of pMDPs and contains proofs for expected rewards.
It also supports expected-time properties for parametric continuous-time MDPs (via the embedded pMDP).
Novel heuristics have been developed to improve the iterative synthesis loop.
All presented techniques, models, and specifications are realised in the state-of-the-art tool \prophesy\footnote{\prophesy is available on \url{https://github.com/moves-rwth/prophesy}.}.

\section{Preliminaries}
\label{sec:preliminaries}

\subsection{Basic notations}
\label{sec:numbers}
We denote the set of real numbers by $\R$, the rational numbers by $\Q$, and the natural numbers including $0$ by $\N$.
Let $\Ireal$ denote the \emph{closed interval} of all real numbers between $0$ and $1$, including the bounds; $\Irealo$ denotes the \emph{open interval} of all real numbers between $0$ and $1$ excluding $0$ and $1$.

\index{Power set}\index{Partition}
Let $X,Y$ denote arbitrary sets. If $X\cap Y=\emptyset$, we write $X\uplus
Y$ for the \emph{disjoint union} of the sets $X$ and $Y$. 
We denote the \emph{power set} of $X$ by $2^X=\{X'\mid X'\subseteq X\}$. 
Let $X$ be a finite or countably infinite set. A \emph{probability
distribution} over $\distDom$ is a function $\distFunc\colon 
\distDom\rightarrow\Ireal$ with
$\sum_{\distDomElem\in\distDom}\distFunc(\distDomElem)=\distFunc(\distDom)=1$.

\subsection{Polynomials, rational functions}
Let $V$ denote a finite set of parameters over $\R$ and $\dom(p)\subseteq\R$ denote the domain of parameter $p\in V$.
%
\begin{definition}[Polynomial, rational function]\label{def:polynomial}\label{def:ratfunc}\index{Rational function}\index{Polynomial}
 For a finite set $V=\{\series{p}{n}\}$ of $n$ parameters,  a \emph{monomial} $m$ is
 \[ m = p_{1}^{e_{1}}\cdot\ldots\cdot p_{n}^{e_{n}}\text{ with } e_i\in\N.\]
  Let $\mono[V]$ denote the set of monomials over $V$.
 A \emph{polynomial} $\pol$ (over $V$) with $t$ \emph{terms} is a weighted sum of monomials:
  \begin{align*}
      \pol= \sum_{j=1}^{t}a_j\cdot m_j \text{ with }  a_j\in\Q \setminus \{ 0 \},\; m_j \in \mono[V].
  \end{align*}
  Let $\poly[V]$ be the set of polynomials over $V$. 
A \emph{rational function} $f=\frac{\pol_1}{\pol_2}$ over $V$ is a fraction of polynomials $\pol_1, \pol_2\in \poly[V]$ with $\pol_2 \not\equiv 0$ (where $\equiv$ states equivalence). 
Let $\ratfunc[V]$ be the set of rational functions over $V$.
\end{definition}
A monomial is \emph{linear}, if $\sum_{i=1}^{|V|} e_{i} \leq 1$, and \emph{multi-linear}, if $e_{i} \leq 1$ for all $1 \leq i \leq |V|$.
A polynomial $\pol$ is (multi-)linear, if all monomials occurring in $\pol$ are (multi-)linear.

\emph{Instantiations} replace parameters by constant values in polynomials or rational functions.
\begin{definition}[Parameter instantiations]
  \label{def:valuation}
  A \emph{(parameter) instantiation $u$ of parameters $V$} is a function $u\colon V \rightarrow \R$.
    \end{definition}
We abbreviate the parameter instantiation $u$ with $u(p_i) = a_i \in \R$ by the $n$-dimensional vector $ (a_1,\hdots,a_n) \in \R^n$ for ordered parameters $p_1,\hdots,p_n$.
Applying the instantiation $u$ on $V$ to polynomial $g \in \poly[V]$ yields $\pol[u]$ which is obtained by replacing each $p \in {V}$ in $g$ by $u(p)$, with subsequent application of $+$ and $\cdot$.
For rational function $f=\frac{\pol_1}{\pol_2}$, let $f[u]=\frac{\pol_1[u]}{\pol_2[u]}\in{\R}$ if $\pol_2[u]\not\equiv 0$, and otherwise $f[u] = \undefcustom$.

\subsection{Probabilistic models}
\label{sec:preliminaries:probmodels}
Let us now introduce the probabilistic models used in this paper. 
We first define parametric Markov models and present conditions such that their instantiations result in Markov models with constant probabilities. 
Then, we discuss how to resolve non-determinism in decision processes.
\subsubsection{Parametric Markov models}
The transitions in parametric Markov models are equipped with rational functions over the set of parameters.
Although this is the general setting, for some of our algorithmic techniques we will restrict ourselves to linear polynomials\footnote{Most models use only simple polynomials such as $p$ and $1{-}p$, and benchmarks available e.g., at the \tool{PRISM} benchmark suite~\cite{KNP12b} or at the \tool{PARAM}~\cite{PARAM10} web page are of this form.}.
We consider parametric MCs and MDPs as sub-classes of a parametric version of classical two-player stochastic games~\cite{Sha53}.
The state space of such games is partitioned into two parts, $\spOne$ and $\spTwo$.
At each state, a player chooses an action upon which the successor state is determined according to the (parametric) probabilities.
Choices in $\spOne$ and $\spTwo$ are made by player~$\pOne$ and $\pTwo$, respectively.
pMDPs and pMCs are parametric stochastic one- and zero-player games respectively.
\begin{definition}[Parametric models]\label{def:parametric_model}
A \emph{parametric stochastic game (pSG)} is a tuple $\psgInit$ with a finite set $S$ of states with $S = \spOne\uplus\spTwo$, a finite set $\Var$ of parameters over $\R$, an initial state $\sinit \in S$, a finite set $\Act$ of actions, and a transition function $\probmdp \colon S \times \Act \times S \rightarrow \ratfunc[\Var] \cup \R \cup \{ \undefcustom\}$ with $|\Act(s)|\geq 1$ for all $s \in S$, where
$\Act(s) = \{\act \in \Act \mid \exists s'\in S.\,\probmdp(s,\act,s') \not\equiv 0\}$ is the set of \emph{enabled} actions at state $s$.
\begin{compactitem}
\item
A pSG is a \emph{parametric Markov decision process (pMDP)} if $\spOne=\emptyset$ or $\spTwo=\emptyset$.
\item
A pMDP is a \emph{parametric Markov chain (pMC)} if $|\Act(s)|=1$ for all $s \in S$.
\end{compactitem}
\end{definition}
A parametric state-action \emph{reward function} $\rew \colon S \times \Act \to \ratfunc[V] \cup \R \cup \{\undefcustom\}$ associates rewards with state-action pairs\footnote{Recall that $\undefcustom$ represents, e.g., $\nicefrac{1}{0}$.}.
It is assumed that deadlock states are absent, i.e., $\Act(s)\neq\emptyset$ for all $s \in S$.
Entries in $\R \cup \{\undefcustom\}$ in the co-domains of the functions $\probmdp$ and $\rew$ ensure that the model is closed under instantiations, see Definition~\vref{def:evpMC} below. Throughout the rest of this paper, we silently assume that any given pSGs only uses constants from $\Q$ and rational functions $\ratfunc[V]$, but no elements from $\R \setminus \Q$ or  $\undefcustom$.
A model is called \emph{parameter-free} if all its transition probabilities are constant.

A pSG intuitively works as follows.
In state $s \in \spOne$, player $\pOne$ non-deterministically selects an action  $\act \in \Act(s)$.
With (parametric) probability $\probmdp(s,\act,s')$ the play then evolves to state $s'$. 
On leaving state $s$ via action $\alpha$, the reward $\rew(s, \alpha)$ is earned.
If $s \in \spTwo$, the choice is made by player $\pTwo$, and as for player $\pOne$, the next state is determined in a probabilistic way.
As by assumption no deadlock states occur, this game goes on forever.
A pMDP is a game with one player, whereas a pMC has no players; a pMC thus evolves in a fully probabilistic way.
Let $\pdtmc$ denote a pMC, $\pmdp$ a pMDP, and $\psg$ a pSG. 
	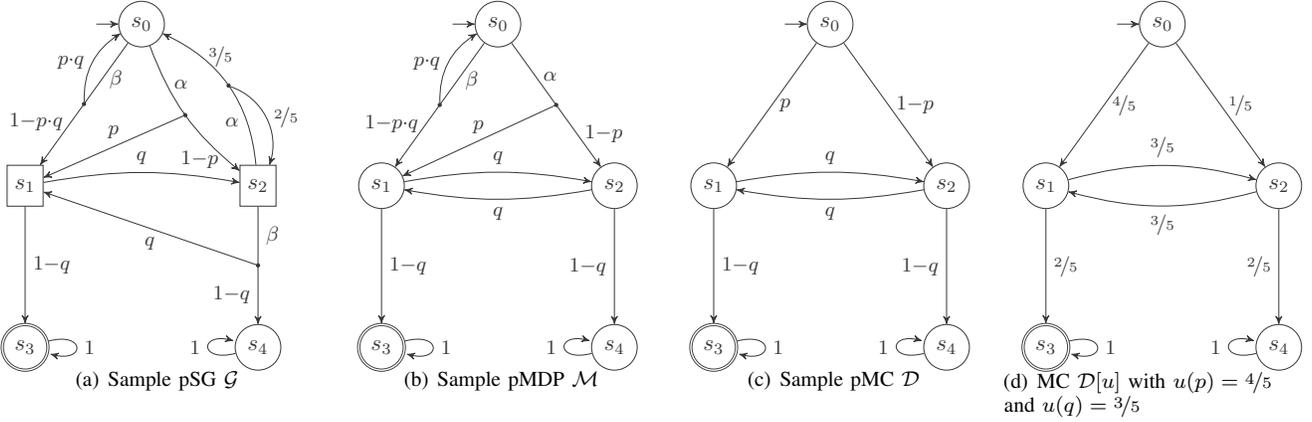
\begin{figure*}[tbh]
		\centering
		\subfigure[Sample pSG $\psg$]{
			\scalebox{.86}{
				\begin{tikzpicture}[scale=1, nodestyle/.style={draw,circle},baseline=(s0)]
    
    \node [nodestyle] (s0) at (0,0) {$s_0$};
    \node [nodestyle, rectangle, minimum height=18pt] (s1) [on grid, below left=2.5cm and 1.8cm of s0] {$s_1$};
    \node [nodestyle, rectangle, minimum height=18pt] (s2) [on grid, below right=2.5cm and 1.8cm of s0] {$s_2$};
    \node [nodestyle, accepting] (s3) [on grid, below=2.5cm of s1] {$s_3$};
    \node [nodestyle] (s4) [on grid, below=2.5cm of s2] {$s_4$};
    
    \draw ($(s0)-(0.7,0)$) edge[->] (s0);
    \draw (s0) edge[->] node[pos=0.3, right] {\small$\beta$} node[draw, circle,  inner sep=0.5pt, fill] (s0beta) {} node [pos=0.65, left] {\small$1{-}p{\cdot}q$} (s1);
    \draw (s0beta) edge[->, bend left=30] node[left] {\small$p{\cdot}q$} (s0);
    \draw (s0) edge[->, bend right=15] node[pos=0.25, right] {\small$\alpha$} node[draw, circle,  inner sep=0.5pt, fill] (s0alpha) {} node [pos=0.9, left] {\small$1{-}p$} (s2);
    \draw (s0alpha) edge[->] node[above] {\small$p$} (s1);

     \draw (s1) edge[->,bend left=10] node[above] {\small$q$} (s2);
     \draw (s1) edge[->] node[right] {\small$1{-}q$} (s3);
    
    \draw (s2) edge[->, bend right] node[pos=0.25, left] {\small$\alpha$} node[draw, circle,  inner sep=0.5pt, fill] (s2alpha) {} node [pos=0.75, right] {\small$\nicefrac{3}{5}$} (s0);
     \draw (s2alpha) edge[bend left=45, ->] node[right] {\small$\nicefrac 2 5$} (s2);
    \draw (s2) edge[->] node[pos=0.25, right] {\small$\beta$} node[draw, circle,  inner sep=0.5pt, fill] (s2beta) {} node [pos=0.75, left] {\small$1{-}q$} (s4);
     \draw (s2beta) edge[->] node[below] {\small$q$} (s1);
    
    \draw (s3) edge[loop right, ->] node[auto] {\small$1$} (s3);
    
    \draw (s4) edge[loop left, ->] node[auto] {\small$1$} (s3);

\end{tikzpicture}
			}
			\label{fig:models:psg}
		}
		\subfigure[Sample pMDP $\pmdp$]{ 
			\scalebox{.86}{
				\begin{tikzpicture}[scale=1, nodestyle/.style={draw,circle},baseline=(s0)]
    
    \node [nodestyle] (s0) at (0,0) {$s_0$};
    \node [nodestyle] (s1) [on grid, below left=2.5cm and 1.8cm of s0] {$s_1$};
    \node [nodestyle] (s2) [on grid, below right=2.5cm and 1.8cm of s0] {$s_2$};
    \node [nodestyle, accepting] (s3) [on grid, below=2.5cm of s1] {$s_3$};
    \node [nodestyle] (s4) [on grid, below=2.5cm of s2] {$s_4$};
    
    \draw ($(s0)-(0.7,0)$) edge[->] (s0);
    \draw (s0) edge[->] node[pos=0.3, right] {\small$\beta$} node[draw, circle,  inner sep=0.5pt, fill] (s0beta) {} node [pos=0.65, left] {\small$1{-}p{\cdot}q$} (s1);
    \draw (s0beta) edge[->, bend left=30] node[left] {\small$p{\cdot} q$} (s0);
    \draw (s0) edge[->, bend right=0] node[pos=0.25, right] {\small$\alpha$} node[draw, circle,  inner sep=0.5pt, fill] (s0alpha) {} node [pos=0.75, right] {\small$1{-}p$} (s2);
    \draw (s0alpha) edge[->] node[above] {\small$p$} (s1);
    
     \draw (s1) edge[bend left=10, ->] node[above] {\small$q$} (s2);
     \draw (s1) edge[->] node[right] {\small$1{-}q$} (s3);
     
     \draw (s2) edge[bend left=10, ->] node[below] {\small$q$} (s1);
     \draw (s2) edge[->] node[left] {\small$1{-}q$} (s4);
    
    \draw (s3) edge[loop right, ->] node[auto] {\small$1$} (s3);
    
    \draw (s4) edge[loop left, ->] node[auto] {\small$1$} (s3);

\end{tikzpicture}
			}
			\label{fig:models:pmdp}
		}
		\subfigure[Sample pMC $\pdtmc$]{ 
			\scalebox{.86}{
				\begin{tikzpicture}[scale=1, nodestyle/.style={draw,circle},baseline=(s0)]
    
    \node [nodestyle] (s0) at (0,0) {$s_0$};
    \node [nodestyle] (s1) [on grid, below left=2.5cm and 1.8cm of s0] {$s_1$};
    \node [nodestyle] (s2) [on grid, below right=2.5cm and 1.8cm of s0] {$s_2$};
    \node [nodestyle, accepting] (s3) [on grid, below=2.5cm of s1] {$s_3$};
    \node [nodestyle] (s4) [on grid, below=2.5cm of s2] {$s_4$};
    
    \draw ($(s0)-(0.7,0)$) edge[->] (s0);
    \draw (s0) edge[->] node[right] {\small$p$} (s1);
    \draw (s0) edge[->] node[right] {\small$1{-}p$} (s2);

     \draw (s1) edge[bend left=10, ->] node[above] {\small$q$} (s2);
     \draw (s1) edge[->] node[right] {\small$1{-}q$} (s3);
    
     \draw (s2) edge[bend left=10, ->] node[below] {\small$q$} (s1);
     \draw (s2) edge[->] node[left] {\small$1{-}q$} (s4);
    
    \draw (s3) edge[loop right, ->] node[auto] {\small$1$} (s3);
    
    \draw (s4) edge[loop left, ->] node[auto] {\small$1$} (s3);

\end{tikzpicture}
			}
			\label{fig:models:pmc}
		}
		\subfigure[{MC $\pdtmc[u]$} with $u(p)=\nicefrac{4}{5}$\newline and $u(q)=\nicefrac{3}{5}$]{
	  \centering
    	\scalebox{.86}{
      \begin{tikzpicture}[scale=1, nodestyle/.style={draw,circle},baseline=(s0)]
    
    \node [nodestyle] (s0) at (0,0) {$s_0$};
    \node [] (leftdummy)  [on grid, left=1.8cm of s0] {};
    \node [] (rightdummy) [on grid, right=1.8cm of s0] {};
    \node [nodestyle] (s1) [on grid, below=2.5cm of leftdummy] {$s_1$};
    \node [nodestyle] (s2) [on grid, below=2.5cm of rightdummy] {$s_2$};
    \node [nodestyle, accepting] (s3) [on grid, below=2.5cm of s1] {$s_3$};
    \node [nodestyle] (s4) [on grid, below=2.5cm of s2] {$s_4$};
    
    \draw ($(s0)-(0.7,0)$) edge[->] (s0);
    \draw (s0) edge[->] node[right] {\small$\nicefrac 4 5$} (s1);
    \draw (s0) edge[->] node[right] {\small$\nicefrac 1 5$} (s2);

     \draw (s1) edge[bend left=15, ->] node[above] {\small$\nicefrac 3 5$} (s2);
     \draw (s1) edge[->] node[right] {\small$\nicefrac 2 5$} (s3);
    
     \draw (s2) edge[bend left=15, ->] node[below] {\small$\nicefrac 3 5$} (s1);
     \draw (s2) edge[->] node[left] {\small$\nicefrac 2 5$} (s4);
    
    \draw (s3) edge[loop right, ->] node[auto] {\small$1$} (s3);
    
    \draw (s4) edge[loop left, ->] node[auto] {\small$1$} (s3);

\end{tikzpicture}
    }
    \label{fig:pmc:Du}
  }
		\caption{The considered types of parametric probabilistic models (a)--(c) and an instantiated model (d).}
		\label{fig:models}
	\end{figure*}
\begin{example}\label{ex:models}
Figure~\vref{fig:models}(a)--(c) depict a pSG, a pMDP, and a pMC respectively over parameters $\Var = \{p,q\}$.
The states of the players $\pOne$ and $\pTwo$ are drawn as circles and rectangles, respectively.
The initial state is indicated by an incoming arrow without source.
We omit actions in state $s$ if $|\Act(s)| = 1$.
In state $s_0$ of Figure~\vref{fig:models}(a), player $\pOne$ can select either action $\alpha$ or $\beta$.
On selecting $\alpha$, the game moves to state $s_1$ with probability $p$, and to $s_2$ with probability $1{-}p$. 
In state $s_2$, player $\pTwo$ can select $\alpha$ or $\beta$; in $s_1$ there is a single choice only.
\end{example}

A \emph{transition} $(s,\act,s')$ exists if $\probmdp(s,\act,s') \not\equiv 0$. 
As pMCs have a single enabled action at each state, we omit this action and just write $\probmdp(s,s')$ for $\probmdp(s,\act,s')$ if $\Act(s) = \{\alpha\}$.
A state $s'$ is a \emph{successor of $s$}, denoted $s' \in \successor(s)$, if $\probmdp(s, \alpha, s') \not\equiv 0$ for some $\alpha$; in this case, $s \in \predecessor(s')$ is a predecessor of $s'$.
\begin{remark}
Parametric stochastic games are the most general model used in this paper.
They subsume pMDPs and pMCs and parameter-free SGs, which are used throughout this paper.
We concisely introduce the formal foundations on this general class and indicate how these apply  to subclasses.
Most algorithmic approaches in this paper are not directly applicable to pSGs, but tailored to either pMDPs or pMCs. 
This is indicated when introducing these techniques.
\end{remark}
\begin{definition}[Stochastic game]
  \label{def:sg}
A pSG $\psgInit$ is a \emph{stochastic game (SG)} if $\probmdp \colon S \times \Act \times S \rightarrow [0,1]$ and $\sum_{s'\in S}\probmdp(s,\act,s') = 1$ for all $s \in S$ and $\act \in \Act(s)$.
\end{definition}
A state-action reward function $\rew \colon S \times \Act \to \R_{\geq 0}$ associates (non-negative, finite) rewards to outgoing actions.
Analogously, Markov chains (MCs) and Markov decision processes (MDPs) are defined as special cases of pMCs and pMDPs, respectively.
We use $\dtmc$ to denote a MC, $\mdp$ for an MDP and $\sg$ for an SG.

\subsubsection{Paths and reachability}
An \emph{infinite path} of a pSG $\psg$ is an infinite sequence $\pi = s_0 \act_0 s_1 \act_1 \ldots$ of states $s_i\in S$ and actions $\act_i\in\Act(s_i)$ with $\probmdp(s_i,\alpha_i,s_{i+1})\not\equiv 0$ for $i\geq0$.
A \emph{finite path} of a pSG $\psg$ is a non-empty finite prefix $s_0 \act_0 \ldots s_n$ of an infinite path $s_0 \act_0 \ldots s_n \act_n\ldots$ of $\psg$ for some $n\in \N$.
Let $\paths[\psg]$ denote the set of all finite or infinite paths of $\psg$ while $\pathsfin[\psg]\subseteq\paths[\psg]$ denotes the set of all finite paths. For paths in (p)MCs, we omit the actions.
The set $\paths[\psg](s)$ contains all paths that start in state $s\in S$.
For a finite path $\pi\in\pathsfin[\psg]$, $\last(\pi)=s_n$ denotes the last state of $\pi$.
The \emph{length} $|\pi|$ of a path $\pi$ is $|\pi|=n$ for $\pi\in\pathsfin[\psg{}]$ and $|\pi|=\infty$ for infinite paths.
The \emph{accumulated reward} along the finite path $s_0 \act_0 \ldots \act_{n-1} s_n$ is given by the sum of the rewards $\rew(s_i,\act_i)$ for $0 \leq i < n$.

We denote the set of states that can reach a set of states $T$ as follows: $\finally T = \{s\in S\mid \exists \pi\in\pathsfin[\psg](s).\ \last(\pi)\in T\}$.
A set of states $T \subseteq S$ is \emph{reachable} from
$s\in S$, written $s\in \finally T$, iff there is a path from $s$ to some $s'\in T$.
A state $s$ is \emph{absorbing} iff $\probmdp(s,\act,s)=1$ for all $\act\in\Act(s)$.
\begin{example}
	The pMC in Figure~\vref{fig:models:pmc} has a path $\pi = s_0s_1s_3s_3$ with $|\pi| = 3$.
	Thus $s_0 \in \finally \{ s_3 \}$.
	There is no path from $s_4$ to $s_3$, so $s_4 \not\in \finally \{ s_3 \}$. 
	States $s_3$ and $s_4$ are the only absorbing states.
\end{example}
\subsubsection{Model instantiation}
Instantiated parametric models are obtained by instantiating the rational functions in all transitions as in Definition~\vref{def:valuation}.
\begin{definition}[Instantiated pSG]
\label{def:evpMC}
For a pSG $\psgInit{}$ and instantiation $u$ of $V$, the \emph{instantiated pSG at $u$} is given by $\psg[u]=(S,\sinit,\Act,\probmdp[u])$ with $\probmdp[u](s,\act,s')=\probmdp(s,\act,s')[u]$
  for all $s, s' \in S$ and $\act\in\Act$.
\end{definition}
The instantiation of the parametric reward function $\rew$ at $u$ is $\rew[u]$ with $\rew[u](s,\act) = \rew(s,\act)[u]$ for all $s\in S, \act\in\Act$.
Instantiating pMDP $\pmdp$ and pMC $\pdtmc$ at $u$ is denoted by $\pmdp[u]$ and $\pdtmc[u]$, respectively.
\begin{remark}
The instantiation of a pSG at $u$ is a pSG, but not necessarily an SG. 
This is due to the fact that an instantiation does not ensure that $\probmdp(s, \act, \cdot)$ is a probability distribution. 
In fact, instantiation yields a transition function of the form $\probmdp\colon S\times\Act\times S\rightarrow \R \cup \{ \undefcustom \}$. 
Similarly, there is no guarantee that the rewards $\rew[u]$ are non-negative.
Therefore, we impose restrictions on the parameter instantiations.
\end{remark}
\begin{definition}[Well-defined instantiation]
  \label{def:well_valuation}
An instantiation $u$ is \emph{well-defined} for a pSG $\psg$ if the pSG $\psg[u]$ is an SG.
\end{definition}
The reward function $\rew$ is well-defined on $u$ if it does only associate non-negative reals to state-action pairs.
\begin{example}\label{ex:welldefined_valuation}
Consider again the pMC in Figure~\vref{fig:models:pmc}.
The instantiation $u$ with $u(p)=\nicefrac 4 5$ and $u(q)=\nicefrac 3 5$ is well-defined and induces the MC $\pdtmc[u]$ depicted in Figure~\vref{fig:pmc:Du}.
\end{example}

From now on, we silently assume that every pSG we consider has at least one well-defined instantiation. This condition can be assured through checking the satisfiability of the conditions in Def. \vref{def:sg}, which we discuss in Section~\ref{sec:ensuringgraphpres}.

Our methods necessitate instantiations that are not only well-defined, but also preserve the topology of the pSG. In particular, we are interested in the setting where reachability between two states coincides for the pSG and the set of instantiations $u$ we consider. We detail this discussion in Section~\ref{sec:ensuringgraphpres}.
\begin{definition}[Graph preserving]\label{def:graph_preserving}
A well-defined instantiation $u$ for pSG $\psgInit$ is \emph{graph preserving} if for all $s,s' \in S$ and $\act\in\Act$, 
\[
\probmdp(s,\act,s') \not\equiv 0 \implies \probmdp(s,\act,s')[u] \in \R\setminus\{0\}.
\]
\end{definition}
\begin{example}
The well-defined instantiation $u$ with $u(p)=1$ and $u(q)=\nicefrac 3 5$ for the pMC in Figure~\vref{fig:models:pmc} is not \emph{graph preserving}.	
\end{example}
\subsubsection{Resolving non-determinism}
\label{sec:schedulers}
\emph{Strategies}\footnote{Also referred to as policies, adversaries, or schedulers.} resolve the non-deterministic choices in stochastic games with at least one player.
For the objectives considered here, it suffices to consider so-called \emph{deterministic} strategies~\cite{Var85}; more general strategies can be found in~\cite[Ch.\ 10]{BK08}.
We define strategies for pSGs and assume well-defined instantiations as in Definition~\vref{def:well_valuation}.
\begin{definition}[Strategy]
\label{def:strategy}
A (deterministic) \emph{strategy} $\sched_i$ for player $i\in\{\pOne,\pTwo\}$ in a pSG $\psg$ with state space $S = \spOne \uplus  \spTwo$ is a function
\[
\sched_i\colon \{\pi\in \pathsfin[\psg] \mid\last(\pi)\in S_i\}\to \Act
\]
such that $\sched_i(\pi)\in\Act(\last(\pi))$.
Let $\Sched^\psg$ denote the set of strategies $\sched = (\schedOne,\schedTwo)$ for pSG $\psg$ and $\Sched^{\psg}_i$ the set of strategies of player $i$.
\end{definition}
A pMDP has only a player-$i$ strategy for the player with $S_i \neq \emptyset$; in this case the index $i$ is omitted.
A player-$i$ strategy $\sched_i$ is \emph{memoryless} if $\last(\pi)=\last(\pi')$ implies $\sched_i(\pi)=\sched_i(\pi')$ for all finite paths $\pi,\pi'$. 
A memoryless strategy can thus be written in the form $\sched_i\colon S_i\to \Act$.
A pSG-strategy $\sched=(\schedOne,\schedTwo)$ is memoryless if both $\schedOne$ and $\schedTwo$ are memoryless.

\begin{remark}
From now on, we only consider memoryless strategies and refer to them as strategies.
\end{remark}

\noindent A strategy~$\sigma$ for a pSG resolves all non-determinism and results in an \emph{induced pMC}.
\begin{definition}[Induced pMC]\label{def:induced_dtmc} 
The \emph{pMC $\psg^\sched$ induced by strategy $\sched = (\schedOne,\schedTwo)$} on pSG $\psgInit$ equals $(S, \Var, \sinit, P^\sched)$ with:
	\begin{align*}
		P^\sched(s,s')=
		\begin{cases} 
		\probmdp(s,\schedOne(s),s') \quad \mbox{if } s \in \spOne \\
			\probmdp(s,\schedTwo(s),s') \quad \mbox{if } s\in \spTwo.
		\end{cases} 
	\end{align*}
	\end{definition}
\begin{example}
Let $\sched$ be a strategy for the pSG $\psg$ in Figure~\vref{fig:models}(a) with $\schedOne(s_0) = \alpha$ and $\schedTwo(s_2)=\beta$.
The induced pMC $\psg^{\sched}$ equals pMC $\pdtmc$ in Figure~\vref{fig:models}(c).
Analogously, imposing strategy $\sched'$ with $\sched'(s_0) = \alpha$ on the pMDP in Figure~\vref{fig:models}(b) yields $\pmdp^{\sched'} = \pdtmc$.
\end{example}
The notions of strategies for pSGs and pMDPs and of induced pMCs naturally carry over to non-parametric models; e.g., the MC $\sg^\sigma$ is induced by strategy $\sched\in\Sched^\sg$ on SG $\sg$. 

\subsection{Specifications and solution functions}

\subsubsection{Specifications}
\label{subsec:specifications}
Specifications constrain the measures of interest for (parametric) probabilistic models.
Before considering parameters, let us first consider MCs.
Let $\DtmcInit$ be an MC and $T\subseteq S$ a set of \emph{target states} that (without loss of generality) are assumed to be absorbing.
Let $\finally T$ denote the path property to reach $T$\footnote{Thereby overloading the earlier notation to denote the set of states for which there exists a path on which this property holds.}. 
Furthermore, the \emph{probability measure} $\pr_s$ over sets of paths can be defined using a cylinder construction with $\pr_s(s_0\alpha_0\ldots s_n)=\Pi_{i=0}^{n-1}\probmdp(s_i,\alpha_i,s_{i+1})$, see~\cite[Ch.\ 10]{BK08}.

\noindent We consider three kinds of specifications:
\begin{enumerate}
\item
\emph{Unbounded probabilistic reachability}\ A specification 
$\p_{\leq \lambda}(\finally\, T)$ asserts that the probability to reach $T$ from the initial state $\sinit$ shall be at most $\lambda$, where $\lambda \in \mathbb{Q} \cap [0,1]$.
More generally, specification $\varphi^r$ is satisfied by MC $\dtmc$, written:
\[ 
\dtmc \models \p_{\sim\lambda}(\finally\, T) \quad \mbox{iff} \quad \pr^{\dtmc}_{\sinit}(\finally\, T) \sim \lambda,
\]
where $\pr_{\sinit}^\dtmc(\finally\, T)$ is the probability mass of all infinite paths that start in $\sinit$ and visit any state from $T$.
\item
\emph{Bounded probabilistic reachability}\ In addition to reachability, these specifications impose a bound on the maximal number of steps until reaching a target state.
Specification $\varphi^b = \p_{\sim\lambda}(\finally^{\leq n}\, T)$ asserts that in addition to $\p_{\sim\lambda}(\finally\, T)$, states in $T$ should be reached within $n \in \mathbb{N}$ steps.
The satisfaction of $\p_{\sim\lambda}(\finally^{\leq n}\, T)$ is defined similar as above.
\item 
\emph{Expected reward until a target}\ The specification $\e_{\leq\kappa}(\finally\, T)$ asserts that the expected reward until reaching a state in $T$ shall be at most $\kappa \in \mathbb{R}$. 
Let $\er_{\sinit}^{\dtmc}(\finally\, T)$ denote the expected accumulated reward until reaching a state in $T \subseteq S$ from state $\sinit$. We obtain this reward by multiplying the probability of every path reaching $T$ with the accumulated reward of that path, up until reaching $T$.  Details are given in \cite[Chapter 10]{BK08}. \footnote{As standard, if $\pr_{\sinit}^\dtmc(\finally\, T)<1$ then we set $\er_{\sinit}^{\dtmc}(\finally\, T) \colonequals \infty$. The rationale is that an infinite amount of reward is collected on visiting a state (with positive reward) infinitely often from which all target states are unreachable.}. Then we define
\[
\dtmc \models \e_{\sim\kappa}(\finally\, T) 
\quad \mbox{iff} \quad 
\er_{\sinit}^{\dtmc}(\finally\, T)\sim \kappa, 
\]
We do not treat the accumulated reward to reach a target within $n$ steps, as this is not a very useful measure. In case there is a possibility to not reach the target within $n$ steps, this yields $\infty$.
\end{enumerate}
We omit the superscript $\dtmc$ if it is clear from the context. 
We write $\neg\varphi$ to invert the relation: $\dtmc \models \neg\p_{\leq\lambda}(\finally\, T)$ is thus equivalent to $\dtmc \models \p_{>\lambda}(\finally\, T)$.
An SG $\sg$ satisfies specification $\varphi$ under strategy $\sigma$ if the induced MC $\sg^{\sched} \models \varphi$.
Unbounded reachability and expected rewards are prominent examples of \emph{indefinite-horizon} properties -- they measure behaviour up-to some specified event (the horizon) which may be reached after arbitrarily many steps.

\begin{remark}
\label{rem:noboundedreach}
Bounded reachability in MDPs can be reduced to unbounded reachability by a technique commonly referred to as \emph{unrolling}~\cite{DBLP:conf/formats/AndovaHK03}. 
For performance reasons, it is sometimes better to avoid this unrolling, and present dedicated approaches.
\end{remark}

\subsubsection{Solution functions}
Computing (unbounded) reachability probabilities and expected rewards for MCs reduces to solving linear equation systems~\cite{BK08} over the field of reals (or rationals). 
For parametric MCs, we obtain a linear equation system over the field of the rational functions over $V$ instead. 
The solution to this equation system is a rational function. 
(See Examples~\ref{ex:solution_function_synth} and \ref{ex:nonlinear} on pages \pageref{ex:solution_function_synth} and \pageref{ex:nonlinear}).
More details on the the solution function and the equation system follow in Section~\ref{sec:solution_fct} and Section~\ref{sec:exact_mc}, respectively.
\begin{definition}[Solution functions]\label{def:param_solutions}
For a pMC $\PdtmcInit$, $T \subseteq S$ and $n\in\N$, a \emph{solution function} for a specification $\varphi$ is a rational function 
\[
\begin{array}{rcl}
f_{\pdtmc,T}^r \in \ratfunc[V] & \mbox{ for } & \varphi=\p_{\sim\lambda}(\finally\, T)\\
f_{\pdtmc,T,n}^b \in \ratfunc[V] & \mbox{ for } & \varphi=\p_{\sim\lambda}(\finally^{\leq n}\, T) \mbox{ , and}\\ 
f_{\pdtmc,T}^e\in\ratfunc[V] & \mbox{ for } & \varphi=\e_{\sim\kappa}(\finally\, T),
\end{array}
\]
such that for every well-defined graph-preserving instantiation $u$:
\begin{align*}
	f_{\pdtmc,T}^r[u] & \ = \ \pr^{\pdtmc[u]}_{\sinit}(\finally\, T),\\
	f_{\pdtmc,T,n}^b[u] & \ = \ \pr^{\pdtmc[u]}_{\sinit}(\finally^{\leq n}\, T) \mbox{, and} \\
	f_{\pdtmc,T}^e[u] & \ = \ \er_{\sinit}^{\pdtmc[u]}(\finally\, T).
\end{align*}
\end{definition}
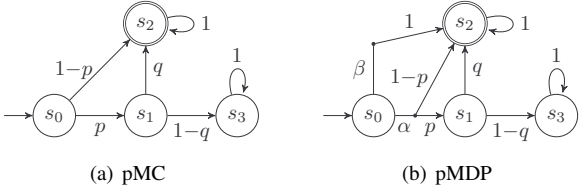
\begin{figure}[t]
\centering
	\subfigure[pMC]{
		\scalebox{1}{
			\begin{tikzpicture}
		\node[draw, circle, , initial, initial text=] (s0) {$s_0$} ;
		\node[draw, circle, , right=0.8cm of s0] (s1) {$s_1$} ;
		\node[draw, circle, , above=0.8cm of s1, accepting] (s2) {$s_2$} ;
		\node[draw, circle, , right=0.8cm of s1] (s3) {$s_3$} ;
		
		\draw[->] (s0) -- node[below] {$p$} (s1);
		\draw[->] (s1) -- node[right] {$q$} (s2);
		\draw[->] (s0) -- node[left] {$1{-}p$} (s2);
		\draw[->] (s1) -- node[below] {$1{-}q$} (s3);
		\draw[->] (s2) edge[loop right] node {$1$} (s2);
		\draw[->] (s3) edge[loop above] node {$1$} (s3);
	\end{tikzpicture}
		}
		\label{Fig:SimpleAcycPMC}
	}
	\subfigure[pMDP]{
		\scalebox{1}{
			\begin{tikzpicture}
		\node[draw, circle, , initial, initial text=] (s0) {$s_0$} ;
		\node[draw, circle, , right=0.8cm of s0] (s1) {$s_1$} ;
		\node[draw, circle, , above=0.8cm of s1, accepting] (s2) {$s_2$} ;
		\node[draw, circle, , right=0.8cm of s1] (s3) {$s_3$} ;
		\node[draw, circle,  inner sep=0.5pt, fill, right=0.3cm of s0] (s0alpha)  {};
		\node[draw, circle,  inner sep=0.5pt, fill, above=0.8cm of s0] (s0beta) {} ;
		
		\draw[-] (s0) -- node[below] {$\alpha$} (s0alpha);
		\draw[-] (s0) -- node[left] {$\beta$} (s0beta);
		\draw[->] (s0beta) -- node[above] {$1$} (s2);
		\draw[->] (s0alpha) -- node[below] {$p$} (s1);
		\draw[->] (s1) -- node[right] {$q$} (s2);
		\draw[->] (s0alpha) -- node[left] {$1{-}p$} (s2);
		\draw[->] (s1) -- node[below] {$1{-}q$} (s3);
		\draw[->] (s2) edge[loop right] node {$1$} (s2);
		\draw[->] (s3) edge[loop above] node {$1$} (s3);
\end{tikzpicture}
		}
		\label{Fig:SimpleAcycPMDP}
	}
	\caption{Two sample parametric models.}
	\label{fig:toyexamples}
\end{figure}
\begin{example}\label{ex:solutionfunction}
Consider the reachability probability to reach $s_2$ for the pMC in Figure~\vref{Fig:SimpleAcycPMC}. 
Any instantiation $u$ with $u(p),u(q)\in(0,1)$ is well-defined and graph-preserving. 
As the only two finite paths to reach $s_2$ are $s_0 s_2$ and $s_0 s_1 s_2$,  we have $f_{\pdtmc,\{s_2\}}^r = 1-p + p \cdot q$. 
\end{example}
For pSGs (and pMDPs), the solution function depends on the resolution of non-determinism by strategies, \ie, they are defined on the induced pMCs. 
Formally, a \emph{solution function} for a pSG $\psg$, a reachability specification $\varphi^r=\p_{\leq\lambda}(\finally\, T)$, and a strategy $\sched\in\Sched^\psg$ is a function
$ f^r_{\psg,\sched,T} \in\ratfunc[V]$ such that for each well-defined graph-preserving instantiations $u$ it holds:
\begin{align*}
	f^r_{\psg,\sched,T}[u] = \pr^{\psg^\sched[u]}_{\sinit}(\finally\, T).
\end{align*}
These notions are defined analogously for bounded reachability (denoted $f^b_{\psg,\sched,T,n}$)
and expected reward (denoted $f^e_{\psg,\sched,T}$) specifications. 

\begin{example}
\label{Ex:SolutionFunctionsSimpleAcycPMDP}
For the pMDP in Figure~\vref{Fig:SimpleAcycPMDP}, the solution functions for reaching $s_2$ are $1{-}p + p \cdot q$, for the strategy $\sigma_\alpha=\{ s_0 \mapsto \alpha \}$, and $1$ for the strategy $\sigma_\beta = \{ s_0 \mapsto \beta \}$.
\end{example}

\begin{remark}
\label{rem:solfuncgp}
We define solution functions only for graph-preserving valuations. For the more general well-defined solutions, a similar definition can be given~\cite{DBLP:phd/dnb/Junges20} where (solution) functions are no longer rational functions but instead a collection of solution functions obtained on the graph-preserving subsets. In particular, unless a pMC is acyclic, such a function is only semi-continuous~\cite{DBLP:journals/jcss/JungesK0W21}. A key reason for the discontinuity is the change of states that are in $\lozenge T$, e.g., consider instantiations with $q=1$ in Figure~\vref{fig:models:pmc}. We provide the decomposition into graph-preserving subsets in Section~\ref{sec:decompose}.
\end{remark}

\subsection{Constraints and formulas}
\label{sec:preliminaries:etr}
We consider \emph{(polynomial) constraints} of the form $g \sim g'$ with $g,g' \in \poly[V]$ and $\sim \in \: \{<,\leq,=,\geq,>\}$.
We denote the set of all constraints over $V$ with $\mathcal{C}[V]$.
A constraint $g \sim g'$ can be equivalently formulated as $g - g' \sim 0$.
A \emph{formula} $\psi$ over a set of polynomial constraints is recursively defined: Each polynomial constraint is a formula, and the Boolean combination of formulae is also a formula.
\begin{example} Let $p,q$ be variables.
	$1-p\cdot q > 0$ and $p^2 < 0$ are constraints, $\neg \left(p^2 < 0\right)$ and $\left(1-p\cdot q > 0\right) \lor \left(p^2 < 0\right)$ are formulae. 
\end{example}
The semantics of constraints are standard: i.e., an instantiation $u$ satisfies $g \sim g'$ if $g[u] \sim g'[u]$. An instantiation satisfies $\psi \land \psi'$ if $u$ satisfies both $\psi$ and $\psi'$.
The semantics for other Boolean connectives are defined analogously. 
Moreover, we will write $g \neq g'$ to denote the formula $g < g' \lor g > g'$.

Checking whether there exists an instantiation that satisfies a formula is equivalent to checking membership of the \emph{existential theory of the reals}~\cite{BPR06}.
Such a check can be automated using SMT-solvers capable of handling quantifier-free non-linear arithmetic over the reals~\cite{demoura_nlsat}, such as~\cite{dMB08,DBLP:conf/sat/CorziliusKJSA15}.

Statements of the form $f \sim f'$ with $f,f' \in \ratfunc[V]$ are not necessarily polynomial constraints: however, we are not interested in instantiations $u$ with $f[u] = \undefcustom$, and thus later (in Section~\ref{subsec:handlinginequalities}) we can transform such constraints into formulae over polynomial constraints.

\section{Formal Problem Statements}
\label{sec:regionsMC}

This section formalises the three problem statements mentioned in the introduction: the verification problem and two synthesis problems.
We start off by making precise what regions are and how to represent them.
We then define what it means for a region to satisfy a given specification.
This puts all in place to making the three problem statements precise.
Finally, it surveys the verification approaches that are detailed later in the paper.

\subsection{Regions}
\label{sec:regionsintro}
Instantiated parametric models are amenable to standard probabilistic model checking. 
However, \emph{sampling} an instantiation is very restrictive---verifying an instantiated model gives results for a single point in the (uncountably large) parameter space.
A more interesting problem is to determine which parts of the parameter space give rise to a model that complies with the specification.
Such sets of parameter values are, inspired by their geometric interpretation, called \emph{regions}.
Regions are solution sets of conjunctions of constraints over the set $V$ of parameters.
\begin{definition}[Region]\label{def:region}
  A \emph{region} $R$ over $V$ is a set of instantiations of $V$ (or dually a subset of $\R^{|V|}$) for which there exists a set $C(R) \subseteq \mathcal{C}[V]$
   of polynomial constraints such that for their conjunction $\Upphi(R)=\bigwedge_{c \in C(R)} c$ we have 
\[ 
R \ = \ \{ u \mid \Upphi(R)[u]\}. 
\]
We call $C(R)$ the \emph{representation} of $R$.
%
\end{definition}
Any region which is a subset of a region $R$ is called a \emph{subregion} of $R$.
%
\begin{example}
Let the region $R$ over $V=\{ p, q\}$ be described by
\[
C(R)=\{ p^2 + q^2 - 1 \leq 0 ,\  p+q-1 \leq 0\}.
\]
Thus, $R = \{\, u \mid (p^2{+}q^2{-}1)[u] \leq 0 \land (p{+}q{-}1)[u] \leq 0 \, \}$.
The region $R$ contains the instantiation $u = (\nicefrac 2 5, \nicefrac 3 5)$ as $(\nicefrac 2 5)^2 + (\nicefrac 3 5)^2 - 1 \leq 0$ and $\nicefrac 2 5+\nicefrac 3 5 - 1 \leq 0$. 
The instantiation $u' = (\nicefrac 1 2, \nicefrac 3 5) \not\in R$ as $\nicefrac 1 2 + \nicefrac 3 5 - 1 >  0$.
Regions do not have to describe a contiguous area of the parameter space; e.g., consider the region $R'$ described by $\{{-}p^2 + 1 < 0\}$ is $R'=(-\infty,-1]\cup[1,+\infty)$.
\end{example}
Regions are semi-algebraic sets~\cite{BPR06} which yield the theoretical formalisation of notions such as distance, convexity, etc. 
It also ensures that regions are well-behaved: Informally, a region in the space $\R^n$ is given by a finite number of connected semialgebraic sets (\emph{cells}\footnote{Connected here intuitively refers to the fact that you can draw a path from two points in a cell that never leaves the cell.}), and (the boundaries of) each cell can be described by a finite set of polynomials.
The size $\| R \|$ of a region $R$ is given by the Lebesgue measure. 
All regions are Lebesgue measurable.

A region is called well-defined if all its instantiations are well defined.
\begin{definition}[Well-defined region]\label{def:well-defined-_region}
Region $R$ is \emph{well defined} for pSG $\psg$ if for all $u\in R$, $u$ is a well-defined valuation for $\psg$.
\end{definition}

\subsection{Angelic and demonic satisfaction relations}

As a next step towards our formal problem statements, we have to define what it means for a region to satisfy a specification.
We first introduce two satisfaction relations---angelic and demonic---for parametric Markov models for a single instantiation.
We then lift these two notions to regions.

\begin{definition}[Angelic and demonic satisfaction relations]\label{def:satRel}
For pSG $\psg$, well-defined instantiation $u$, and specification $\varphi$, the \emph{satisfaction relations $\models_a$ and $\models_d$} are defined by:
\begin{align*}
\psg, u \models_a \varphi \quad \mbox{ iff } \quad \exists \sched \in \Sched^{\psg}.\ \psg[u]^\sched \models \varphi & \quad (\text{angelic})\\
\psg, u \models_d \varphi \quad \mbox{ iff } \quad \forall \sched \in \Sched^{\psg}.\ \psg[u]^\sched \models \varphi & \quad(\text{demonic}).
\end{align*}
\end{definition}
\noindent
The angelic relation $\models_a$ refers to the \emph{existence} of a strategy to fulfil the specification $\varphi$, whereas the demonic counterpart $\models_d$ requires \emph{all} strategies to fulfil $\varphi$.
Observe that $\psg, u \not\models_a \varphi$ if and only if $\psg, u \models_d \neg\varphi$.
Thus, demonic and angelic can be considered to be dual. By $\models_\heartsuit$ we denote the dual of $\models_\clubsuit$, that is, if $\clubsuit=a$ then $\heartsuit=d$ and vice versa.
For pMCs, the relations $\models_a$ and $\models_d$ coincide and the subscripts $a$ and $d$ are omitted.
\begin{example}
 Consider the pMDP $\pmdp$ in Figure~\vref{Fig:SimpleAcycPMDP}, instantiation $u = (\nicefrac 1 2, \nicefrac 1 2)$ and $\varphi = \p_{> \nicefrac 4 5}(\finally \{s_2\})$.
We have $\pmdp, u \models_a \varphi$, as for strategy $\sched_\beta = \{ s_0 \mapsto \beta \}$ the state $s_2$ is reached with probability one; thus, $\pmdp[u]^{\sigma_\beta} \models \varphi$.
However, $\pmdp, u \not\models_d \varphi$, as for strategy $\sched_\alpha = \{ s_0 \mapsto \alpha \}$, we have $(1{-}p+p \cdot q)[u] = \nicefrac 3 4 \not > \nicefrac 4 5$; thus, $\pmdp[u]^{\sigma_\alpha} \not\models \varphi$. 
By duality, $\pmdp, u \models_a \neg\varphi$. 
\end{example}

We now lift these two satisfaction relations to regions.
The aim is to consider specifications $\varphi$ that hold \emph{for all instantiations} represented by a region $R$ of a parametric model $\psgGeneric$.
This is captured by the following satisfaction relation.
\begin{definition}{\bf (Satisfaction relation for regions)}\label{def:satRelReg}
For pSG $\psg$, well-defined region $R$, and specification $\varphi$, the relation $\models_\clubsuit$, $\clubsuit \in \{a, d\}$, is defined as:
\begin{align*}
\psg, R \models_\clubsuit \varphi \quad \mbox{ iff } \quad 
\psg, u \models_\clubsuit \varphi \text{ for all } u\in R.
\end{align*}
\end{definition}
\noindent Before we continue, we note the difference between $\psg, R \not\models_\clubsuit \varphi$ and $\psg, R \models_\clubsuit\neg\varphi$: 
\[ \psg, R \models_\clubsuit \neg\varphi\text{ implies }\psg, u  \models_\clubsuit \neg \varphi\text{ for \emph{all} }u\in R,\] whereas in constrast, 
\[ \psg, R \not\models_\clubsuit \varphi\text{ implies }\psg, u \not\models_\clubsuit \varphi\text{ for \emph{some} }u \in R.\]
\begin{definition}[Accepting/rejecting/inconsistent region]
A well-defined region $R$ is \emph{accepting} (for $\psg$, $\varphi$, $\clubsuit$) if $\psg, R \models_\clubsuit \varphi$.  
Region $R$ is \emph{rejecting} (for $\psg$, $\varphi$, $\clubsuit$) if $\psg, R \models_{\heartsuit} \neg\varphi$.
Region $R$ is \emph{inconsistent} if it is neither accepting nor rejecting.
\end{definition}
By the duality of $\models_a$ and $\models_d$, a region is thus rejecting iff $\forall u\in R.~\psg, u \not\models_{\clubsuit} \varphi$.
Note that this differs from $\psg, R \not\models_\clubsuit \varphi$.
\begin{example}
Reconsider the pMDP in Figure~\vref{Fig:SimpleAcycPMDP}, with $R = [\nicefrac 2 5, \nicefrac 1 2] \times [\nicefrac 2 5, \nicefrac 1 2]$ and $\varphi = \p_{>\nicefrac 4 5}(\finally \{s_2\})$. The corresponding solution functions are given in Example~\vref{Ex:SolutionFunctionsSimpleAcycPMDP}.
It follows that:
\begin{compactitem}
\item  
$\pmdp, R \models_a \varphi$, as for strategy $\sched_\beta = \{ s_0 \mapsto \beta \}$, we have $\pmdp^{\sigma_\beta}, u \models \varphi$ for all $u\in R$.
\item 	
$\pmdp, R \not\models_d \varphi$, as for strategy $\sched_\alpha = \{ s_0 \mapsto \alpha \}$, $\pmdp^{\sigma_\alpha}, u \not\models \varphi$ for $u = (\nicefrac 1 2, \nicefrac 1 2)$. 
\item 
$\pmdp, R \models_a \neg\varphi$ using strategy $\sched_\alpha$.
\end{compactitem}
Regions can be inconsistent \wrt a relation, and consistent \wrt its dual relation. 
The region $(0,1) \times (0,1)$ is inconsistent for $\pmdp$ and $\models_d$, as for both $\varphi$ and $\neg\varphi$, there is a strategy that is not accepting. 
For $\models_a$, there is a single strategy which accepts $\varphi$; other strategies do not affect the relation.

As an example of an accepting region under the demonic relation, consider $R' = [\nicefrac 4 5, \nicefrac 9 {10}] \times [\nicefrac 2 5, \nicefrac 9 {10}]$. 
We have $\pmdp, R' \models_d \varphi$, as for both strategies, the induced probability is always exceeding $\nicefrac 4 5$.
\end{example}

\subsection{Formal problem statements}
\label{subsec:formalprobstatements}

We are now in a position to formalise the two synthesis problems and the verification problem from the introduction, page~\pageref{informal_verification}.
We present the formal problem statements in the order of treatment in the rest of the paper.
\begin{center}
\smallskip\noindent\fcolorbox{black}{black!10}{%
\parbox{0.95\columnwidth}
{\textbf{The formal synthesis problem.}
Given pSG $\psg$, specification $\varphi$, and well-defined region $R$, the \emph{synthesis problem} is to partition $R$ into $R_a$ and $R_r$ such that:
	\[
		 \psg, R_a \models_\clubsuit \varphi 
		 \quad \text{ and } \quad
		 \psg, R_r \models_\heartsuit \neg\varphi.
	\]
	This problem is the topic of \textbf{Section~\ref{sec:solution_fct}}.

	}}
\end{center}

\begin{remark}
The solution function for pMCs precisely describes how (graph-preserving) instantiations map to the relevant measure.
Therefore, comparing the solution function with the threshold divides the parameter space into an accepting region $R_a$ and a rejecting region $R_r$ and defines the exact result for the formal synthesis problem.
Recall also Example~\vref{ex:solution_function_synth}.
\end{remark}

\begin{center}
\smallskip\noindent\fcolorbox{black}{black!10}{%
    \parbox{0.95\columnwidth}
    {\textbf{The formal verification problem.}
Given pSG $\psg$, specification $\varphi$, and well-defined region $R$, the \emph{verification problem} is to check whether: 
\begin{align*}
			   &	\psg, R \models_\clubsuit \varphi & \ (R \text{ is accepting})\\
	\text{or } &\psg, R \models_\heartsuit \neg\varphi & \ (R \text{ is rejecting})\\
	\text{or } &\psg, R \not\models_\clubsuit \varphi \ \land \ \psg, R \not\models_\heartsuit \neg\varphi & \ (R \text{ is inconsistent})
\end{align*}
where $\models_\heartsuit$ denotes the dual satisfaction relation of $\models_\clubsuit$.

\noindent This problem is the topic of \textbf{Section~\ref{sec:exact_mc}--\ref{sec:nondet}}.
}}
\end{center}

The verification procedure allows us to utilise an approximate synthesis problem in which verification procedures are used as a backend.

\begin{center}
\smallskip\noindent\fcolorbox{black}{black!10}{%
    \parbox{0.95\columnwidth}
    {\textbf{The formal approximate synthesis problem.}
Given pSG $\psg$, specification $\varphi$, percentage $c$, and well-defined region $R$, the \emph{approximate synthesis problem} is to partition $R$ into regions $R_a$, $R_o$, and $R_r$ such that:
	\[
		 \psg, R_a \models_\clubsuit \varphi 
		 \quad \text{ and } \quad
		 \psg, R_r \models_\heartsuit \neg\varphi,
	\]
	where $R_a \uplus R_r$ cover at least $c\%$ of the region $R$.    
	
	\noindent This problem is the topic of \textbf{Section~\ref{sec:psp}}.
	}}    
\end{center} 
Note that no requirements are imposed on the (unknown, open) region $R_o$. 

\begin{remark}
\label{rem:robust}
By definition, the angelic satisfaction relation for region $R$ and pSG $\psg$ is equivalent to:
\[ 
\psg, R \models_a \varphi \quad \text{if and only if} \quad
\forall u \in R.~\exists \sched \in \Sched^\psg.~\psg^\sched, u \models \varphi. 
\]
An alternative notion in parameter synthesis is the existence of a \emph{robust} strategy:
\[ 
\exists \sched \in \Sched^\psg.~\forall u \in R.~ \psg^\sched, u \models \varphi. 
\]
Note the swapping of quantifiers compared to $\models_a$.
That is, $\psg, R \models_a \varphi$ considers potentially different strategies for different parameter instantiations $u \in R$. 
The notion of robust strategies leads to a series of quite orthogonal challenges.
For instance, the notion is not compositional, i.e., if robust strategies exist in $R_1$ and $R_2$, then we \emph{cannot} conclude the existence of a robust strategy in $R_1 \cup R_2$. 
Moreover, memoryless strategies are not sufficient, see~\cite{DBLP:conf/qest/ArmingBCKS18}.
Robust strategies are outside the scope of this paper and are only shortly mentioned in 
Section~\ref{sec:nondet}.
\end{remark}

\subsection{A bird's eye view on the verification procedures}
In the later sections, we will present several techniques that decide the verification problem for pMCs and pMDPs.
(Recall that stochastic games were only used to define the general setting.)

The verification problem is used to analyse the regions of interest. 
The assumption that this region contains only well-defined instantiations is therefore natural. 
It can be checked algorithmically as described in Section~\ref{sec:ensuringgraphpres} below. 
Many verification procedures require that the region is graph preserving. A decomposition result of well-defined into graph-preserving regions is given in Section~\ref{sec:decompose}.

Section~\ref{sec:exact_mc} presents two verification procedures. 
The first one directly solves the non-linear equation system, see Example~\vref{ex:nonlinear}, as an SMT query.
The second procedure
reformulates the SMT query using the solution function.
While this reformulation drastically reduces the number of variables in the query, it requires an efficient computation of the solution function, as described in Section~\ref{sec:solution_fct}.

Section~\ref{sec:approx_mc} covers an \emph{approximate} and more efficient verification procedure, called \emph{parameter lifting}, which is tailored to multi-linear functions and closed rectangular regions.
Under these mild restrictions, the verification problem for pMCs (pMDPs) can be approximated using a sequence of \emph{standard verification analyses on non-parametric} MDPs (SGs) of similar size, respectively.
The key steps here are to relax the parameter dependencies, and consider lower- and upper-bounds of parameters as worst and best cases. 
\section{Regions}
\label{sec:regions} 

Section~\ref{sec:regionsintro} already introduced regions. 
This section details specific region structures such as linear, rectangular and graph-preserving regions.
It then presents procedures to check whether a region is graph preserving. 
Finally, we describe how well-defined but not graph-preserving regions can be turned into several regions that are graph preserving.

\subsection{Regions with specific structure}

As defined before, a region $R$ is a (typically uncountably infinite) set of parameter valuations described by a set $C(R)$ of polynomial constraints.
Two classes of regions are particularly relevant: linear and rectangular regions. 

\begin{definition}[Linear region]
A region with representation $C(R)$ is \emph{linear} if for all $\pol \sim 0 \in C(R)$, the polynomial $\pol$ is linear.
\end{definition}
Linear regions describe convex polytopes. We refer to the vertices (or angular points) of the polytope as the \emph{region vertices}. 
\begin{definition}[Rectangular region]\label{def:rectangularRegion}
A region $R$ with representation
\[
C(R) \ = \ \bigcup_{i=1}^{|V|}  \{ \, {-}p_i + a_i \unlhd_i^1 0, p_i + b_i \unlhd_i^2 0 \, \}
\]
with $a_i \leq b_i \in \Q$ and $\unlhd^j_i \in \{ <, \leq \}$ for $0 < i \leq |V|$ and $j \in \{ \, 1,2 \, \}$ is called \emph{rectangular}.
A rectangular region is \emph{closed} if 
all inequalities $\unlhd_i^j$ in the constraints in $C(R)$ are non-strict.
\end{definition}
Rectangular regions are hyper-rectangles and a subclass of linear regions.
A \emph{closed} rectangular region $R$ can be represented as $R = \bigtimes_{p \in V} [a_p, b_p]$ with parameter intervals $[a_p, b_p]$ described by the bounds $a_p$ and $b_p$ for all $p \in V$. 
For a region $R$, we refer to the \emph{bounds} of parameter $p$ by $B_R(p) = \{ a_p, b_p \}$ and to the \emph{interval} of parameter $p$ by $I_R(p) = [a_p,b_p]$. We may omit the subscript $R$, if it is clear from the context.
For a rectangular region $R$, the size $\|R\|$ equals $\prod_{p \in V} (b_p - a_p)$.

Regions represent sets of instantiations $\psg[u]$ of a pSG $\psg$. 
The notion of graph-preservation from Definition~\vref{def:graph_preserving} lifts to regions in a straightforward manner:
\begin{definition}[Graph-preserving region]\label{def:graphpres_region}
Region $R$ is \emph{graph preserving} for pSG $\psg$ if for all $u\in R$, $u$ is a graph-preserving valuation for $\psg$.
\end{definition}
By this definition, all instantiations from graph-preserving regions have the same topology as the parametric model, cf.~Remark~\vref{rem:graphpreserving} below.
In addition, all such instantiations are well-defined.

\begin{example}
\label{ex:regions}
Let $\pdtmc$ be the pMC in Figure~\vref{fig:models:pmc}, $R = [\nicefrac 1 {10}, \nicefrac 4 5] \times [\nicefrac 2 5, \nicefrac 7 {10}]$ be a (closed rectangular) region, and instantiation $u = (\nicefrac 4 5, \nicefrac 3 5) \in R$.
Figure~\vref{fig:pmc:Du} depicts the instantiation $\pdtmc[u]$, an MC with the same topology as $\pdtmc$.
As the topology is preserved for all possible instantiations $\pdtmc[u']$ with $u' \in R$, the region $R$ is graph preserving.
The region $R'=[0,1] \times [0,1]$ is not graph preserving as, \eg, the instantiation $(0,0) \in R'$ results in an MC that has no transition from state $s_1$ to $s_2$.
\end{example}

\begin{remark}\label{rem:graphpreserving}
Graph-preserving regions have the nice property that if \[ \exists u \in R, \psg,u  \models_\clubsuit \p_{=1}(\finally\, T)\text{ implies }\psg, R \models_\clubsuit \p_{=1}(\finally\, T).\]
This property can be checked by standard graph analysis~\cite[Ch.\ 10]{BK08}. 
It is thus straightforward to check $\psg, R \models_\clubsuit \p_{=1}(\finally T)$, an important precondition for computing expected rewards.
In the rest of this paper when considering expected rewards, it is assumed that within a region the probability to reach a target is one.
\end{remark}

The following two properties of regions are frequently (and often implicitly) used in this paper.
\begin{lemma}[Characterisation for inconsistent regions]
For any inconsistent region $R$ it holds that $R = R_a \cup R_r$ for some accepting $R_a \neq \emptyset$ and rejecting $R_r \neq \emptyset$.
\end{lemma}
\begin{lemma}[Compositionality]
Region $R = R_1 \cup R_2$ is accepting (rejecting) if and only if both $R_1$ and $R_2$ are accepting (rejecting).
\end{lemma}
\noindent
The statements follow from the universal quantification over all instantiations in the definition of $\models_\clubsuit$.

\subsection{Checking whether a region is graph preserving}
\label{sec:ensuringgraphpres}
The verification problem for region $R$ requires $R$ to be well-defined.
We first address the problem on how to check this condition.
In fact, we present a procedure to check graph preservation which is slightly more general and useful later, see also Remark~\vref{rem:graphpreserving}.
To show that region $R$ is not graph preserving, a point in $R$ suffices that violates the conditions in Definition~\vref{def:graph_preserving}.
Using the representation of region $R$, the implication
\[ \Upphi(R)  \implies R \text{ graph preserving}\]
needs to be valid since any violating assignment corresponds to a non-graph-preserving instantiation inside $R$. 
Technically, we consider satisfiability of the conjunction of:
\begin{compactitem}
\item the inequalities $C(R)$ representing the candidate region, and
\item a disjunction of (in)equalities describing violating graph-preserving.
\end{compactitem}
This conjunction is satisfiable if and only if the region is not graph preserving.

\subsubsection{Graph preservation for polynomial transition functions} 
Let us consider the above for pSGs with polynomial transition functions. 
The setting for pSGs with rational functions is discussed at the end of this section.
The following constraints \eqref{eq:graph_preserving_polynomial_start}--\eqref{eq:graph_preserving_polynomial_end}, which we denote $\textsl{GP}$, capture the notion of graph preservation:
\begin{align}
& \bigwedge_{\substack{s,s'\in S,\act\in\Act(s)\\\probmdp(s,\act,s') \not\equiv 0}}  0 \leq \probmdp(s,\act,s')  \leq 1  \, \label{eq:graph_preserving_polynomial_start}\\
   \land &\bigwedge_{s\in S,\act\in\Act(s)} \sum_{s'\in S} \probmdp(s,\act,s') = 1 \\
   \land & \bigwedge_{s\in S,\act\in\Act(s)} \rew(s,\act) \geq 0 \label{eq:well_defined_polynomial_end}\\
   \land & \bigwedge_{\substack{s,s'\in S,\act\in\Act(s)\\\probmdp(s,\act,s') \not\equiv 0}} 0 < \probmdp(s,\act,s') . \label{eq:graph_preserving_polynomial_end}
\end{align}
The constraints ensure that (1) all non-zero entries are evaluated to a probability, (2) transition probabilities are probability distributions, (3) rewards are non-negative, and (4) non-zero entries remain non-zero.
The constraints \eqref{eq:graph_preserving_polynomial_start}--\eqref{eq:well_defined_polynomial_end} suffice to ensure well-definedness.
The constrains (1)--(4) can be simplified to:
\begin{align*}
	& \bigwedge_{\substack{s,s'\in S,\act\in\Act(s)\\\probmdp(s,\act,s') \not\equiv 0 }} & &  \probmdp(s,\act,s') > 0  \\
   \land &\bigwedge_{s\in S,\act\in\Act(s)} & & \sum_{s'\in S} \probmdp(s,\act,s') = 1 \\
   \land & \bigwedge_{s\in S,\act\in\Act(s)} & & \rew(s,\act) \geq 0.
\end{align*}

\begin{example}
Recall the pMC from Figure~\vref{fig:models:pmc}. 
\begin{align*}\textsl{GP} = \quad & p > 0 \,\land\, 1{-}p > 0  \, \land \, p{+}1{-}p = 1  \ \land \ q > 0 \, \land \, 1{-}q > 0 \, \land \, q{+}1{-}q = 1. 
\end{align*}
This equation simplifies to $0 < p < 1 \land 0 < q < 1$.
To check whether the region $R$ described by $\Upphi(R) = \nicefrac 1 {{10}} \leq p \leq \nicefrac 4 5 \land \nicefrac 2 5 \leq q \leq \nicefrac 7 {10}$ is graph preserving, we check whether the conjunction $\Upphi(R) \land \neg \textsl{GP}$ is satisfiable, with
\[ \neg \textsl{GP} \ = \ p \leq 0 \lor p \geq 1 \lor q \leq 0 \lor q \geq 1. \]
As the conjunction is not satisfiable, the region $R$ is graph preserving. 
Contrary, $R' = [0,1] \times [0,1]$ is not graph preserving as $u = (0,0)$ satisfies the conjunction $\Upphi(R') \land \neg \textsl{GP}$.
\end{example}
Satisfiability of $\textsl{GP}$, or equivalently, deciding whether a region is graph preserving, is as hard as the existential theory of the reals~\cite{BPR06}, if no assumptions are made about the transition probability and reward functions.
This checking can be automated using SMT-solvers capable of handling quantifier-free non-linear arithmetic over the reals~\cite{demoura_nlsat}. 
The complexity drops to polynomial time once both the region $R$ and all transition probability (and reward) functions are linear as linear programming has a polynomial complexity and the formula is then a disjunction over linear programs (with trivial optimisation functions).

\subsubsection{Graph preservation for rational transition functions}\label{subsec:handlinginequalities}
In case the transition probability and reward function of a pSG are not polynomials, the left-hand side of the statements in \eqref{eq:graph_preserving_polynomial_start}--\eqref{eq:graph_preserving_polynomial_end} are not polynomials, and the statements would not be constraints. 
We therefore perform the following transformations on \eqref{eq:graph_preserving_polynomial_start}--\eqref{eq:graph_preserving_polynomial_end}:

\begin{itemize}
\item Transforming equalities:
\[
\frac{g_1}{g_2} = c \quad \mbox{becomes} \quad g_1 - c \cdot g_2 = 0 \land g_2 \neq 0
\quad \mbox{ with } c \in \Q. 
\]
%
%
\item Transforming inequalities $\unrhd\, \in \{ >, \geq \}$:
\begin{align}
\frac{g_1}{g_2} \unrhd c \: \mbox{ becomes } \:
g_2 \neq 0 \land \Big( (g_2  > 0 \land g_1 \unrhd\,c \cdot g_2)  \lor	(g_2 < 0 \land g_1 \not\unrhd\;c \cdot g_2 )\Big)\nonumber
\end{align}
with $c \in \Q$, and $\not\unrhd$ equals $<$ for $\not >$ and $\leq$ for $\not \geq$.
\item Transforming $<, \leq$ is analogous. 
\item Transforming $g \neq g'$ (i.e., $g < g' \lor g > g'$) involves transforming both disjuncts.
\end{itemize}
The result is a formula with polynomial constraints that correctly describes graph preservation (or well-definedness).
\begin{example}
	Consider a state with outgoing transition probabilities $q$ and $\frac{p}{1+p}$.
	The graph preservation statements are (after some simplification):
	\[  q > 0 \text{ and } \frac{p}{1+p} > 0 \text{ and } q +  \frac{p}{1+p} = 1. \]
	Transforming the second item as explained above yields:
	\begin{align*}
 	 1+p \neq 0 \land \Big((1+p > 0 \land p > 0) \lor (1+p < 0 \land p < 0)  \Big)
	\end{align*}
	while transforming the third item yields:
	\begin{align*}
	(1+p \neq 0) \land q \cdot (1{+}p) -1 = 0.	
	\end{align*}
Finally, we obtain the following formula (after some further simplifications):
\[ q>0 \;\land\; \left( p>0 \lor p<-1 \right) \;\land\; q \cdot (1+p) - 1  = 0.  \]
\end{example}

\subsection{Reduction to graph-preserving regions}
\label{sec:decompose}
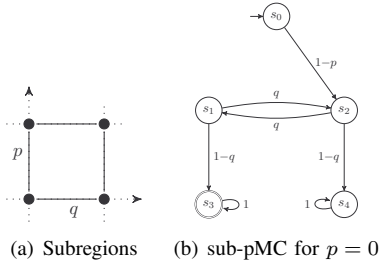
\begin{figure}
\centering
\subfigure[Subregions]{
\scalebox{2}{
\centering
\begin{tikzpicture}
	\node[circle, draw, fill, scale=0.4] at (0,0) {};
	\node[circle, draw, fill, scale=0.4] at (1,0) {};
	\node[circle, draw, fill, scale=0.4] at (0,1) {};
	\node[circle, draw, fill, scale=0.4] at (1,1) {};
	\draw[-] (0,0.1) -- (0,0.9);
	\draw[-] (1,0.1) -- (1,0.9);
	\draw[-] (0.1,0) -- (0.9,0);
	\draw[-] (0.1,1) -- (0.9,1);
	\draw[dotted,->] (0,-0.3) -- node[left, scale=0.7] {$p$} (0,1.5);
	\draw[dotted,->] (-0.3,0) -- node[below, scale=0.7] {$q$} (1.5,0);
	
	\draw[dotted,-] (1,-0.3) -- (1,1.3);
	\draw[dotted,-] (-0.3,1) -- (1.3,1);
	
\end{tikzpicture}
}
\label{fig:subregions:partitioning}
}
\subfigure[sub-pMC for $p=0$]{
\scalebox{1}{
\begin{tikzpicture}[scale=1, nodestyle/.style={draw,circle}]
    \draw[white, use as bounding box] (-2.7,-5.5) rectangle (2.7,0.3); 
    
    \node [nodestyle] (s0) at (0,0) {$s_0$};
    \node [nodestyle] (s1) [on grid, below left=2.5cm and 1.8cm of s0] {$s_1$};
    \node [nodestyle] (s2) [on grid, below right=2.5cm and 1.8cm of s0] {$s_2$};
    \node [nodestyle, accepting] (s3) [on grid, below=2.5cm of s1] {$s_3$};
    \node [nodestyle] (s4) [on grid, below=2.5cm of s2] {$s_4$};
    
    \draw ($(s0)-(0.7,0)$) edge[->] (s0);
    \draw (s0) edge[->] node[right] {\small$1{-}p$} (s2);

     \draw (s1) edge[bend left=10, ->] node[above] {\small$q$} (s2);
     \draw (s1) edge[->] node[right] {\small$1{-}q$} (s3);
    
     \draw (s2) edge[bend left=10, ->] node[below] {\small$q$} (s1);
     \draw (s2) edge[->] node[left] {\small$1{-}q$} (s4);
    
    \draw (s3) edge[loop right, ->] node[auto] {\small$1$} (s3);
    
    \draw (s4) edge[loop left, ->] node[auto] {\small$1$} (s3);
    
\end{tikzpicture}	
}
\label{fig:subregions:subpmc}
}

\caption{Ensuring graph-preservation on subregions.}
\label{fig:subregions}
\end{figure}

In this section, we show how we can partition a well-defined region into a set of graph-preserving regions. This is useful, e.g., as we only define solution functions for graph-preserving  regions. The decomposition in this section allows to define solution functions on each of these partitions, see also  Remark~\vref{rem:solfuncgp}.
Before we illustrate the decomposition, we define sub-pSGs: Given two pSGs $\psgInit$ and $\psgInit[']$, $\psg'$ is a \emph{sub-pSG} of $\psg$ if $S' \subseteq S$, $V'\subseteq V$, $\sinit'=\sinit \in S'$, $\Act' \subseteq \Act$, and $\probmdp'(s,\act,s')\in\{ \probmdp(s,\act,s'), 0 \}$ for all $s,s'\in S'$ and $\alpha\in\Act'$.
Note that for a given state $s\in S$ and action $\act\in\Act(s)$, the sub-pSG might not contain $s$ or $\act$ might not be enabled in $s$, but it is also possible that the sub-pSG omits some but not all successors of $\act$ in $s$.
\begin{example}
Reconsider the pMC $\pdtmc$ from Figure~\vref{fig:models:pmc}, and let $R=[0,1] \times [0,1]$, which is well-defined but not graph preserving.
Region $R$ can be partitioned into $9$ regions, see Figure~\ref{fig:subregions:partitioning} where each dot, line segment, and the inner region are subregions of $R$. 
All subregions are graph preserving on some sub-pMC of $\pdtmc$.
Consider, e.g., the line-region $R' = \{ u \in R \mid p[u] = 0 \}$.
The subregion $R'$ is not graph preserving on pMC $\pdtmc$, as the transition $s_0 \xrightarrow{p} s_1$ vanishes when $p=0$.
However, $R'$ is graph preserving on the sub-pMC $\pdtmc'$ in Figure~\ref{fig:subregions:subpmc}, which is obtained from $\pdtmc$ by removing the transitions on the line-region $p{=}0$.
\end{example}

Let us formalise the construction from this example.
For a given well-defined region $R$, and pSG $\psg$, let $\mathcal{Z}_R$ describe the set of constraints:
\[
\begin{array}{l@{}l}
  \{ \probmdp(s,\act,s') {=} 0\ |\  & s, s' \in S\wedge \act \in \Act(s)\wedge \probmdp(s,\act,s')\not\equiv 0\wedge \exists u \in R.\,\probmdp(s,\act,s')[u] = 0 \}.
\end{array}
\]
For $X \subseteq \mathcal{Z}_R$, the subregion $R_X \subseteq R$ is defined as:
\[
\Upphi(R_X) \ = 
   \Upphi(R) \land 
   \bigwedge_{c \in X} c \land 
   \bigwedge_{c \in \mathcal{Z}_R \setminus X} \neg c. 
\]
It follows that $X$ uniquely characterises which transition probabilities in $\psg$ are set to zero.
In fact, each instance in $R_X$ is graph preserving for the unique sub-pSG $\psg'$ of $\psg$ obtained from $\psg$ by removing all zero-transitions in $R_X$.
The pSG $\psg'$ is well-defined as $R$ on $\psg$ is well-defined. 
By construction, it holds that $\psg[u] = \psg'[u]$ for all instantiations $u \in R'$.

\section{Exact Synthesis by Computation of the Solution Function}\label{sec:solution_fct}
This section discusses \emph{how to compute the solution function}. 
The solution function for pMCs describes the exact accepting and rejecting regions, as discussed in Section~\ref{subsec:formalprobstatements}\footnote{for pMDPs, one may compute a solution function for every strategy, but this has little practical relevance}. This section thus provides an algorithmic approach to the exact synthesis problem. 
In Section~\ref{sec:exact_mc}, we will also see that the solution function may be beneficial for the performance of SMT-based (region) verification.

The original approach to compute the solution function of pMCs is via \emph{state elimination}~\cite{Daws04,param_sttt}, and is analogous to the computation of regular expressions from nondeterministic finite automata (NFAs)~\cite{HMU-Automata}. 
It is suitable for a range of indefinite-horizon properties. 
The core idea behind state elimination and the related approaches presented here is based on two operations:
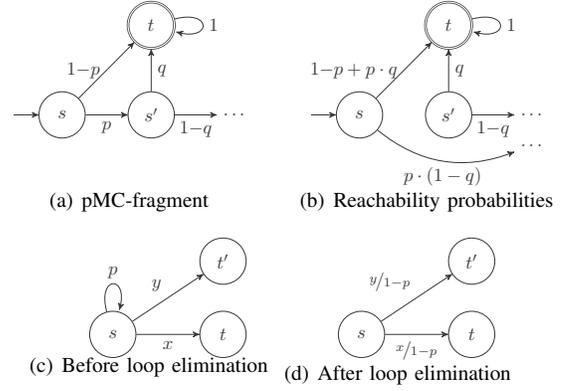
\begin{figure}
\centering
\subfigure[pMC-fragment]{
\scalebox{1}{
	\begin{tikzpicture}[baseline=(s0)]
		\node[state, , initial, initial text=] (s0) {$s$} ;
		\node[state, right=0.8cm of s0] (s1) {$s'$} ;
		\node[state, , above=0.8cm of s1, accepting] (s2) {$t$} ;
		\node[ right=0.8cm of s1] (s3) {$\hdots$} ;
		\node[below=0.3cm of s3] (s4) {\phantom{$\hdots$}} ;

		\draw[->] (s0) -- node[below] {$p$} (s1);
		\draw[->] (s1) -- node[right] {$q$} (s2);
		\draw[->] (s0) -- node[left] {$1{-}p$} (s2);
		\draw[->] (s1) -- node[below] {$1{-}q$} (s3);
		\draw[->] (s2) edge[loop right] node {$1$} (s2);
		\draw[->,white] (s0) edge[bend right] node[below] {\phantom{$p \cdot (1-q)$}} (s4);
	\end{tikzpicture}
	}
	\label{fig:solutionfunction:shortcut}
}
\subfigure[Reachability probabilities]{
\scalebox{1}{
	\begin{tikzpicture}[baseline=(s0)]
		\node[state , initial, initial text=] (s0) {$s$} ;
		\node[state , right=0.8cm of s0] (s1) {$s'$} ;
		\node[state , above=0.8cm of s1, accepting] (s2) {$t$} ;
		\node[right=0.8cm of s1] (s3) {$\hdots$} ;
		\node[below=0.3cm of s3] (s4) {$\hdots$} ;
		
		\draw[->] (s1) -- node[right] {$q$} (s2);
		\draw[->] (s0) -- node[left] {$1{-}p + p \cdot q$} (s2);
		\draw[->] (s1) -- node[below] {$1{-}q$} (s3);
		\draw[->] (s2) edge[loop right] node {$1$} (s2);
		\draw[->] (s0) edge[bend right] node[below] {$p \cdot (1-q)$} (s4);
	\end{tikzpicture}
	}
	\label{fig:solutionfunction:shortcut_applied}
}

\subfigure[Before loop elimination]{
\scalebox{1}{
	\begin{tikzpicture}[baseline=(s1)]
	\node[state] (s1) {$s$};
	\node[state, right=1.2cm of s1] (s2) {$t$};
	\node[state, above=0.5cm of s2] (s3) {$t'$};
	\node[left=0.6cm of s1] (bbl) {};
	\node[right=0.9cm of s1] (bbr) {};
	\draw[->] (s1) edge[loop above] node[auto] {$p$} (s1);
	\draw[->] (s1) edge node[below] {$x$} (s2);
	\draw[->] (s1) edge node[auto] {$y$} (s3);
\end{tikzpicture}
	}
	\label{fig:solutionfunction:with_loop}
}
\subfigure[After loop elimination]{
\scalebox{1}{
	\begin{tikzpicture}[baseline=(s1)]
	\node[state] (s1) {$s$};
	\node[state, right=1.2cm of s1] (s2) {$t$};
	\node[state, above=0.5cm of s2] (s3) {$t'$};
	\node[left=0.6cm of s1] (bbl) {};
	\node[right=0.9cm of s1] (bbr) {};
	\draw[->] (s1) edge node[below] {$\nicefrac{x}{1-p}$} (s2);
	\draw[->] (s1) edge node[auto] {$\nicefrac{y}{1-p}$} (s3);
\end{tikzpicture}
	}
	\label{fig:solutionfunction:without_loop}
}
\caption{Essential ideas for state elimination}
\end{figure}

\begin{itemize}
\item \emph{Adding short-cuts:}
	Consider the pMC-fragment in Figure~\vref{fig:solutionfunction:shortcut}. 
	The reachability probabilities from any state to $t$ are as in Figure~\ref{fig:solutionfunction:shortcut_applied}, where we replaced the transition from $s$ to $s'$ by shortcuts from $s$ to $t$ and all other successors of $s'$, bypassing $s'$. 
	By successive application of shortcuts, any path from the initial state to the target state eventually has length $1$.
\item \emph{Elimination of self-loops:}
	A prerequisite for introducing a short-cut is that the bypassed state is loop-free.
	Recall that the probability of staying forever in a non-absorbing state is zero, and justifies elimination of self-loops by rescaling all other outgoing transitions, as depicted in the transition from Figure~\ref{fig:solutionfunction:with_loop} to Figure~\ref{fig:solutionfunction:without_loop}.
\end{itemize}

The remainder of this section is organised as follows: Section~\ref{subsec:stateelimobs} recaps the original state elimination approach in Section~\ref{subsec:stateelim}, albeit slightly rephrased. The algorithm is given for (indefinite) reachability probabilities, expected rewards, and bounded reachability probabilities.
In the last part, we present alternative, equivalent formulations which sometimes allow for superior performance. In particular, Section~\ref{sec:lineq} clarifies the relation to solving a linear equation system over a field of rational functions, and Section~\ref{sec:sucshortcut} discusses a variation of state elimination applicable to pMCs described by multi-terminal binary decision diagrams.


\subsection{Algorithm based on state elimination}
\label{subsec:stateelim}
Let $T \subseteq S$ be a set of target states and assume w.\,l.\,o.\,g.\, that all states in $T$ are absorbing and that $\sinit \not\in T$.  

\subsubsection{Reachability probabilities}
We describe the algorithm to compute reachability probabilities based on state elimination in Algorithm~\vref{alg:elim}. 
In the following, $\probdtmc$ is the transition matrix.
The function \textbf{eliminate\_selfloop}$(\probdtmc, s)$ rescales all outgoing probabilities of a non-absorbing state $s$ by eliminating its self-loop.
The function \textbf{eliminate\_transition}($\probdtmc, s_{1}, s_{2}$) adds a shortcut from $s_1$ to the successors of $s_2$.
Both operations preserve reachability to $T$. 
The function \textbf{eliminate\_state}$(\probdtmc, s)$ ``bypasses'' a state $s$ by adding shortcuts from all its predecessors.
More precisely, we eliminate the incoming transitions of $s$, and after all incoming transitions are removed, the state $s$ is unreachable. 
It is thereby effectively removed from the model.

After removing all non-absorbing, non-initial states $S^?$, the remaining model contains only self-loops at the absorbing states and transitions emerging from the initial state. 
Eliminating the self-loop on the initial state (by rescaling) yields a pMC. In this pMC, after a single step, an absorbing state is reached. These absorbing states are either a target or a sink. The solution function is then the sum over all (one-step) transition probabilities to target states.

\begin{algorithm}[t]
\caption{State elimination for pMCs}
  \begin{newalgorithm}{reachability}{pMC $\PdtmcInit$, $T\subseteq S$}
    \State $S^{?}\coloneqq\{s\in S\mid s\neq\sinit \land s \in \finally T   \setminus T \}$\\
    \While{$S^{?}\neq\emptyset$}
    	\State select $s \in S^{?}$\\
    	\State \textbf{eliminate\_selfloop}($\probmdp, s$)\\
        \State \textbf{eliminate\_state}($\probmdp, s$) \\
        \State $S^{?} := S^{?} \setminus \{s\}$ \\
    \EndWhile
    \State \textbf{eliminate\_selfloop}($\probmdp, \sinit$) \label{alg:elim:bypass_init} \\
    \Comment{All $S^?$ eliminated. Only direct transitions to target.}\\
    \Return{$\sum\limits_{t\in T}\probmdp(\sinit,t)$}
  \end{newalgorithm}
  \label{alg:elim}

\begin{newalgorithm}{eliminate\_selfloop}{$\probdtmc, s \in S$}
    \Require $\probdtmc(s,s) \neq 1$\\
      \ForEachx {$\sOut\in\successor(s), s \neq \sOut$}
        \Statex $\probdtmc(s,\sOut) \coloneqq \frac{\probdtmc(s,s_2)}{1-\probdtmc(s,s)}$ \\
      \EndFor
      \Statex $\probdtmc(s, s) \coloneqq 0$ \\
    \EndIf
    
  \end{newalgorithm}

\begin{newalgorithm}{eliminate\_transition}{$\probdtmc, \sIn \in S, s \in S$}
    \setcounter{linenno}{\value{oldlinenumber}}
     \Require $s_1 \in \predecessor(s)$, $\probdtmc(s,s) = 0$\\
     \ForEachx {$\sOut\in\successor(s)$}
        \Statex $\probdtmc(\sIn,\sOut) \coloneqq \probdtmc(\sIn,\sOut)+\probdtmc(\sIn,s)\cdot \probdtmc(s,\sOut)$ \\
      \EndFor
      \Statex $\probdtmc(\sIn, s) \coloneqq 0$ \\
    \end{newalgorithm}

  \begin{newalgorithm}{eliminate\_state}{$\probdtmc, s \in S$}
  \Require $\probdtmc(s,s) = 0$\\
    \ForEachx {$\sIn \in \predecessor(s)$} \label{alg:elim:loop}
      \Statex \textbf{eliminate\_transition}($\probdtmc, \sIn, s$)\\
    \EndFor
      \end{newalgorithm}

\end{algorithm}

\begin{figure*}[t]
\centering
\subfigure[pMC]{
\scalebox{\picscale}{
      \begin{tikzpicture}[scale=1, nodestyle/.style={draw,circle},baseline=(s0)]
    
    \node [nodestyle] (s0) at (0,0) {$s_0$};
    \node [] (leftdummy)  [on grid, left=1.2cm of s0] {};
    \node [] (rightdummy) [on grid, right=1.2cm of s0] {};
    \node [nodestyle] (s1) [on grid, below=\distforsone of leftdummy] {$s_1$};
    \node [nodestyle] (s2) [on grid, below=\distforsone of rightdummy] {$s_2$};
    \node [nodestyle, accepting] (s3) [on grid, below=\distsonesthree of s1] {$s_3$};
    \node [nodestyle] (s4) [on grid, below=\distsonesthree of s2] {$s_4$};
    
    \draw ($(s0)-(0.7,0)$) edge[->] (s0);
    \draw (s0) edge[->] node[right] {\scriptsize$p$} (s1);
    \draw (s0) edge[->] node[right] {\scriptsize$1{-}p$} (s2);

     \draw (s1) edge[bend left=15, ->] node[above] {\scriptsize$q$} (s2);
     \draw (s1) edge[->] node[right] {\scriptsize$1{-}q$} (s3);
    
     \draw (s2) edge[bend left=15, ->] node[below] {\scriptsize$q$} (s1);
     \draw (s2) edge[->] node[left] {\scriptsize$1{-}q$} (s4);
    
    \draw (s3) edge[loop right, ->] node[auto] {\scriptsize$1$} (s3);
    
    \draw (s4) edge[loop left, ->] node[auto] {\scriptsize$1$} (s3);

\end{tikzpicture}
    }
    \label{fig:se:D}
    }
    \subfigure[Eliminating  $s_1 \rightarrow s_2$]{
\scalebox{\picscale}{
\begin{tikzpicture}[scale=1, nodestyle/.style={draw,circle},baseline=(s0)]
    
    \node [nodestyle] (s0) at (0,0) {$s_0$};
    \node [] (leftdummy)  [on grid, left=1.2cm of s0] {};
    \node [] (rightdummy) [on grid, right=1.2cm of s0] {};
    \node [nodestyle] (s1) [on grid, below=\distforsone of leftdummy] {$s_1$};
    \node [nodestyle] (s2) [on grid, below=\distforsone of rightdummy] {$s_2$};
    \node [nodestyle, accepting] (s3) [on grid, below=\distsonesthree of s1] {$s_3$};
    \node [nodestyle] (s4) [on grid, below=\distsonesthree of s2] {$s_4$};
    
    \draw ($(s0)-(0.7,0)$) edge[->] (s0);
    \draw (s0) edge[->] node[right] {\scriptsize$p$} (s1);
    \draw (s0) edge[->] node[right] {\scriptsize$1{-}p$} (s2);

	 \draw (s1) edge[->, loop left] node[left] {\scriptsize$q^2$} (s1); 
     \draw (s1) edge[->] node[right] {\scriptsize$1{-}q$} (s3);
     \draw (s1) edge[->] node[below] {\scriptsize$q{-}q^2$} (s4);
    
     \draw (s2) edge[bend left=15, ->] node[below] {\scriptsize$q$} (s1);
     \draw (s2) edge[->] node[left] {\scriptsize$1{-}q$} (s4);
    
    \draw (s3) edge[loop right, ->] node[auto] {\scriptsize$1$} (s3);
    
    \draw (s4) edge[loop left, ->] node[auto] {\scriptsize$1$} (s3);

\end{tikzpicture} 	
}
\label{fig:se:es1s2}
}
    \subfigure[Eliminating  $s_0 \rightarrow s_2$]{
\scalebox{\picscale}{
\begin{tikzpicture}[scale=1, nodestyle/.style={draw,circle},baseline=(s0)]
    
    \node [nodestyle] (s0) at (0,0) {$s_0$};
    \node [] (leftdummy)  [on grid, left=1.2cm of s0] {};
    \node [] (rightdummy) [on grid, right=1.2cm of s0] {};
    \node [nodestyle] (s1) [on grid, below=\distforsone of leftdummy] {$s_1$};
    \node [nodestyle] (s2) [on grid, below=\distforsone of rightdummy] {$s_2$};
    \node [nodestyle, accepting] (s3) [on grid, below=\distsonesthree of s1] {$s_3$};
    \node [nodestyle] (s4) [on grid, below=\distsonesthree of s2] {$s_4$};
    
    \draw ($(s0)-(0.7,0)$) edge[->] (s0);
    \draw (s0) edge[->] node[right, pos=0.25] {\scriptsize$p{+}q{-}pq$} (s1);
    \draw (s0) edge[->, bend left=60] node[left, pos=0.45] {\scriptsize$(1{-}p)(1{-}q)$} (s4);
    
	 \draw (s1) edge[->, loop left] node[left] {\scriptsize$q^2$} (s1); 
     \draw (s1) edge[->] node[right] {\scriptsize$1{-}q$} (s3);
     \draw (s1) edge[->] node[below] {\scriptsize$q{-}q^2$} (s4);
    
     \draw (s2) edge[bend left=15, ->] node[below] {\scriptsize$q$} (s1);
     \draw (s2) edge[->] node[left] {\scriptsize$1{-}q$} (s4);
    
    \draw (s3) edge[loop right, ->] node[auto] {\scriptsize$1$} (s3);
    
    \draw (s4) edge[loop left, ->] node[auto] {\scriptsize$1$} (s3);

\end{tikzpicture} 	
}
\label{fig:se:es0s2}
}
    \subfigure[Remove unreachable state $s_2$]{
\scalebox{\picscale}{
\begin{tikzpicture}[scale=1, nodestyle/.style={draw,circle},baseline=(s0)]
    
    \node [nodestyle] (s0) at (0,0) {$s_0$};
    \node [] (leftdummy)  [on grid, left=1.2cm of s0] {};
    \node [] (rightdummy) [on grid, right=1.2cm of s0] {};
    \node [nodestyle] (s1) [on grid, below=\distforsone of leftdummy] {$s_1$};
    \node [nodestyle, accepting] (s3) [on grid, below=\distsonesthree of s1] {$s_3$};
    \node [nodestyle] (s4) [on grid, below=\distsonesthree of s2] {$s_4$};
    
    \draw ($(s0)-(0.7,0)$) edge[->] (s0);
    \draw (s0) edge[->] node[right, pos=0.25] {\scriptsize$p{+}q{-}pq$} (s1);
    \draw (s0) edge[->, bend left=45] node[left, pos=0.55] {\scriptsize$(1{-}p)(1{-}q)$} (s4);
    
	 \draw (s1) edge[->, loop left] node[left] {\scriptsize$q^2$} (s1); 
     \draw (s1) edge[->] node[right] {\scriptsize$1{-}q$} (s3);
     \draw (s1) edge[->] node[below] {\scriptsize$q{-}q^2$} (s4);
    
    \draw (s3) edge[loop right, ->] node[auto] {\scriptsize$1$} (s3);
    
    \draw (s4) edge[loop left, ->] node[auto] {\scriptsize$1$} (s3);

\end{tikzpicture} 	
}
\label{fig:se:es2}
}
   \subfigure[Eliminate loop on $s_1$]{
\scalebox{\picscale}{
\begin{tikzpicture}[scale=1, nodestyle/.style={draw,circle},baseline=(s0)]
    
    \node [nodestyle] (s0) at (0,0) {$s_0$};
    \node [] (leftdummy)  [on grid, left=1.2cm of s0] {};
    \node [] (rightdummy) [on grid, right=1.2cm of s0] {};
    \node [nodestyle] (s1) [on grid, below=\distforsone of leftdummy] {$s_1$};
    \node [nodestyle, accepting] (s3) [on grid, below=\distsonesthree of s1] {$s_3$};
    \node [nodestyle] (s4) [on grid, below=\distsonesthree of s2] {$s_4$};
    
    \draw ($(s0)-(0.7,0)$) edge[->] (s0);
    \draw (s0) edge[->] node[right, pos=0.25] {\scriptsize$p{+}q{-}pq$} (s1);
    \draw (s0) edge[->, bend left=45] node[left, pos=0.55] {\scriptsize$(1{-}p)(1{-}q)$} (s4);
    
     \draw (s1) edge[->] node[right] {\scriptsize$\frac{1}{1+q}$} (s3);
     \draw (s1) edge[->] node[below] {\scriptsize$\frac{q}{1+q}$} (s4);
    
    \draw (s3) edge[loop right, ->] node[auto] {\scriptsize$1$} (s3);
    
    \draw (s4) edge[loop left, ->] node[auto] {\scriptsize$1$} (s3);

\end{tikzpicture} 	
}
\label{fig:se:es1s1}
}
   \subfigure[Eliminate $s_1$]{
\scalebox{\picscale}{
\begin{tikzpicture}[scale=1, nodestyle/.style={draw,circle},baseline=(s0)]
    
    \node [nodestyle] (s0) at (0,0) {$s_0$};
    \node [] (leftdummy)  [on grid, left=1.2cm of s0] {};
    \node [] (rightdummy) [on grid, right=1.2cm of s0] {};
    \node [nodestyle, accepting] (s3) [on grid, below=\distsonesthree of s1] {$s_3$};
    \node [nodestyle] (s4) [on grid, below=\distsonesthree of s2] {$s_4$};
    
    \draw ($(s0)-(0.7,0)$) edge[->] (s0);
    \draw (s0) edge[->] node[left, pos=0.4] {\scriptsize$(p{+}q{-}pq) \cdot \frac{1}{1+q}$} (s3);
    \draw (s0) edge[->] node[right, pos=0.45] {\scriptsize$1-\left((p{+}q{-}pq) \cdot \frac{1}{1+q}\right)$} (s4);

    \draw (s3) edge[loop right, ->] node[auto] {\scriptsize$1$} (s3);
    
    \draw (s4) edge[loop left, ->] node[auto] {\scriptsize$1$} (s3);

\end{tikzpicture} 	
}
\label{fig:se:final}
}
\caption{State elimination exemplified}
\end{figure*}

\begin{example}
Consider again the pMC from Example~\vref{ex:models}, also depicted in Figure~\vref{fig:se:D}. Assume state $s_2$ is to be eliminated. 
Applying the function \textbf{eliminate\_state}($\probmdp, s_2$), we first eliminate the transition $s_1 \rightarrow s_2$, which yields Figure~\ref{fig:se:es1s2}, and subsequently eliminate the transition $s_0 \rightarrow s_2$ (Figure~\ref{fig:se:es0s2}). 
State $s_2$ is now unreachable, so we can eliminate $s_2$, reducing computational effort when eliminating state $s_1$.
For state $s_1$, we first eliminate the self-loop (Figure~\ref{fig:se:es1s1}) and then eliminate the transition $s_0 \rightarrow s_1$. The final result, after additionally removing the now unreachable $s_1$, is depicted in Figure~\ref{fig:se:final}.
The result, i.e., the probability to eventually reach $s_3$ from $s_0$ in the original model, can now be read from the single transition between these two states.   
\end{example}

As for computing of regular expressions from NFAs,
the order in which the states are eliminated is essential.
Computing an optimal order with respect to minimality of the result, however, is already NP-hard for acyclic NFAs, see~\cite{DBLP:journals/fuin/Han13}.
For state elimination on pMCs, the analysis is more intricate, as the cost of every operation crucially depends on the size and the structure of the rational functions.
We briefly discuss the implemented heuristics in Section~\ref{sec:impl:stateelim}.

\begin{remark}
The elimination of self-loops yields a rational function. 
In order to keep these functions as small as possible, it is natural to eliminate common factors of the numerator and the denominator.
Such a reduction, however, involves the computation of greatest common divisors (gcds). 
This operation is expensive for multivariate polynomials.
In~\cite{jansen-et-al-qest-2014}, data structures to avoid their computation are introduced,
in~\cite{gandalf-journal} a method is presented that mostly avoids introducing common factors.
\end{remark}

\subsubsection{Expected rewards}
The state elimination approach can also be adapted to compute \emph{expected rewards}~\cite{param_sttt}. 
When eliminating a state $s$, in addition to adjusting the probabilities of the transitions from all predecessors $s_1$ of $s$ to all successors $s_2$ of $s$, it is also necessary to ``summarise'' the reward that would have been gained from $s_1$ to $s_2$ via $s$. 
The presentation in \cite{param_sttt} describes these operations on so-called transition rewards.
Observe that for the analysis of expected rewards in MCs, we can always reformulate transition rewards in terms of state rewards.
We preprocess pMCs to only have rewards at the states: this adjustment simplifies the necessary operations considerably.

The treatment of the expected reward computation is easiest from an adapted (and more performant) implementation of state elimination, as outlined in Algorithm~\vref{alg:elim_rephrased}.
Here, we eliminate the probabilities to reach a target state in exactly one step, and collect these probabilities in a vector $x$ which we refer to as \emph{one-step-probabilities}. 
Then, we proceed similar as before.
However, the elimination of a transition from $s_1$ to $s$ now has two effects: 
it updates the probabilities within the non-target states as before,
and (potentially) updates the probability $x(s_1)$ to reach the target within one step from $s_1$ (with the probability that the target was reached via $s$ in two steps).
Upon termination of the outer loop, the vector $x$ contains the probabilities from all states to reach the target, that is, $x(s_i) = x_{s_i}$.

Finally, when considering rewards, the \emph{one-step-probabilities} contain initially the rewards for the states. 
Eliminating a transition then moves the (expected) reward to the predecessors by the \emph{same} sequence of arithmetic operations.

\begin{algorithm}[t]
\caption{State elimination with one-step probabilities}

  \begin{newalgorithm}{reachability}{pMC $\PdtmcInit$, $T \subseteq S$}
    \State $S^{?}\coloneqq\{s\in S\mid  s \in \finally T   \setminus T \}$\\
    \Comment $x\colon S^? \rightarrow [0,1]$\\
    \Statex $x(s) := \sum_{t\in T} \probmdp(s,t)$ for each $s \in S^{?}$ \\
    \Statex $\probmdp(s,t) \coloneqq 0$ for all $s \in S, t \in T$\\
    \While{$S^{?}\neq\emptyset$}
    	\State \textbf{eliminate\_state}($\probmdp,x, s$) for some $s \in S^{?}$ \\
        \State $S^{?} := S^{?} \setminus \{s\}$ \\
    \EndWhile
    \Comment{All $S^?$ eliminated. One-step probability is reachability probability.}\\
    \Return{$x(\sinit)$}
  \end{newalgorithm}

\begin{newalgorithm}{eliminate\_transition}{$\probmdp, x,\sIn \in S, s \in S$}
     \Comment{Algorithm modifies $\probmdp$}
    \setcounter{linenno}{\value{oldlinenumber}}
     \Require $s_1 \neq s$, $\probmdp(s,s) \neq 1$\\
     \Statex $x(s_1) \coloneqq x(s_1) + \frac{\probdtmc(\sIn,s)\cdot x(s)}{1-\probmdp(s,s)}$ \\
     \ForEachx {$\sOut\in\successor(s), s \neq \sOut$}
     	\Statex $\probmdp(\sIn,\sOut) \coloneqq \probmdp(\sIn,\sOut)+\frac{P(\sIn,s)\cdot \probmdp(s,\sOut)}{1-\probmdp(s,s)}$ \\
      \EndFor
      \Statex $\probdtmc(\sIn, s) \coloneqq 0$ \\
    \end{newalgorithm}

  \begin{newalgorithm}{eliminate\_state}{$\probmdp, x, s \in S$}
   \Comment{Algorithm modifies $\probmdp$}
  \Require $\probmdp(s,s) = 0$
    
    \ForEachx {$\sIn \in \predecessor(s)$}
      \Statex \textbf{eliminate\_transition}($\probmdp, x, \sIn, s$)\\
    \EndFor
      \end{newalgorithm}

  \label{alg:elim_rephrased}
\end{algorithm}

\subsubsection{Bounded reachability}
As discussed in Remark~\vref{rem:noboundedreach},
bounded reachability can typically be considered by an unfolding of the Markov model and considering an unbounded reachability property on that (acyclic) unfolding. 
In combination with state elimination, that yields the creation of many states that are eliminated afterwards, and does not take into account any problem-specific properties.
Rather, and analogous to the parameter-free case~\cite{BK08},
it is better to do the adequate matrix-vector multiplication (\# number of steps often).
The matrix originates from the transition matrix, 
the vector (after $i$ multiplications) encodes the probability to reach a state within $i$ steps.

\subsection{Algorithm based on solving the linear equation system}
\label{sec:lineq}
The following set of equations is a straightforward adaption of the Bellman linear equation system for MCs found in, e.g.,~\cite{Put94,BK08} to pMCs.
For each state $s$, a variable $x_s$ is used to express the probability $\pr_s(\finally T)$ to reach a state in $T$ from the state $s$. 
Recall that we overloaded $\finally T$ to also denote the set of states from which $T$ is reachable (with positive probability).
Analogously, we use $\neg\finally T$ to denote the set of states from which $T$ is not reachable, \ie, $\neg\finally T=S\setminus\finally T$. We have:

\begin{align}
\label{eq:pmcencodingstart}
	x_s &= 0  &\quad & \forall s \in \neg\lozenge T \\
	x_s &= 1  &\quad & \forall s \in T \\
    x_s &= \sum_{s' \in S} \probmdp(s,s') \cdot x_{s'} & \quad  & \forall s \in \lozenge T \setminus T.
\label{eq:pmcencodingend}    
\end{align}

This system of equations has a unique solution for every well-defined parameter instantiation. 
In particular, the set of states satisfying $\neg\lozenge T$ is the same for all well-defined graph-preserving parameter instantiations, as instantiations that maintain the graph of the pMC do not affect the reachability of states in $T$.

For pMCs, the coefficients are no longer from the field of the real numbers, but rather from the field of rational functions. 
\begin{example}
\label{example:linear_equation_system}
Consider the equations for the pMC from Figure~\vref{fig:se:D}.
\begin{align*}
x_0 = &~ p \cdot x_1 + (1-p) \cdot x_2 \\
x_1 = &~ q \cdot x_2 + (1-q) \cdot x_3 \\
x_2 = &~ q \cdot x_1 + (1-q) \cdot x_4 \\
x_3 = &~ 1 \\
x_4 = &~ 0.
\end{align*}
Bringing the system in normal form yields:
 \begin{align*}
x_0 - p \cdot x_1 - (1-p) \cdot x_2 = &~0 \\
x_1 - q \cdot x_2 - (1-q) \cdot x_3= &~0 \\
-q \cdot x_1  + x_2 - (1-q) \cdot x_4= &~0  \\
x_3 = &~1 \\
x_4 = &~0.
\end{align*}
 Adding $q$ times the second equation to the third equation (concerning state $s_2$) brings the left-hand side matrix in upper triangular form: 
 \begin{align*}
x_0 - p \cdot x_1 - (1-p) \cdot x_2 = &~0 \\
x_1 - q \cdot x_2 - (1-q) \cdot x_3= &~0 \\
(1 - q^2) \cdot x_2 -  q(1-q) \cdot x_3 - (1-q) \cdot x_4= &~0  \\
x_3 = &~1\\
x_4 = &~0 .
\end{align*}
The equation system yields the same result as the elimination of the transition from $s_2$ to $s_1$ (notice the symmetry between $s_1$ and $s_2$).
\end{example}

The example illustrates that there is no elementary advantage in doing state elimination over resorting to solving the linear equation sytem by (some variant of) Gaussian elimination. 
If we are only interested in the probability from the initial state, we do not need to solve the full equation system. 
The state-elimination algorithm, in which we can remove unreachable states, optimises for this observation, in contrast to (standard) linear equation solving.
As in state elimination, the elimination order of the rows has a significant influence.

\subsection{Algorithm based on set-based transition elimination}\label{sec:sucshortcut}
To succinctly represent large state spaces, Markov chains are often represented by multi-terminal binary decision diagrams (or variants thereof)~\cite{DBLP:conf/icalp/BaierCHKR97}. 
Such a \emph{symbolic representation} handles sets of states instead of single states (and thus also sets of transitions), and thereby exploits symmetries and similarities in the underlying graph of a model.
To support efficient elimination, we describe how to eliminate sets of transitions at once.
  The method is similar to the Floyd-Warshall algorithm for all-pair shortest paths~\cite{CLRS}.
  The transition matrix contains one-step probabilities for every pair of source and target states.
  Starting with a self-loop-free pMC (obtained by eliminating all self-loops from the original pMC), we iterate two operations until convergence.
  By doing a matrix-matrix multiplication, we effectively eliminate all transitions emanating from all non-absorbing states \emph{simultaneously}.
  As this step may reintroduce self-loops, we eliminate them in a second step.
  As before, eventually only direct transitions to absorbing states remain, which effectively yield the unbounded reachability probabilities.
  The corresponding pseudo-code is given in Algorithm~\vref{alg:succshort}.
  
  The approach of this algorithm can conveniently be explained in the equation system representation.
  Let us therefore conduct one step of the algorithm as an example, where we use the observation that the matrix-matrix multiplication corresponds to replacing the variables $x_s$ by their defining equations in all \emph{other} equations.
  
 
\begin{example}
Reconsider the equations from Example~\vref{example:linear_equation_system}:
\begin{align*}
x_0 = &~ p \cdot x_1 + (1-p) \cdot x_2 \\
x_1 = &~ q \cdot x_2 + (1-q) \cdot x_3 \\
x_2 = &~ q \cdot x_1 + (1-q) \cdot x_4 \\
x_3 = &~ 1 \\
x_4 = &~ 0. 
\end{align*}
Using the equations for $x_0, x_1, x_2$ to replace their occurrences in all \emph{other} equations yields:
\begin{align*}
x_0 = &~ p \cdot (q \cdot x_2 + (1-q) \cdot x_3) + (1-p)(q \cdot x_1 + (1-q) \cdot x_4) \\
x_1 = &~ q \cdot (q \cdot x_1 + (1-q) \cdot x_4) + (1-q) \cdot x_3 \\
x_2 = &~ q \cdot (q \cdot x_2 + (1-q) \cdot x_3) + (1-q) \cdot x_4 \\
x_3 = &~ 1 \\
x_4 = &~ 0 
\end{align*}
which simplifies to
\begin{align*}
x_0 = &~ (1-p) \cdot q \cdot x_1 + p \cdot q \cdot x_2 + p \cdot (1-q) \cdot x_3 \\ &\quad+ (1-p)(1-q)\cdot x_4 \\
x_1 = &~ \frac{1}{1+q} \cdot x_3 + \frac{q}{1+q} \cdot x_4 \\
x_2 = &~ \frac{q}{1+q} \cdot x_3 + \frac{1}{1+q} \cdot x_4 \\
x_3 = &~ 1 \\
x_4 = &~ 0. 
\end{align*}
We depict the pMC which corresponds to this equation system in Figure~\vref{fig:succshort_1}. Again, notice the similarity to state elimination. For completeness, the result after another iteration is given in Figure~\vref{fig:succshort_2}.
\end{example}
\begin{figure}
\centering
\subfigure[After first iteration]{
\scalebox{\picscale}{
\begin{tikzpicture}[scale=1, nodestyle/.style={draw,circle},baseline=(s0)]
    
    \node [nodestyle] (s0) at (0,0) {$s_0$};
    \node [] (leftdummy)  [on grid, left=1.2cm of s0] {};
    \node [] (rightdummy) [on grid, right=1.2cm of s0] {};
    \node [nodestyle] (s1) [on grid, below=\distforsone of leftdummy] {$s_1$};
    \node [nodestyle] (s2) [on grid, below=\distforsone of rightdummy] {$s_2$};
    \node [nodestyle, accepting] (s3) [on grid, below=\distsonesthree of s1] {$s_3$};
    \node [nodestyle] (s4) [on grid, below=\distsonesthree of s2] {$s_4$};
    
    \draw ($(s0)-(0.7,0)$) edge[->] (s0);
    \draw (s0) edge[->] node[right,pos=0.85] {\scriptsize$(1{-}p)\cdot q$} (s1);
    \draw (s0) edge[->] node[left,pos=0.4] {\scriptsize$p\cdot q$} (s2);

     \draw (s1) edge[->] node[right] {\scriptsize$\frac{1}{1+q}$} (s3);
     \draw (s1) edge[->] node[above,pos=0.25] {\scriptsize$\frac{q}{1+q}$} (s4);
    
     \draw (s2) edge[->] node[above,pos=0.25] {\scriptsize$\frac{q}{1+q}$} (s3);
     \draw (s2) edge[->] node[left] {\scriptsize$\frac{1}{1+q}$} (s4);
    
    \draw (s3) edge[loop right, ->] node[auto] {\scriptsize$1$} (s3);
    
    \draw (s4) edge[loop left, ->] node[auto] {\scriptsize$1$} (s3);

     \draw (s0) edge[->, bend right=60] node[left,pos=0.1] {\scriptsize$p\cdot(1{-}q)$} (s3);
     \draw (s0) edge[->, bend left=60] node[right,pos=0.1] {\scriptsize$(1{-}p)\cdot(1{-}q)$} (s4);
\end{tikzpicture}
}	

\label{fig:succshort_1}
}
\subfigure[After second iteration]{
\scalebox{\picscale}{
\begin{tikzpicture}[scale=1, nodestyle/.style={draw,circle},baseline=(s0)]
    
    \node [nodestyle] (s0) at (0,0) {$s_0$};
    \node [] (leftdummy)  [on grid, left=1.2cm of s0] {};
    \node [] (rightdummy) [on grid, right=1.2cm of s0] {};
    \node [nodestyle] (s1) [on grid, below=\distforsone of leftdummy] {$s_1$};
    \node [nodestyle] (s2) [on grid, below=\distforsone of rightdummy] {$s_2$};
    \node [nodestyle, accepting] (s3) [on grid, below=\distsonesthree of s1] {$s_3$};
    \node [nodestyle] (s4) [on grid, below=\distsonesthree of s2] {$s_4$};
    
    \draw ($(s0)-(0.7,0)$) edge[->] (s0);
    \draw (s1) edge[->] node[right] {\scriptsize$\frac{1}{1+q}$} (s3);
    \draw (s1) edge[->] node[above,pos=0.25] {\scriptsize$\frac{q}{1+q}$} (s4);
    
    \draw (s2) edge[->] node[above,pos=0.25] {\scriptsize$\frac{q}{1+q}$} (s3);
    \draw (s2) edge[->] node[left] {\scriptsize$\frac{1}{1+q}$} (s4);
    
    \draw (s3) edge[loop right, ->] node[auto] {\scriptsize$1$} (s3);
    
    \draw (s4) edge[loop left, ->] node[auto] {\scriptsize$1$} (s3);

    \draw (s0) edge[->, bend right=60] node[right, pos=0.3] {\scriptsize$(p{+}q{-}pq){\cdot}\frac{1}{1+q}$} (s3);
    \draw (s0) edge[->, bend left=60] node[left, pos=0.45] {\scriptsize$1-(p{+}q{-}pq){\cdot}\frac{1}{1+q}$} (s4);
\end{tikzpicture}
}
\label{fig:succshort_2}	
}
\caption{One step of set-based transition elimination exemplified}
\end{figure}

\begin{algorithm}[t]
\caption{Set-based transition elimination for pMCs}

  \begin{newalgorithm}{reachability}{pMC $\PdtmcInit$, $T\subseteq S$}
    \State $S^{?}\coloneqq\{s\in S\mid s\neq\sinit \land s \in \finally T   \setminus T \}$\\
     \ForEachx{ $s \in S^?$ }
        \Comment can be done in parallel for all $s$ \\
        \State \textbf{eliminate\_selfloop}($\probmdp,s$)  \\
     \EndFor
    \While{$\exists s, s' \in S^?.~\probmdp(s,s') \neq 0$}
        \ForEachx {$s \in S^?, s' \in S$} 
        	\Comment can be done in parallel for all $s, s'$ \\
        	\State $\probmdp'(s,s') := \sum_{s''} \probmdp(s,s'') \cdot \probdtmc(s'',s')$\\
        \EndFor
        \ForEachx{ $s \in S^?$ }
        \Comment can be done in parallel for all $s$ \\
        \State \textbf{eliminate\_selfloop}($\probmdp',s$)  \\
        \EndFor
        \State $\probmdp := \probmdp'$\\
    \EndWhile
    \Comment{All $S^?$ eliminated. Only direct paths to target.} \\
    \Return{$\sum\limits_{t\in T}\probmdp(\sinit,t)$}
    
  \end{newalgorithm}
  \label{alg:succshort}
\end{algorithm}

The correctness follows from the following argument:
After every iteration, the equations describe a pMC over the same state space as before.
As all absorbing states have defining equations $x_i \in \{0,1\}$, the equation system is known to have a unique solution \cite{BK08}.
Moreover, as the equation system in iteration $i$ implies the equation system in iteration $i+1$, they preserve the same (unique) solution.

\section{SMT-based region verification}\label{sec:exact_mc}

In this section, we discuss a complete procedure to verify regions by encoding them as queries for an SMT solver, or more precisely, in the existential theory of the reals (the \texttt{QF\_NRA} theory in the SMT literature). 
We first introduce the constraints for verifying regions on pMCs in Section~\ref{subsec:satpmc}. The constraints are either based on the equation system encoding from Section~\ref{sec:lineq} or use the solution function, which yields an equation system with less variables at the cost of precomputing the solution function. 
In Section~\ref{subsec:satpmdp}, we then introduce the encodings for region verification on pMDPs  under angelic and demonic strategies.

Throughout the section, we focus on unbounded reachability, that is, we assume $\varphi =\p_{\leq \lambda}(\finally T)$. As expected rewards can be described by a similar equation system, lifting the concepts is straightforward.
We assume a graph-preserving region~$R$: Assuming that $R$ is graph preserving eases the encodings significantly, but is not strictly necessary: 
In~\cite[Ch.~4]{DBLP:phd/dnb/Junges20}, we provide encodings for well-defined regions $R$.

\begin{figure}[t]
\centering
	\subfigure[pMC]{
		\scalebox{1}{
			\begin{tikzpicture}
		\node[draw, circle, , initial, initial text=] (s0) {$s_0$} ;
		\node[draw, circle, , right=0.8cm of s0] (s1) {$s_1$} ;
		\node[draw, circle, , above=0.8cm of s1, accepting] (s2) {$s_2$} ;
		\node[draw, circle, , right=0.8cm of s1] (s3) {$s_3$} ;
		
		\draw[->] (s0) -- node[below] {$p$} (s1);
		\draw[->] (s1) -- node[right] {$q$} (s2);
		\draw[->] (s0) -- node[left] {$1{-}p$} (s2);
		\draw[->] (s1) -- node[below] {$1{-}q$} (s3);
		\draw[->] (s2) edge[loop right] node {$1$} (s2);
		\draw[->] (s3) edge[loop above] node {$1$} (s3);
	\end{tikzpicture}
		}
		\label{Fig:SimpleAcycPMCRepeated}
	}
	\subfigure[pMDP]{
		\scalebox{1}{
			\begin{tikzpicture}
		\node[draw, circle, , initial, initial text=] (s0) {$s_0$} ;
		\node[draw, circle, , right=0.8cm of s0] (s1) {$s_1$} ;
		\node[draw, circle, , above=0.8cm of s1, accepting] (s2) {$s_2$} ;
		\node[draw, circle, , right=0.8cm of s1] (s3) {$s_3$} ;
		\node[draw, circle,  inner sep=0.5pt, fill, right=0.3cm of s0] (s0alpha)  {};
		\node[draw, circle,  inner sep=0.5pt, fill, above=0.8cm of s0] (s0beta) {} ;
		
		\draw[-] (s0) -- node[below] {$\alpha$} (s0alpha);
		\draw[-] (s0) -- node[left] {$\beta$} (s0beta);
		\draw[->] (s0beta) -- node[above] {$1$} (s2);
		\draw[->] (s0alpha) -- node[below] {$p$} (s1);
		\draw[->] (s1) -- node[right] {$q$} (s2);
		\draw[->] (s0alpha) -- node[left] {$1{-}p$} (s2);
		\draw[->] (s1) -- node[below] {$1{-}q$} (s3);
		\draw[->] (s2) edge[loop right] node {$1$} (s2);
		\draw[->] (s3) edge[loop above] node {$1$} (s3);
\end{tikzpicture}
		}
		\label{Fig:SimpleAcycPMDPRepeated}
	}
	\caption{Toy-examples (repeated from Figure~\vref{fig:toyexamples})}
\end{figure}
\subsection{Satisfiability checking for pMC region checking}
\label{subsec:satpmc}
Recall from Section~\ref{sec:lineq} the equation system for pMCs, exemplified by the following running example.
\begin{example}
\label{ex:equationsystem}
Reconsider the pMC $\pdtmc$ from Figure~\vref{Fig:SimpleAcycPMC}, repeated in Figure~\vref{Fig:SimpleAcycPMCRepeated} for convenience.
 The concrete equation system of \eqref{eq:pmcencodingstart}--\eqref{eq:pmcencodingend} on page \pageref{eq:pmcencodingstart} for reaching $T=\{s_2\}$, using $x_i$ to denote $x_{s_i}$, is given by:
\begin{align*}
x_0 &= p \cdot x_1 + (1{-}p) \cdot x_2 \\
x_1 &= q \cdot x_2 + (1{-}q) \cdot x_3 \\
x_2 &= 1 \\
x_3 &= 0.
\end{align*}
\end{example}

The conjunction of the equation system for the pMC,  \eqref{eq:pmcencodingstart}--\eqref{eq:pmcencodingend} on page \pageref{eq:pmcencodingstart}, is an implicitly existential quantified formula to which we refer by $\Upphi(\pdtmc)$---consider the remark below. 
By construction, this formula is satisfiable.
\begin{remark}
\label{rem:ratfunctransitions}
If transitions in the pMC are not polynomial but rational functions, the equations are not polynomial constraints, hence their conjunction is not a formula (Section~\ref{sec:preliminaries:etr}). Instead, each $x = \sum \probdtmc(s,s')$ has to be transformed by the rules in Section~\ref{subsec:handlinginequalities}: then, their conjunction is a formula.
This transformation can always be applied, in particular, in the equalities we are never interested in the evaluation of instantiations $u \in R$ with $\probdtmc(s,s')[u] = \bot$: 
Recall that 
we are interested in analysing this equation system on a \emph{well-defined} parameter region $R$:
Therefore, for any $u \in R$, $\probdtmc(s,s')[u] \neq \bot$ for each $s, s'\in S$.
Thus, when $\Upphi(\pdtmc)$ is used in conjunction with $\Upphi(R)$, we do not need to consider this special case.
\end{remark}
%
%

We consider the conjunction of the equation system, a property and a region. Concretely, let us first consider the conjunction of:
\begin{compactitem}
\item the equation system $\Upphi(\pdtmc)$,
\item a comparison of the initial state $\sinit$ with the threshold $\lambda$, and
\item a formula $\Upphi(R)$ describing the parameter region $R$.
\end{compactitem}
Satisfiability of this conjunction means that---for some parameter instantiation within the region $R$---the reachability probability from the initial state $\sinit$ satisfies the bound. Unlike $\Upphi(\pdtmc)$, this conjunction may be unsatisfiable.

\begin{example}\label{ex:formulaforsatisfiability}
We continue with Example~\vref{ex:equationsystem}. Let $\varphi =\p_{\leq 0.4}(\finally \{ s_2 \})$ and $R = \{ (p,q) \in [0.4, 0.6] \times [0.2, 0.5]\}$.
We have $\Upphi(R)=0.4 \leq p \land p \leq 0.6 \land 0.2 \leq q \land q \leq 0.5$.
We obtain the following conjunction:
\begin{align}\label{eq:ex_eqsystem1}
\Upphi(\pdtmc) \land x_0 &\leq 0.4 \land \Upphi(R)
\end{align}
where $\Upphi(\pdtmc)$ is the conjunction of the equation system, i.e.: 
\begin{align*}
\Upphi(\pdtmc) &=&  
\Big( x_0 &= p \cdot x_1 + (1{-}p) \cdot x_2 &\land \\
& & x_1 &= q \cdot x_2 + (1{-}q) \cdot x_3 &\land \\
& & x_2 &= 1~\land~x_3 = 0 & \Big).
\end{align*}
Formula~\eqref{eq:ex_eqsystem1} is unsatisfiable, thus, no instance of $p$ and $q$ within the region $R$ induces a reachability probability of at most $\nicefrac{2}{5}$.
\end{example}

Towards region verification, consider that the satisfaction relations $\models_a$\footnote{Recall that $\models_d$ coincides with $\models_a$ for pMCs.} as defined in Definition~\vref{def:satRel}, we have to certify that all parameter values within a region yield a reachability probability that satisfies the threshold.
Thus, we have to quantify over all instantiations $u$, (roughly) leading to a formula of the form $\forall u \hdots \models \varphi$.
By negating this statement, we obtain the proof obligation $\neg \exists u \hdots  \models \neg\varphi$: no parameter value within the region $R$ satisfies the negated comparison with the initial state. This intuition leads to the following conjunction of:
\begin{compactitem}
\item the equation system $\Upphi(\pdtmc)$,
\item a comparison of the initial state with the threshold, by inverting the given threshold-relation, and
\item a formula $\Upphi(R)$ describing the parameter region.
\end{compactitem}
This conjunction is formalised in the following definition.
\begin{definition}[Equation system formula]
Let $\pdtmc$ be a pMC, $\varphi =\p_{\sim \lambda}(\finally T)$, and $R$ a region. The \emph{equation system formula} is given by: 
	\[ \Upphi(\pdtmc) \land x_{\sinit} \not\sim \lambda \land \Upphi(R). \]
\end{definition}

\begin{theorem}	
The equation system formula is unsatisfiable iff $\pdtmc,R\models\varphi$.
\end{theorem}
Otherwise, a satisfying solution is a \emph{counterexample}.
\begin{example}\label{ex:formulaformodelchecking}
 We continue Example~\vref{ex:formulaforsatisfiability}. We invert the relation $x_0 \leq 0.4$ and obtain:  \begin{align*}
\Upphi(\pdtmc) \land x_0 &> 0.4 \land \Upphi(R).
\end{align*}
By SMT-checking, we determine that the formula is satisfiable, e.g., with $p = 0.5$ and $q = 0.3$. Thus, $\pdtmc,R\not\models\varphi$.
If we consider instead the region $R' = \{ (p,q) \in [0.8, 0.9] \times [0.1, 0.2]\}$ with $\Upphi(R')=0.8 \leq p \land p \leq 0.9 \land 0.1 \leq q \land q \leq 0.2$, we obtain:
 \begin{align*}
\Upphi(\pdtmc) \land x_0 &> 0.4 \land \Upphi(R')
\end{align*}
which is unsatisfiable. 
Hence, no point in $R'$ induces a probability larger than $\nicefrac{2}{5}$ and, equivalently, all points in $R'$ induce a probability of at most $\nicefrac{2}{5}$. 
Thus, $\pdtmc,R'\models\varphi$.
\end{example}

We observe that the number of variables in this encoding is $|\states| + |\Var|$.  In particular, we are often interested in systems with at least thousands of states. The number of variables is therefore often too large for SMT-solvers dealing with non-linear real arithmetic. 
However, many of the variables are auxiliary variables that encode the probability to reach target states from each individual state.
We can get rid of these variables by replacing the full equation system by the solution function (Definition~\vref{def:param_solutions}). 
\begin{definition}[Solution function formula]
	Let $\pdtmc$ be a pMC, $\varphi =\p_{\sim \lambda}(\finally T)$, and $R$ a region. The \emph{solution function formula}\footnote{Remark~\vref{rem:ratfunctransitions} applies also here.} is given by: 
	\[ f^r_{\pdtmc,T} \not\sim \lambda \land \Upphi(R). \] 
\end{definition}
\begin{corollary}
	The solution function formula is unsatisfiable iff $\pdtmc,R\models\varphi$.
\end{corollary}
	
\begin{example}
We consider the same scenario as in Example~\vref{ex:formulaforsatisfiability}. The solution function is given in Example~\vref{ex:solutionfunction}. 
The solution function formula is:
\begin{align*}
1 - p + p\cdot q > 0.4 \land \Upphi(R).
\end{align*}
\end{example}
By construction, the equation system formula and the solution function formula for pMC $\pdtmc$ and reachability property $\varphi$ are equisatisfiable.

\subsection{Existentially quantified formula for parametric MDPs}\label{sec:eqfpmdp}
\label{subsec:satpmdp}
We can also utilise an SMT solver to tackle the verification problem on pMDPs.
For parametric MDPs, we distinguish between the angelic and the demonic case, cf.\ Definition~\vref{def:satRelReg}. 
 We use the fact that optimal strategies for unbounded reachability objectives are memoryless and deterministic~\cite{Put94}. 
\subsubsection{Demonic strategies}
The satisfaction relation $\models_d$ is defined by two universal quantifiers, $\forall u \forall \sched \hdots \models \varphi$.
We therefore try to refute satisfiability of $\exists u \exists \sched \hdots \models \neg\varphi$. 
Put in a game-theoretical sense, the same player can choose both the parameter instantiation $u$ and the strategy $\sched$ to resolve the non-determinism.
We generalise the set of linear equations from the pMC to an encoding for pMDPs, where we define a disjunction over all possible nondeterministic choices:
\begin{align}
	& x_s = 0  &\quad & \forall s \in \neg\lozenge T\label{eq:gener_lin_eq1} \\
	& x_s = 1  &\quad & \forall s \in T\label{eq:gener_lin_eq2} \\
	& \bigvee_{\act\in \Act(s)} \Big( x_s = \sum_{s' \in S} \probmdp(s,\act,s') \cdot x_{s'}\Big) & \quad  & \forall  s \in \lozenge T \setminus T.\label{eq:gener_lin_eq3}
	\end{align}
We denote the conjunction of \eqref{eq:gener_lin_eq1}--\eqref{eq:gener_lin_eq3} as $\Upphi_d(\pmdp)$ for pMDP $\pmdp$\footnote{Recall again Remark~\vref{rem:ratfunctransitions}.}.
Instead of a single equation for the probability to reach the target from state $s$, we get one equation for each action. 
The solver can now freely choose which (memoryless deterministic) strategy it uses to refute the property.
\begin{definition}[Demonic equation system formula]
	Let $\pmdp$ be a pMDP, $\varphi =\p_{\leq \lambda}(\finally T)$, and $R$ a region. The \emph{demonic equation system formula} is given by: 
	\[ \Upphi_d(\pmdp) \land x_{\sinit} > \lambda \land \Upphi(R). \]
\end{definition}
\begin{theorem}
	The demonic equation system formula is unsatisfiable iff $\pmdp,R\models_d \varphi$.
\end{theorem}

\begin{example}
\label{ex:demonicsystem}
Let $\pmdp$ be the pMDP from Figure~\vref{Fig:SimpleAcycPMDPRepeated}. 
Let $R, \varphi$ be as in Example~\vref{ex:formulaforsatisfiability}. 
The demonic equation system formula is \[  \Upphi_d(\pmdp) \land x_0 > 0.4 \land \Upphi(R) \] with $\Upphi(R)$ as before, and 
\begin{align*}
\Upphi_d(\pmdp)&=& 	\Big( \big( x_0 &= p \cdot x_1 + (1{-}p) \cdot x_2 \quad\lor\quad
 x_0 = x_2\big) &\land \\
& & x_1 &= q \cdot x_2 + (1{-}q) \cdot x_3 &\land \\
& & x_2 &= 1~\land~x_3 = 0 & \Big).
\end{align*}
\end{example}

Similarly, when using the (potentially exponential) set of solution functions, we let the solver choose:
\begin{definition}[Demonic solution function formula]
		Let $\pmdp$ be a pMDP, $\varphi =\p_{\sim \lambda}(\finally T)$, and $R$ a region. The \emph{demonic solution function formula} is given by: 
	\[ \bigvee_{\sched \in \Sched^\pmdp} f^r_{\pmdp^\sched,T} \not\sim  
	 \lambda \land \Upphi(R). \] 
\end{definition}
\begin{corollary}
	The demonic solution function formula is unsatisfiable iff $\pmdp,R\models_d \varphi$.
\end{corollary}
As the set of solution functions can be exponential, the demonic solution function formula can grow exponentially.
\begin{example}
The demonic solution function formula for $\pmdp, \varphi, R$ as in Example~\vref{ex:demonicsystem}, is given by:
\begin{align*}
\Big(1 &> 0.4 \lor
1 - p + p\cdot q > 0.4 \Big) \land \Upphi(R).
\end{align*}\end{example}

\subsubsection{Angelic strategies}
The satisfaction relation $\models_a$ has two different quantifiers, $\forall u \exists \sched \hdots \models \varphi$. 
Again, we equivalently try to refute the satisfiability of $\exists u \forall \sched \hdots \models \neg\varphi$.
The quantifier alternation can be circumvented by lifting the linear programming (LP) formulation for MDPs \cite{Put94}, where for each nondeterministic choice an upper bound on the probability variables $x_s$ is obtained:
\begin{align}
 	& x_s = 0  &\quad & \forall s \in \neg\lozenge T\label{eq:lifting_lp_1} \\
	& x_s = 1  &\quad & \forall s \in T\label{eq:lifting_lp_2} \\
	& \bigwedge_{\act\in \Act(s)} \Big( x_s \leq \sum_{s' \in S} \probmdp(s,\act,s') \cdot x_{s'}\Big) & \quad  & \forall s \in \lozenge T \setminus T .\label{eq:angelic_scheduler}
\end{align}
Intuitively, the conjunction in constraint \eqref{eq:angelic_scheduler} eliminates the freedom of choosing any strategy from the solver and forces it to use the strategy that minimises the reachability probability.
This means that the constraint system is only satisfiable if \emph{all} strategies violate the probability bound.
We denote the conjunction of \eqref{eq:lifting_lp_1}--\eqref{eq:angelic_scheduler} as $\Upphi_a(\pmdp)$.
Notice that, as for parameter-free MDPs, the optimisation objective of the LP formulation can be substituted by a constraint on probability in the initial state.
\begin{definition}[Angelic equation system formula]
	Let $\pmdp$ be a pMDP, $\varphi =\p_{\leq \lambda}(\finally T)$, and $R$ a region. The \emph{angelic equation system formula} is given by: 
	\[ \Upphi_a(\pmdp) \land x_{\sinit} > \lambda \land \Upphi(R). \]
\end{definition}
\begin{theorem}
	The angelic equation system formula is unsatisfiable iff $\pmdp,R\models_a \varphi$.
\end{theorem}
\begin{example}
Let $\pmdp, \varphi, R$ as in Example~\vref{ex:demonicsystem}.
The angelic equation system formula is given by \[  \Upphi_a(\pmdp) \land x_0 > 0.4 \land \Upphi(R) \] with
\begin{align*}
\Upphi_a(\pmdp)&=& 	\Big( \big( x_0 &\leq p \cdot x_1 + (1{-}p) \cdot x_2 \land x_0 \leq x_2\big) &\land \\
& & x_1 &\le q \cdot x_2 + (1{-}q) \cdot x_3 &\land \\
& & x_2 &= 1~\land~x_3 = 0 & \Big).
\end{align*}
\end{example}

When using the set of solution functions, all strategies have to be considered. Again, for most pMDPs, this set is prohibitively large.
\begin{definition}[Angelic solution function formula]
		Let $\pmdp$ be a pMDP, $\varphi =\p_{\leq \lambda}(\finally T)$, and $R$ a region. The \emph{angelic solution function formula} is given by: 
	\[ \bigwedge_{\sched \in \Sched^\pmdp} f^r_{\pmdp^\sched,T} > \lambda \land \Upphi(R). \] 
\end{definition}
\begin{corollary}
	The angelic solution function formula is unsatisfiable iff $\pmdp,R\models_a \varphi$.
\end{corollary}

\begin{example}
The angelic solution function formula for $\pmdp, \varphi, R$ as in Example~\vref{ex:demonicsystem} is given by:
\begin{align*}
\Big(1 &> 0.4 \land
1 - p + p\cdot q > 0.4\Big) \land \Upphi(R).
\end{align*}\end{example}

\section{Model-checking-based Region Verification of Parametric MCs}\label{sec:approx_mc}

This section discusses an abstraction (and refinement) procedure for region verification of pMCs. 
Intuitively, in order to bound the probability in a region from above, we bound the value induced by any instantation from above. We aim to do this by finding an instantiation that maximises the reachability probability in the region. This problem is particularly hard, as there are dependencies between the different parameters: 
\begin{figure}[t]
  \centering
  \subfigure[$\pdtmc$]{
  \centering
    \scalebox{\picscale}{
      \begin{tikzpicture}[scale=1, nodestyle/.style={draw,circle},baseline=(s0)]
    
    \node [nodestyle] (s0) at (0,0) {$s_0$};
    \node [] (leftdummy)  [on grid, left=1.2cm of s0] {};
    \node [] (rightdummy) [on grid, right=1.2cm of s0] {};
    \node [nodestyle] (s1) [on grid, below=\distforsone of leftdummy] {$s_1$};
    \node [nodestyle] (s2) [on grid, below=\distforsone of rightdummy] {$s_2$};
    \node [nodestyle, accepting] (s3) [on grid, below=\distsonesthree of s1] {$s_3$};
    \node [nodestyle] (s4) [on grid, below=\distsonesthree of s2] {$s_4$};
    
    \draw ($(s0)-(0.7,0)$) edge[->] (s0);
    \draw (s0) edge[->] node[right] {\scriptsize$p$} (s1);
    \draw (s0) edge[->] node[right] {\scriptsize$1{-}p$} (s2);

     \draw (s1) edge[bend left=15, ->] node[above] {\scriptsize$q$} (s2);
     \draw (s1) edge[->] node[right] {\scriptsize$1{-}q$} (s3);
    
     \draw (s2) edge[bend left=15, ->] node[below] {\scriptsize$q$} (s1);
     \draw (s2) edge[->] node[left] {\scriptsize$1{-}q$} (s4);
    
    \draw (s3) edge[loop right, ->] node[auto] {\scriptsize$1$} (s3);
    
    \draw (s4) edge[loop left, ->] node[auto] {\scriptsize$1$} (s3);

\end{tikzpicture}
    }
    \label{fig:pmcrel:D}
  }
   \subfigure[$\subst{\pdtmc}{R}$]{
  \centering
    \scalebox{\picscale}{
      \begin{tikzpicture}[scale=1, nodestyle/.style={draw,circle},baseline=(s0)]
   
    \node [nodestyle] (s0) at (0,0) {$s_0$};
    \node [] (leftdummy)  [on grid, left=1.2cm of s0] {};
    \node [] (rightdummy) [on grid, right=1.2cm of s0] {};
    \node [nodestyle] (s1) [on grid, below=\distforsone of leftdummy] {$s_1$};
    \node [nodestyle] (s2) [on grid, below=\distforsone of rightdummy] {$s_2$};
    \node [nodestyle, accepting] (s3) [on grid, below=\distsonesthree  of s1] {$s_3$};
    \node [nodestyle] (s4) [on grid, below=\distsonesthree of s2] {$s_4$};
    
    \draw ($(s0)-(0.7,0)$) edge[->] (s0);
    \draw (s0) edge[->, bend left=0, dashed] node[left] {\scriptsize$\nicefrac{1}{10}$} (s1);
    \draw (s0) edge[->, bend right=0, dashed] node[right] {\scriptsize$\nicefrac{9}{10}$} (s2);
    \draw (s0) edge[->, bend right=40] node[left] {\scriptsize$\nicefrac{4}{5}$} (s1);
    \draw (s0) edge[->, bend left=40] node[right] {\scriptsize$\nicefrac{1}{5}$} (s2);

     \draw (s1) edge[bend left=40, ->, dashed] node[above] {\scriptsize$\nicefrac{2}{5}$} (s2);
     \draw (s1) edge[->, bend left=15, dashed] node[right] {\scriptsize$\nicefrac{3}{5}$} (s3);
     \draw (s1) edge[bend left=10, ->] node[above] {\scriptsize$\nicefrac{7}{10}$} (s2);
     \draw (s1) edge[->, bend right=15] node[left] {\scriptsize$\nicefrac{3}{10}$} (s3);
    
     \draw (s2) edge[bend left=40, ->, dashed] node[below] {\scriptsize$\nicefrac{2}{5}$} (s1);
     \draw (s2) edge[->, bend right=15, dashed] node[left] {\scriptsize$\nicefrac{3}{5}$} (s4);
     \draw (s2) edge[bend left=10, ->] node[below] {\scriptsize$\nicefrac{7}{10}$} (s1);
     \draw (s2) edge[->, bend left=15] node[right] {\scriptsize$\nicefrac{3}{10}$} (s4);
    
    \draw (s3) edge[loop right, ->] node[auto] {\scriptsize$1$} (s3);
    
    \draw (s4) edge[loop left, ->] node[auto] {\scriptsize$1$} (s3);
\end{tikzpicture}
    }
    \label{fig:output}
  }
  \subfigure[$\rel{\pdtmc}$]{
  \centering
    \scalebox{\picscale}{
      \begin{tikzpicture}[scale=1, nodestyle/.style={draw,circle},baseline=(s0)]

    \node [nodestyle] (s0) at (0,0) {$s_0$};
    \node [] (leftdummy)  [on grid, left=1.2cm of s0] {};
    \node [] (rightdummy) [on grid, right=1.2cm of s0] {};
    \node [nodestyle] (s1) [on grid, below=\distforsone of leftdummy] {$s_1$};
    \node [nodestyle] (s2) [on grid, below=\distforsone of rightdummy] {$s_2$};
    \node [nodestyle, accepting] (s3) [on grid, below=\distsonesthree of s1] {$s_3$};
    \node [nodestyle] (s4) [on grid, below=\distsonesthree of s2] {$s_4$};
    
    \draw ($(s0)-(0.7,0)$) edge[->] (s0);
    \draw (s0) edge[->] node[right] {\scalebox{0.8}{$p^{s_0}$}} (s1);
    \draw (s0) edge[->] node[right] {\scalebox{0.8}{$1{-}p^{s_0}$}} (s2);

     \draw (s1) edge[bend left=15, ->] node[above] {\scalebox{0.8}{$q^{s_1}$}} (s2);
     \draw (s1) edge[->] node[right] {\scalebox{0.8}{$1{-}q^{s_1}$}} (s3);
    
     \draw (s2) edge[bend left=15, ->] node[below] {\scalebox{0.8}{$q^{s_2}$}} (s1);
     \draw (s2) edge[->] node[left] {\scalebox{0.8}{$1{-}q^{s_2}$}} (s4);
    
    \draw (s3) edge[loop right, ->] node[auto] {\scriptsize$1$} (s3);
    
    \draw (s4) edge[loop left, ->] node[auto] {\scriptsize$1$} (s3);
\end{tikzpicture}
    }
    \label{fig:pmcrel:Dprime}
  }
  \caption{A pMC $\pdtmc$ and its substitution $\subst{\pdtmc}{R}$ and its relaxation $\rel{\pdtmc}$.}
  \label{fig:pmcrel}
\end{figure}
\begin{example}
\label{ex:tradeoffs}
	Consider the pMC $\pdtmc$ in Figure~\vref{fig:pmcrel:D}---repeating Figure~\vref{fig:models:pmc}--- and region $R = [\nicefrac{1}{10}, \nicefrac{4}{5}] \times [\nicefrac{2}{5}, \nicefrac{7}{10}]$. We again aim to reach $s_3$.
	We make two observations:
	$s_4$ is the only state from which we cannot reach $s_3$, furthermore, $s_4$ is only reachable via $s_2$. Hence, it is best to avoid $s_2$.
From state $s_0$, it is thus beneficial if the transition probability to $s_2$ is as small as possible. Equivalently, it is beneficial if $p$ is as large as possible, as this minimises the probability of reaching $s_2$ and as $p$ does not occur elsewhere.
Now we consider state $s_1$: As we want to reach $s_3$, the value of $q$ should be preferably low.
However, $q$ occurs also at transitions leaving $s_2$.
 From $s_2$, $q$ should be assigned a high value as we want to avoid $s_4$.
In particular, the optimal value for $q$ depends on the probability that we ever visit $s_2$, which is directly influenced by the value of $p$.
\end{example}
In a nutshell, the abstraction we propose in this section ignores the dependencies between the same occurence of a parameter. 
Conveniently, the abstraction transforms a pMC into an (parameter-free!) MDP whose minimal (maximal) reachability probability under-approximates (over-approximates) the reachability probability of the pMC. This result is formalised in Theorem~\ref{th:dtmcApprox}, below.

\begin{example}
Consider the pMC in Figure~\vref{fig:pmcrel:D} and a region $R = [\nicefrac{1}{10}, \nicefrac{4}{5}] \times [\nicefrac{2}{5}, \nicefrac{7}{10}]$. 
The method creates the MDP in Figure~\ref{fig:output}, where different types of arrows reflect different actions.
The MDP is created 
by adding in each state two actions: 
One reflecting the lower bound of the parameter range, one reflecting the upper bound. 
Model checking on this MDP yields a maximal probability of $\nicefrac{47}{60}$.
From this result, we infer that $\max_{u \in R} \reachPrT[{\pdtmc[u]}] \leq \nicefrac{47}{60}$. 	
\end{example}

The essence of this construction is to consider parameter values as a local, discrete choice that we can capture with nondeterminism. To support the discretisation, we must ensure that the optimal values are taken at the bounds of the region. While this is not true in general due to the nonlinearity of the solution function, creating a suitable over-approximation, called the relaxation, enforces this property, as we show in Theorem~\ref{lem:optimalValuation}, also below.

In the remainder of this section, we first clarify helpful assumptions on the type of pMCs we support in Section~\ref{sec:plapmcprelims}. 
We then construct  so-called relaxed pMCs in Section~\ref{sec:regionsMCs:pmcs}. In Section~\ref{sec:regionsMCs:substituting}, we translate relaxed pMCs to parameter-free MDPs to allow off-the-shelf MDP analysis for region verification of pMCs.

\subsection{Preliminaries}
\label{sec:plapmcprelims}
We formalise the perspective that underpins our approach to region verification and introduce some assumptions.
\subsubsection{A Perspective for Region Verification}
	The probability $\reachPr{\pdtmc}{T}$ can be expressed as a rational function $f = \nicefrac{g_1}{g_2}$ with polynomials $g_1, g_2$ due to Definition~\vref{def:param_solutions}.
	Recall that we assume region $R$ to be graph preserving. 
	Therefore, $g_2[u] \neq 0$ for all $u \in R$ and $f$ is continuous on any closed region $R$.
	Hence, there is an instantiation $u \in R$ that induces the maximal (or minimal) reachability probability: 
	\begin{align*}
	&\sup_{u\in R}\reachPr{\pdtmc[u]}{T}  = 
	\max_{u\in R} \reachPr{\pdtmc[u]}{T} 
	\quad\text{and}\quad \inf_{u\in R} \reachPr{\pdtmc[u]}{T}  = \min_{u\in R}\reachPr{\pdtmc[u]}{T}.	
	\end{align*}
To infer that $R$ is accepting (i.e. all instantiations $u \in R$ induce probabilities at most $\lambda$), it suffices to show that the \emph{maximal} reachability probability over all instantiations is at most  $\lambda$:  
\begin{align*}
\pdtmc, R \models \reachProplT &\iff \big(\max_{u\in R} \reachPr{\pdtmc[u]}{T} \big) \le \lambda, \text{ and } \\ 
\pdtmc, R \models \neg \reachProplT &\iff \big(\min_{u\in R} \reachPr{\pdtmc[u]}{T} \big) > \lambda .
\end{align*}
One way to determine the maximum reachability probability is to first determine which $u \in R$ induces the maximum, and then compute the probability on the instantiated model $\pdtmc[u]$. While we only discuss upper-bounded specifications here, the results can be analogously described for lower-bounded specifications.

\begin{example}\label{ex:usemaximum}
 Consider $\pdtmc$ depicted in Figure~\vref{Fig:SimpleAcycPMCRepeated}, $\varphi =\p_{\leq \nicefrac{9}{10}}(\finally \{ s_2 \})$, and $R' = \{ (p,q) \in [\nicefrac{2}{5}, \nicefrac{3}{5}] \times [\nicefrac{1}{5}, \nicefrac{1}{2}]\}$ as in Example~\vref{ex:formulaforsatisfiability}.
The maximum is obtained at $u = (\nicefrac{2}{5},\nicefrac{1}{2})$ (via some oracle).
We have  $\pdtmc[u] \models \p_{\leq \nicefrac{9}{10}}(\finally \{ s_2 \})$, and thus, $\pdtmc, R' \models  \p_{\leq \nicefrac{9}{10}}(\finally \{ s_2 \})$.
\end{example}
However, constructing an oracle that determines the $u$ that induces the maximum is difficult in general. 
We focus on the essential idea an therefore make the following  assumptions throughout the rest of this section: 

\begin{assumption}
\begin{compactitem}
\item We restrict the (graph-preserving) region $R$ to be
(i)~rectangular, and
(ii)~closed.
This restriction makes the bounds of the parameters independent of other parameter instantiations, and ensures that the maximum over the region exists. 
\item
We restrict the pMC $\pdtmc$ to be \emph{locally monotone}--explained in Section~\ref{subsec:locmonotone}--  to exclude difficulties from analysing single transitions. 
\end{compactitem}
\label{assumptionpmc}
\end{assumption}
The first assumption can be a nuisance. In particular, it is not always clear how to create an adequate closed region from an open region. 
The second assumption is very mild and can be accomodated for using adequate preprocessing~\cite[Section 5.1]{DBLP:phd/dnb/Junges20} that introduced additional states.
\subsubsection{Locally Monotone pMCs}
\label{subsec:locmonotone}
 Recall that the solution function is nonlinear. We aim to approximate this $u$ and therefore want to exploit the structure of the pMC.
Therefore, we want to make an assumption on the transition relation.

\begin{example}
Consider a three-state pMC where the probability from initial state $\sinit$ to target state $t$ is a non-linear, non-monotone transition function, as, e.g., the transition probability from $s_0$ to $s_3$ of the pMC in Figure~\vref{fig:se:final}.
Finding the maximum requires an analysis of the derivative of the solution function, and is (approximately) as hard as the exact verification problem.
\end{example}
Instead, we assume monotonic transition probabilities, and consider a slightly restricted class of pMCs.
\begin{definition}[Locally monotone pMCs]
	A pMC $\PdtmcInit$ is \emph{locally monotone} iff for all $s \in S$ there is a multilinear polynomial $\pol_s \in \poly[V] $ satisfying 
	\begin{displaymath}
		\probdtmc(s, s')  \in \left\{ \nicefrac{f}{\pol_s} \mid  f\in \poly[V] \text{ is multilinear}  \right\}
	\end{displaymath}
 for all $s' \in S$.
\end{definition}
Locally monotone pMCs include most pMCs from the literature~\cite{QDJJK16}\footnote{It even includes the embedded pMCs of parametric continuous-time Markov chains with multilinear exit rates.}. 
Examples of the egligible transition probabilities are $p,pq,\nicefrac{1}{p}$ and their complements formed by $1-p$ etc.

Thanks to monotonicity, for a locally monotone pMC $\PdtmcInit$, and a closed rectangular region $R$ we have that for all $s,s' \in S:$ \[ \max_{u \in R} \probmdp(s,s') = \max_{u \in B(\Var)} \probmdp(s,s') \] where $B(\Var) = \{ u \mid \forall p \in \Var. u(p) \in B_R(p) \}$, i.e., all maxima of the individual transition probabilities are attained at the bounds of the region. 
However, the restriction to local monotonicity does not immediately overcome the challenge of constructing an oracle. The resulting solution function may still be highly nonlinear. In particular, Example~\vref{ex:tradeoffs} uses a locally monotone pMC and a closed rectangular region. 
However, as the example indicates, trade-offs in locally monotone pMCs occur due to dependencies where parameters occur at multiple states. 

\subsection{Relaxation}
\label{sec:approx_mc:relax}
\label{sec:regionsMCs:pmcs}

The idea of our approach, inspired by~\cite{DBLP:conf/cav/BrimCDS13}, is to drop the aforementioned dependencies between parameters by means of a \emph{relaxation} of the pMC. We want to highlight that this relaxed pMC is very similar to so-called interval MCs, a detailed discussion is given in~\cite[Section 5.1.1.3]{DBLP:phd/dnb/Junges20}.
Intuitively, the relaxation $\rel{\pdtmc}$ is a pMC that arises from $\pdtmc$ to a pMC with the same state space but an updated transition relation. In particular, it introduces a fresh copy of every parameter in every state,  thereby eliminating parameter dependencies between different states (if any).
This step simplifies finding an optimal instantiation (in the relaxation). However, the set of instantiated pMCs grows: some of the instantiations cannot be obtained from the original pMC. In this subsection, we first formalize the relaxation, then clarify the relation between properties being satisfied on the pMC and properties satisfied on the relaxation. We finish the subsection by discussing how to efficiently analyze a relaxed pMC.
\begin{definition}[Relaxation]\label{def:relaxation_pmc}
The \emph{relaxation} of pMC $\PdtmcInit$ is the pMC $\rel{\pdtmc} = (S, \rel[\pdtmc]{V}, \sinit, \probdtmc')$ with
$\rel[\pdtmc]{V}=\{p_i^s \mid p_i \in \Var, s\in S\}$ and  
$\probdtmc'(s,s')=\probdtmc(s,s')[p_1, \dots, p_n / p_1^s, \dots, p_n^s]$.
\end{definition}
We extend an instantiation $u$ for $\pdtmc$ to the \emph{relaxed instantiation} $\rel[\pdtmc]{u}$ for $\rel{\pdtmc}$ by $\rel[\pdtmc]{u}(p_i^s) = u(p_i)$ for every $s$. We have that for all $u$, $\pdtmc[u] = \rel{\pdtmc}[\rel[\pdtmc]{u}]$.
We lift the relaxation to regions such that $B(p_i^s) = B(p_i)$ for all $s$, \ie, $\rel[\pdtmc]{R} = \bigtimes_{p_i^s \in \rel[\pdtmc]{V}} I(p_i)$.
We drop the subscript $\pdtmc$, whenever it is clear from the context.
\begin{example}\label{ex:relaxation}
Figure~\vref{fig:pmcrel:Dprime} depicts the relaxation $\rel{\pdtmc}$ of the pMC $\pdtmc$ from Figure~\vref{fig:pmcrel:D}.
For $R=[\nicefrac{1}{10}, \nicefrac{4}{5}] \times [\nicefrac{2}{5}, \nicefrac{7}{10}]$ and $u=(\nicefrac{4}{5}, \nicefrac{3}{5}) \in R$ from Example~\vref{ex:regions}, we obtain $\rel{R}=[\nicefrac{1}{10}, \nicefrac{4}{5}] \times [\nicefrac{2}{5}, \nicefrac{7}{10}] \times [\nicefrac{2}{5}, \nicefrac{7}{10}]$ and $\rel{u}=(\nicefrac{4}{5}, \nicefrac{3}{5}, \nicefrac{3}{5})$.
An instantiation $\rel{\pdtmc}[\rel{u}]$ corresponds to $\pdtmc[u]$ as depicted in Figure~\vref{fig:pmc:Du}.
The relaxed region $\rel{R}$ contains also instantiations, e.g., $(\nicefrac{4}{5}, \nicefrac{1}{2}, \nicefrac{3}{5})$ which are not realisable in $R$. 
\end{example}
For a pMC $\pdtmc$ and a graph-preserving region $R$, relaxation increases the set of possible instantiations: $\{\pdtmc[u] \mid u \in R\} \subseteq \{\rel{\pdtmc}[u] \mid u \in \rel{R}\}$. 
Thus, the maximal reachability probability over all instantiations of $\pdtmc$ within $R$ is bounded by the maximum over the instantiations of $\rel{\pdtmc}$ within $\rel{R}$. 
\begin{lemma}\label{lem:relaxationOverapproximates}
For pMC $\pdtmc$ and region $R$:
\begin{align*}
 \max_{u\in R}\big( \reachPr{\pdtmc[u]}{T} \big) \  =  \
 \max_{u\in R}\big( \reachPr{\rel{\pdtmc}[\rel{u}]}{T} \big)  \le  \
 \max_{u\in \rel{R}}
 \big( \reachPr{\rel{\pdtmc}[u]}{T} \big).
\end{align*}
\end{lemma}
%
\noindent Consequently, if $\rel{\pdtmc}$ satisfies a reachability property, so does $\pdtmc$.
\begin{corollary}
For pMC $\pdtmc$ and region $R$:
\begin{align*}
\max_{u\in \rel{R}}\big( \reachPr{\rel{\pdtmc}[u]}{T}\big) \le \lambda \text{ implies }\pdtmc, R \models \reachProplT.
\end{align*}
\end{corollary}

We now formalise the earlier observation: \emph{Without parameter dependencies, finding optimal instantiations in a pMC is simpler}.
Although \rel{\pdtmc} has (usually) more parameters than $\pdtmc$, finding an instantiation $u \in \rel{R}$ that maximises the reachability probability is simpler than in $u \in R$:
For any $p_i^s \in \rel{V}$, we can in state $s$ pick a value in $I(p^s_i)$ that maximises the probability to reach $T$ from state $s$.
There is no (negative) effect for the reachability probability at the other states as $p_i^s$ only occurs at $s$.
Optimal instantiations can thus be determined \emph{locally} (at the states).

Furthermore, as both $\pdtmc$ is locally monotone, and there are no parameter dependencies, the maximum reachability probability is relatively easy to find:
We only need to consider instantiations $u$ that set the value of each parameter to either the lowest or highest possible value, \ie, $u(p_i^s) \in B(p_i^s)$ for all $p_i^s \in \rel{V}$:
\begin{theorem}
\label{lem:optimalValuation}
Let $\pdtmc$ be a pMC with states $S$ and $T\subseteq S$ and $R$ a region subject  subject to Assumption~\ref{assumptionpmc}.
 There exists an instantiation $u \in \rel{R}$ satisfying $u(p_i^s) \in B(p_i^s)$ for all $p_i^s\in \rel{V}$ such that: 
 \begin{align*}
 \reachPr{\rel{\pdtmc}[u]}{T} = \max_{v\in \rel{R}}\reachPr{\rel{\pdtmc}[v]}{T}.
 \end{align*}
\end{theorem}
To prove this statement, we consider an instantiation which assigns a value to a parameter strictly between its bounds.
Any such instantiation can be modified such that all parameters are assigned to its bound, 
without decreasing the induced reachability probability. The essential statement is the monotonicity of a parameter without any further dependencies. The number of instantiations that must be analysed is therefore finite, compared for infinitely many candidates for non-relaxed pMCs.

\begin{lemma}\label{lem:monotonicpmc}
Let $\pdtmc$ be a locally monotone pMC with a single parameter $p$ that only occurs at one state $s \in S$, i.e. $\probdtmc(\hat{s},s') \in [0,1]$ for all $\hat{s}, s' \in S$ with $\hat{s} \neq s$. For region $R$ and $T\subseteq S$, the probability $\reachPrT[\pdtmc]$ is monotonic on $R$.
\end{lemma}

\begin{proof}
	W.\,l.\,o.\,g. let $s\notin T$ be the initial state of $\pdtmc$ and let $T$ be reachable from $s$.
	Furthermore, let $\pctlUntil$ denote the standard until-modality and $\neg T$ denote $S\setminus T$.
	Using the characterisation of reachability probabilities as linear equation system (cf.~\cite{BK08}), the reachability probability \wrt $T$ (from the initial state) in $\pdtmc$ is given by:
	\begin{align*}
	& \reachPrT[\pdtmc] \\
	=& \sum_{s'\in S} \probdtmc(s,s') \cdot \reachPrs{\pdtmc}{s'}{T} \\
	=& \sum_{s'\in S} \probdtmc(s,s') \cdot \Big(\pr_{s'}^{\pdtmc}(\neg s \, \pctlUntil \, T) + \pr_{s'}^{\pdtmc}(\neg T \, \pctlUntil \, s) \cdot \reachPrT[\pdtmc] \Big)\\
	=& \sum_{s'\in S}  \probdtmc(s,s')  \cdot \pr_{s'}^{\pdtmc}(\neg s \, \pctlUntil \, T) 
	 + \sum_{s'\in S}     \probdtmc(s,s')  \cdot \pr_{s'}^{\pdtmc}(\neg T \, \pctlUntil \, s) \cdot \reachPrT[\pdtmc].
	\end{align*}
	Transposing the equation yields
	\begin{align*}
	\reachPrT[\pdtmc] = \frac{\sum_{s'\in S}  \probdtmc(s,s')  \cdot \pr_{s'}^{\pdtmc}(\neg s \, \pctlUntil \, T)}{1-\sum_{s'\in S}     \probdtmc(s,s')  \cdot \pr_{s'}^{\pdtmc}(\neg T \, \pctlUntil \, s)}.
	\end{align*}
	The denominator can not be zero as $T$ is reachable from $s$.
	Since $\pdtmc$ is locally monotone, we have $\probdtmc(s,s') = \nicefrac{f_{s'}}{g_s}$ for $s' \in S$ and multilinear functions $f_{s'}, g_s \in \poly[p]$. We obtain:
	\begin{align*}
	\reachPrT[\pdtmc] = \frac{\sum_{s'\in S}   f_{s'}  \cdot \overbrace{\pr_{s'}^{\pdtmc}(\neg s \, \pctlUntil \, T)}^\mathit{constant}}{g_s -\sum_{s'\in S}   f_{s'}  \cdot \underbrace{\pr_{s'}^{\pdtmc}(\neg T \, \pctlUntil \, s)}_\mathit{constant}}.
	\end{align*}
	Hence, $\reachPrT[\pdtmc] = \nicefrac{f_1}{f_2}$ is a fraction of two multilinear functions $f_1,f_2 \in \poly[p]$ and therefore monotonic on $R$.
	 \end{proof}
	 
\begin{proof}[Theorem~\vref{lem:optimalValuation}]
We prove the statement by contraposition.
Let $u\in \rel{R}$ with $\reachPr{\rel{\pdtmc}[u]}{T} = \max_{v\in \rel{R}}\big( \reachPr{\rel{\pdtmc}[v]}{T} \big)$.
For the contraposition, assume that there exists a parameter $p \in \rel{\Var}$ with $u(p) \in I_R(p) \setminus B_R(p)$ such that all instantiations $u' \in \rel{R}$ that set $p$ to a value in $B_R(p)$ induce a smaller reachability probability, i.e. $u'(p) \in B_R(p)$ and $u'(q) = u(q)$ for  $ q \neq p$ implies
\[
\reachPr{\rel{\pdtmc}[u']}{T} < \reachPr{\rel{\pdtmc}[u]}{T}.
\]
Consider the pMC $\hat{\pdtmc} = (S, \{p\}, s, \hat{\probdtmc})$ with the single parameter $p$ that arises from $\rel{\pdtmc}$ by replacing all parameters $q \in \rel{\Var}\setminus\{p\}$ with $u(q)$.
We have $\hat{\pdtmc}[u] = \rel{\pdtmc}[u]$.
Moreover, $\reachPrT[\hat{\pdtmc}]$ is monotonic on $I(p)$ according to Lemma~\vref{lem:monotonicpmc}.
Thus, there is an instantiation $u' \in \rel{R}$ with $u'(p) \in B_R(p)$ and $u'(q) = u(q)$ for $q \neq p$ satisfying
\[
 \reachPr{\hat{\pdtmc}[u]}{T}
\le 
 \reachPr{\hat{\pdtmc}[u']}{T} = \reachPr{\rel{\pdtmc}[u']}{T} .
\]
This contradicts our assumption for parameter $p$.
\end{proof}

\subsection{Replacing parameters by nondeterminism}
\label{sec:regionsMCs:substituting}
In order to determine $\max_{u\in \rel{R}	} \reachPr{\rel{\pdtmc}[u]}{T}$,
it suffices to make a discrete choice over instantiations $u \colon \rel{\Var} \rightarrow \R$ with $u(p_i^s) \in B(p_i)$. 
This choice can be made locally at every state, which brings us to the key idea of \emph{constructing a (non-parametric) MDP out of the pMC $\pdtmc$ and the region $R$}, where nondeterministic choices represent all instantiations that have to be considered.
In the following, it is convenient to refer to the parameters in a given state $s$ by: 
\[V_s=\{\, p \in \Var \mid p \text{ occurs in } \pdtmc(s,s') \text{ for some } s'\in S \, \}.\]

\begin{definition}[Substitution (pMCs)]
	\label{def:approxPMC}
	For pMC $\PdtmcInit$ and region $R$, let the MDP $\subst{\pdtmc}{R} = (S, \sinit, \Act_{\substAbbrev}, \probmdp_{\substAbbrev})$ with 
	\begin{itemize}
	 \item $\Act_{\substAbbrev} = \biguplus_{s \in S} \Act_s$ where 
\[\Act_s = \{u \colon \Var_s \rightarrow \R \mid  \forall p \in \Var_s.\; u(p) \in B(p)\ \}, \text{ and} \]
\item 
\[
\probmdp_{\substAbbrev}(s,u, s') =
\begin{cases}
\probdtmc(s,s')[u] &\text{if } u \in \Act_s,\\
0                    &\text{otherwise.}
\end{cases}
\]
\end{itemize}
 be the \emph{(parameter-)substitution of $\pdtmc$ and $R$}.\end{definition}
Thus, choosing action $u$ in $s$ corresponds to assigning one of the extremal values $B(p_i)$ to the parameters $p_i^s$. The number of outgoing actions from state $s$ is therefore $2^{|\Var_s|}$. 
\begin{example}
\label{ex:approxmdp}
Consider pMC $\pdtmc$ -- depicted in Figure~\vref{fig:pmcrel:D}
-- with $R = [\nicefrac{1}{10}, \nicefrac{4}{5}] \times [\nicefrac{2}{5}, \nicefrac{7}{10}]$ as before. The substitution of $\pdtmc$ and $R$ is shown in Figure~\vref{fig:approxmdp:M}.
In $\pdtmc$, each outgoing transition of states $s_0, s_1, s_2$ is replaced by a nondeterministic choice in MDP $\subst{\pdtmc}{R}$. That is, we either pick the upper or lower bound for the corresponding variable.
The solid (dashed) lines depict transitions that belong to the action for the upper (lower) bound.
For the states $s_3$ and $s_4$, the choice is unique as their outgoing transitions in $\pdtmc$ are constant.
Figure~\vref{fig:approxmdp:Msched} depicts the MC $\subst{\pdtmc}{R}^\sched$ which is induced by the strategy $\sched$ on MDP $\subst{R}{\pdtmc}$ that chooses the upper bounds at $s_0$ and $s_2$, and the lower bound at $s_1$.
Notice that $\subst{\pdtmc}{R}^\sched$ coincides with $\rel{\pdtmc}[v]$ for a suitable instantiation $v$, as depicted in Fig. \vref{fig:pmcrel:Dprime}.
\end{example}
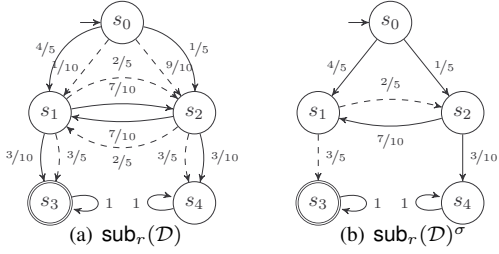
\begin{figure}[t]
  \centering
  \subfigure[$\subst{\pdtmc}{r}$]{
    \scalebox{\picscale}{
      \begin{tikzpicture}[scale=1, nodestyle/.style={draw,circle},baseline=(s0)]
   
    \node [nodestyle] (s0) at (0,0) {$s_0$};
    \node [] (leftdummy)  [on grid, left=1.2cm of s0] {};
    \node [] (rightdummy) [on grid, right=1.2cm of s0] {};
    \node [nodestyle] (s1) [on grid, below=\distforsone of leftdummy] {$s_1$};
    \node [nodestyle] (s2) [on grid, below=\distforsone of rightdummy] {$s_2$};
    \node [nodestyle, accepting] (s3) [on grid, below=\distsonesthree  of s1] {$s_3$};
    \node [nodestyle] (s4) [on grid, below=\distsonesthree of s2] {$s_4$};
    
    \draw ($(s0)-(0.7,0)$) edge[->] (s0);
    \draw (s0) edge[->, bend left=0, dashed] node[left] {\scriptsize$\nicefrac{1}{10}$} (s1);
    \draw (s0) edge[->, bend right=0, dashed] node[right] {\scriptsize$\nicefrac{9}{10}$} (s2);
    \draw (s0) edge[->, bend right=40] node[left] {\scriptsize$\nicefrac{4}{5}$} (s1);
    \draw (s0) edge[->, bend left=40] node[right] {\scriptsize$\nicefrac{1}{5}$} (s2);

     \draw (s1) edge[bend left=40, ->, dashed] node[above] {\scriptsize$\nicefrac{2}{5}$} (s2);
     \draw (s1) edge[->, bend left=15, dashed] node[right] {\scriptsize$\nicefrac{3}{5}$} (s3);
     \draw (s1) edge[bend left=10, ->] node[above] {\scriptsize$\nicefrac{7}{10}$} (s2);
     \draw (s1) edge[->, bend right=15] node[left] {\scriptsize$\nicefrac{3}{10}$} (s3);
    
     \draw (s2) edge[bend left=40, ->, dashed] node[below] {\scriptsize$\nicefrac{2}{5}$} (s1);
     \draw (s2) edge[->, bend right=15, dashed] node[left] {\scriptsize$\nicefrac{3}{5}$} (s4);
     \draw (s2) edge[bend left=10, ->] node[below] {\scriptsize$\nicefrac{7}{10}$} (s1);
     \draw (s2) edge[->, bend left=15] node[right] {\scriptsize$\nicefrac{3}{10}$} (s4);
    
    \draw (s3) edge[loop right, ->] node[auto] {\scriptsize$1$} (s3);
    
    \draw (s4) edge[loop left, ->] node[auto] {\scriptsize$1$} (s3);
\end{tikzpicture}
    }
    \label{fig:approxmdp:M}
  }
  \subfigure[$\subst{\pdtmc}{r}^\sched$]{
    \scalebox{\picscale}{
      \begin{tikzpicture}[scale=1, nodestyle/.style={draw,circle},baseline=(s0)]
   
    \node [nodestyle] (s0) at (0,0) {$s_0$};
    \node [] (leftdummy)  [on grid, left=1.2cm of s0] {};
    \node [] (rightdummy) [on grid, right=1.2cm of s0] {};
    \node [nodestyle] (s1) [on grid, below=\distforsone of leftdummy] {$s_1$};
    \node [nodestyle] (s2) [on grid, below=\distforsone of rightdummy] {$s_2$};
    \node [nodestyle, accepting] (s3) [on grid, below=\distsonesthree of s1] {$s_3$};
    \node [nodestyle] (s4) [on grid, below=\distsonesthree of s2] {$s_4$};
    
    \draw ($(s0)-(0.7,0)$) edge[->] (s0);
    \draw (s0) edge[->,] node[left] {\scriptsize$\nicefrac{4}{5}$} (s1);
    \draw (s0) edge[->, ] node[right] {\scriptsize$\nicefrac{1}{5}$} (s2);

     \draw (s1) edge[bend left=15, ->, dashed] node[above] {\scriptsize$\nicefrac{2}{5}$} (s2);
     \draw (s1) edge[->, dashed] node[right] {\scriptsize$\nicefrac{3}{5}$} (s3);
    
     \draw (s2) edge[bend left=15, ->] node[below] {\scriptsize$\nicefrac{7}{10}$} (s1);
     \draw (s2) edge[->] node[right] {\scriptsize$\nicefrac{3}{10}$} (s4);
    
    \draw (s3) edge[loop right, ->] node[auto] {\scriptsize$1$} (s3);
    
    \draw (s4) edge[loop left, ->] node[auto] {\scriptsize$1$} (s3);
\end{tikzpicture}
    }
    \label{fig:approxmdp:Msched}
  }
  \caption{Illustrating parameter-substitution.}
  \label{fig:approxmdp}
\end{figure}
The substitution encodes the local choices for a relaxed pMC. That is, for an arbitrary pMC, there is a one-to-one correspondence between strategies $\sched$ in the MDP $\subst{\rel{\pdtmc}}{\rel{R}}$ and instantiations $u \in \rel{R}$ for $\rel{\pdtmc}$ with $u(p_i^s) \in B(p_i)$.
For better readability, we will omit the superscripts for sets of strategies $\Sched$.
Combining these observations with Theorem~\vref{lem:optimalValuation}, yields the following. %
 \begin{corollary}\label{cor:dtmcApprox}
For a pMC $\pdtmc$, a graph-preserving region $R$, and a set $T$ of target states of $\pdtmc$:
\begin{align*}
 \max_{u\in R} \reachPrT[{\pdtmc[u]}] \; \le \; 
\smashoperator{\max_{\sched \in \SchedSymbol}} \reachPrT[\subst{\rel{\pdtmc}}{\rel{R}}^\sched] 
\\
 \min_{u\in R} \reachPrT[{\pdtmc[u]}] \; \ge \;
\smashoperator{\min_{\sched \in \SchedSymbol}} \reachPrT[\subst{\rel{\pdtmc}}{\rel{R}}^\sched].
\end{align*}
\end{corollary}
Furthermore, the nondeterministic choices introduced by the substitution only depend on the values $B(p_i)$ of the parameters $p_i$ in $R$. Since the ranges of the parameters $p_i^s$ in $\rel{R}$ agree with the range of $p_i$ in $R$, we have 
\begin{align}
\subst{\rel{\pdtmc}}{\rel{R}} = \subst{\pdtmc}{R} \quad \text{for all graph-preserving } R.
\label{eqn:relaxationNotNecessary}
\end{align} 
A direct consequence of these statements yields:
\begin{theorem}
\label{th:dtmcApprox}
Let $\pdtmc$ be a pMC, $R$ a graph-preserving region, $\varphi$ a reachability property, subject to Assumption~\ref{assumptionpmc}. Then it holds:
\begin{align*}
  \forall \sched \in \Sched.\,~\subst{\pdtmc}{R}^\sched \models \varphi \implies& \pdtmc, R \models \varphi\quad\land \\
 \forall \sched \in \Sched.\,~ \subst{\pdtmc}{R}^\sched \models \neg\varphi \implies& \pdtmc, R \models \neg \varphi .
\end{align*}
\end{theorem}
Hence, we can deduce via Algorithm~\vref{alg:parlifting} whether $\pdtmc, R \models \varphi$ by applying standard techniques for MDP model checking to $\subst{\pdtmc}{R}$, such as value- and policy iteration, cf.~\cite{Put94,BK08}.
We stress that while the relaxation is key for showing the correctness, equation (\vref{eqn:relaxationNotNecessary}) proves that this step does not actually need to be performed.

\begin{example}
  Reconsider Example \vref{ex:approxmdp}. From $\subst{\pdtmc}{R}$ in Figure~\vref{fig:approxmdp:M}, we can derive $\max_{\sched \in \SchedSymbol} \reachPrT[\subst{\pdtmc}{R}^\sched] = \nicefrac{47}{60}$ and, by Theorem~\vref{th:dtmcApprox}, $\pdtmc, R \models \reachProp{\nicefrac{4}{5}}{T}$ follows. Despite the large region $R$, we establish a non-trivial upper bound on the reachability probability over all instantiations in $R$.
\end{example}

 If the over-approximation by region $R$ is too coarse for a conclusive answer, region $R$ can be refined, meaning that we split $R$ into a set of smaller regions\footnote{Strictly speaking, these regions will overlap as we always consider closed regions. This is not a concern for correctness. When splitting, we may take this information into account, see Section~\ref{sec:neighbourhoods}.}~\cite{DBLP:conf/cav/BrimCDS13}. We discuss splitting strategies in Section~\ref{sec:psp}.
Intuitively, as more potential parameter values are excluded by reducing the region size, the actual choice of the parameter value has less impact on reachability probabilities.
The smaller the region gets, the smaller the over-approximation: 
The optimal instantiation on the pMC $\pdtmc$ is over-approximated by some strategy on $\subst{\pdtmc}{R}$. 
The approximation error originates from choices where an optimal strategy on $\subst{\pdtmc}{R}$ chooses actions $u_1$ and $u_2$ at states $s_1$ and $s_2$, respectively, with $u_1(p_i^{s_1}) \neq u_2(p_i^{s_2})$ for some parameter $p_i$, and therefore intuitively disagree on its value.
The probability mass that is affected by these choices decreases the smaller the region is.
For infinitesimally small regions, the error from the over-approximation vanishes, as the actions for the upper and the lower bound of a parameter become equal up to an infinitesimal. 
More formally, the difference in reachability probability between two MCs corresponding to instantiations in a region tends is bounded and tends to zero if the region gets smaller~\cite[Lemma 9]{DBLP:journals/corr/Chonev17}.

\begin{algorithm}[t]
\caption{Parameter lifting}

\begin{newalgorithm}{reachability}{pMC $\pdtmc$, $T\subseteq S$, region $R$, specification $\reachProplT$ }
    \Comment{Check whether $\pdtmc, R \models \reachProplT$} \\
	\State Construct $\subst{\pdtmc}{R}$\\
	\If {$\forall \sched \in \Sched~\subst{\pdtmc}{R} \models \reachProplT$}
	\Comment via standard MDP model checking procedures \\
		\State \textbf{return} true\\
	\ElseIf{$\forall \sched \in \Sched~\subst{\pdtmc}{R} \models \p_{> \lambda}(\finally T)$}
	\Comment via standard MDP model checking procedures \\
	\State \textbf{return} false\\
	\Else
	\State \textbf{return} unknown
	\EndIf
  \end{newalgorithm}
  \label{alg:parlifting}
\end{algorithm}
\subsection{Expected reward properties}
The reduction of bounding reachability probabilities on pMCs  to off-the-shelf MDP model checking can also be applied to bound  expected rewards.
To see this, we have to extend the notion of locally monotone parametric Markov chains.
\begin{definition}[Locally monotone reward pMC]
	A pMC $\PdtmcInit$ with reward function  $\rew \colon S \to \ratfunc[V]$ is \emph{locally monotone} iff for all $s \in S$, there is a multilinear polynomial $\pol_s \in \poly[V] $ with \[
\{\rew(s), \probdtmc(s, s') \mid s' \in S \} \subseteq \left\{ \nicefrac{f}{\pol_s} \mid  f\in \poly[V] \text{ multilinear}  \right\}.\]
\end{definition}

We now generalise relaxation and substitution to the reward models, and obtain analogous results.

\begin{definition}[Substitution for reward pMCs]
	\label{def:rewardapproxPMC}
	Let $\PdtmcInit$ be a pMC, $\rewFunction \colon S \to \ratfunc[V]$ a reward function, $T \subseteq S$ a set of target states, and $R$ a region.
	For $s \in S$, let \[\Var^\rew_s =\Var_s \cup \{p_i \in \Var \mid p_i \text{ occurs in }  \rewFunction(s)\}.\]
	The MDP $\substrew{\pdtmc}{R} = (S, \sinit, \Act^\rew_\substAbbrev, \probmdp^\rew_\substAbbrev)$  with reward function $\rew_\substAbbrev$ is the \emph{(parameter-)substitution of $\pdtmc, \rew$ on $R$}, where 
	\begin{itemize}
	\item $\Act^\rew_\substAbbrev$ and $\probmdp^\rew_\substAbbrev$
		are analogous to Definition~\vref{def:approxPMC}, but over $\Var^\rew_s$.
	\item $\rewFunction_\substAbbrev$ is given by: \[
	(s,u) \mapsto
	\begin{cases}
	\rewFunction(s)[u] &\text{if } u \in \Act^\rew_s,\\
	0                    &\text{otherwise,}
	\end{cases}
	\]
	where $\Act^\rew_s$ is defined analogously to $\Act_s$ in Definition~\ref{def:approxPMC}.
	\end{itemize}
\end{definition}
The reward approximation of a pMC can be used to identify regions as accepting or rejecting for expected reward properties.
\begin{theorem}
	\label{th:rewarddtmcApprox}
	Let $\pdtmc$ be a pMC with locally monotone rewards $\rew$, $R$ a region, and $\varphi$ an expected reward property, subject to Assumption~\ref{assumptionpmc}:
	\begin{align*}
	 \forall \sched \in \Sched.\,~ \substrew{\pdtmc}{R} \models \varphi \text{ implies }& \pdtmc, R \models \varphi  \text{ and }\\
	 \forall \sched \in \Sched.\,~ \substrew{\pdtmc}{R} \models \neg\varphi \text{ implies }& \pdtmc, R \models \neg \varphi.
	\end{align*}
\end{theorem}
\noindent
The proof is analogous to the proof of Theorem~\vref{th:dtmcApprox}.

\section{Model-checking-based Region Verification of Parametric MDPs}
\label{sec:nondet}
In the previous section, we approximated reachability probabilities in (locally-monotone) pMCs by considering the substitution MDP, see Definition~\vref{def:approxPMC}.
The non-determinism in the MDP encodes the finitely many parameter valuations that approximate the reachability probabilities in the pMC. 
By letting an adversary player resolve the non-determinism in the MDP, we obtain bounds on the reachability probabilities in the pMC.
These bounds can efficiently be computed by standard MDP model checking.

In this section, we generalise the approach to pMDPs, which  already contain non-determinism.
The result naturally leads to a 2-player stochastic game: 
One player controls the nondeterminism inherent to the MDP, while the other player controls the (abstracted) parameter values.
Letting the two players adequately minimise and/or maximise the reachability probabilities in the SG yields bounds on the minimal (and maximal) reachability probabilities in the pMDP.
For example, if the player for the original non-determinism maximises and the parameter player minimises, we obtain a lower bound on the maximal probability. 
These bounds can efficiently be computed by standard SG model checking procedures.

In our presentation below, we discuss the interplay of the two sources of non-determinism.
 In particular, we show how the generalisation of the method yields an additional source of (over-)approximation. 
Then, we formalise the construction of the substitution with nondeterminism, analogous to the pMCs from the previous section. 
In particular, Definition~\vref{def:approxPMDP} is analogous to Definition~\vref{def:approxPMC} and Theorem~\vref{th:mdpApprox} is analogous to Theorem~\vref{th:dtmcApprox}. 
We do not repeat relaxation, described in Section~\ref{sec:approx_mc:relax},  as---as also discussed in the previous section---it is not a necessary ingredient for the correctness of the approach.
\subsection{Two types of approximation}
In the following, let $\pmdpInit$ be a pMDP and $R$ a graph-preserving, rectangular, closed region.

\paragraph{Demonic strategies}
We analyse $R$ with respect to the demonic relation $\models_d$. We have: 
\begin{align*}
\pmdp, R \models_d \varphi \iff \forall u \in R.~\forall \sched \in \Sched^{\pmdp}.~\pmdp[u]^\sched \models \varphi.
\end{align*}
The two universal quantifiers can be reordered, and in addition $\pmdp[u]^\sigma = \pmdp^\sigma[u]$. 
We obtain:
\begin{align*}
\pmdp, R \models_d \varphi \iff \forall \sched \in \Sched^{\pmdp}. ~\forall u \in R.~\underbrace{\pmdp^\sched}_{\text{a pMC}}[u] \models \varphi
\end{align*}
Intuitively, the reformulation states that we have to apply pMC region verification on $\pmdp^\sched$ and $R$ for all $\sched \in \Sched^\pmdp$.
We now want to employ parameter lifting for each strategy. 
Thus, we want to consider the verification of the substituted pMCs $\subst{\pmdp^\sched}{R}$. 
As these substituted pMCs share most of their structure, the set of all such substituted pMCs can be concisely represented as an SG, in which both players cooperate (as witnessed by the same quantifiers). 
In the scope of this paper, an SG with cooperating players can be concisely represented as an MDP.
Consequently, for the demonic relation, pMDP verification can be approximated by MDP model checking.

\paragraph{Angelic strategies}
We now turn our attention to the angelic relation $\models_a$, cf.\ Definition~\vref{def:satRelReg}.  
\begin{align*}
\pmdp, R \models_a \varphi \iff \forall u \in R.~\exists \sched \in \Sched^{\pmdp}.~\pmdp[u]^\sched \models \varphi.
\end{align*}
Here, we cannot simply reorder the quantifiers. 
However:
\begin{align*}
	\exists \sched \in \Sched^{\pmdp}.~ \forall u \in R.~\pmdp^\sched[u] \models \varphi
 \implies \pmdp, R \models_a \varphi.
\end{align*}
Now, the left-hand side expresses again that we want to do region verification for pMCs induced by a strategy, as in the demonic case, and that we likewise want to represent by a stochastic game. 
As witnessed by the quantifier alternation, this SG does not reduce to an MDP; the two players have opposing objectives.
Nevertheless, we can efficiently analyse this SG (with a variant of value iteration), and thus the left-hand side of the implication above.

Observe that the over-approximation actually computes a robust strategy, as discussed in Remark~\vref{rem:robust}.
In particular, we now have two sources of approximation:
\begin{itemize}	
\item The approximation that originates from dropping parameter dependencies (as also in the demonic case).
\item The application of the substitution of parameters with non-determinism on robust strategies rather than of the actual angelic relation. 
\end{itemize}
Both over-approximations vanish with declining region size.

\subsection{Replacing parameters by nondeterminism}

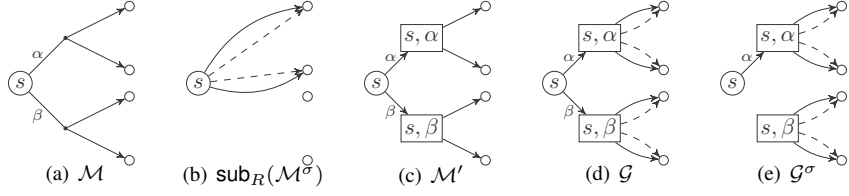
\begin{figure*}[t]
	\centering
	\subfigure[$\pmdp$]{
		\scalebox{0.85}{
			\begin{tikzpicture}[scale=1, nodestyle/.style={draw,circle, inner sep=1.5pt},baseline=(s)]
    
    \node [nodestyle] (s) at (0,0) {$s$};
    \node [draw, fill, circle, inner sep=0.5pt] (sa) [on grid, above right=1cm of s] {};
    \node [draw, fill, circle, inner sep=0.5pt] (sb) [on grid, below right=1cm of s] {};
    \node [nodestyle] (sa1) at ($(sa) + (1,0.5)$) {};
    \node [nodestyle] (sa2) at ($(sa) + (1,-0.5)$) {};
    \node [nodestyle] (sb1) at ($(sb) + (1,0.5)$) {};
    \node [nodestyle] (sb2) at ($(sb) + (1,-0.5)$) {};
    
    \draw (s) edge[] node[above, pos=0.25] {\scriptsize$\alpha$} (sa);
    \draw (s) edge[] node[below, pos=0.25] {\scriptsize$\beta$} (sb);
    
    \draw (sa) edge[->] node[right] {\scriptsize} (sa1);
    \draw (sa) edge[->] node[right] {\scriptsize} (sa2);
    \draw (sb) edge[->] node[right] {\scriptsize} (sb1);
    \draw (sb) edge[->] node[right] {\scriptsize} (sb2);
    
\end{tikzpicture}
		}
		\label{fig:pmdp:M}
	}
	\subfigure[$\subst{\pmdp^\sched}{R}$]{
		\scalebox{0.85}{
			\begin{tikzpicture}[scale=1, nodestyle/.style={draw,circle, inner sep=1.5pt},baseline=(s)]

\node [nodestyle] (s) at (0,0) {$s$};
\node [nodestyle] (sa1) at ($(sa) + (1,0.5)$) {};
\node [nodestyle] (sa2) at ($(sa) + (1,-0.5)$) {};
\node [nodestyle] (sb1) at ($(sb) + (1,0.5)$) {};
\node [nodestyle] (sb2) at ($(sb) + (1,-0.5)$) {};

\draw (s) edge[->, bend left=25] node[right] {\scriptsize} (sa1);
\draw (s) edge[->, bend right=25] node[right] {\scriptsize} (sa2);
\draw (s) edge[->, dashed, bend right=0] node[right] {\scriptsize} (sa1);
\draw (s) edge[->, dashed, bend left=0] node[right] {\scriptsize} (sa2);

\end{tikzpicture}
		}
		\label{fig:pmc:Msched}
	}
	\subfigure[$\pmdp'$]{
		\scalebox{0.85}{
			\begin{tikzpicture}[scale=1, nodestyle/.style={draw,circle, inner sep=1.5pt},baseline=(s)]
    
    \node [nodestyle] (s) at (0,0) {$s$};
    \node [nodestyle, rectangle, minimum height=12pt] (sa) [on grid, above right=1cm of s] {$s,\alpha$};
    \node [nodestyle, rectangle, minimum height=12pt] (sb) [on grid, below right=1cm of s] {$s,\beta$};
    \node [nodestyle] (sa1) at ($(sa) + (1,0.5)$) {};
    \node [nodestyle] (sa2) at ($(sa) + (1,-0.5)$) {};
    \node [nodestyle] (sb1) at ($(sb) + (1,0.5)$) {};
    \node [nodestyle] (sb2) at ($(sb) + (1,-0.5)$) {};
    
    \draw (s) edge[->] node[above, pos=0.25] {\scriptsize$\alpha$} (sa);
    \draw (s) edge[->] node[below, pos=0.25] {\scriptsize$\beta$} (sb);
    
    \draw (sa) edge[->] node[right] {\scriptsize} (sa1);
    \draw (sa) edge[->] node[right] {\scriptsize} (sa2);
    \draw (sb) edge[->] node[right] {\scriptsize} (sb1);
    \draw (sb) edge[->] node[right] {\scriptsize} (sb2);

\end{tikzpicture}
		}
		\label{fig:pmdp:Mp}
	}
	\subfigure[$\subst{\pmdp'}{R}$]{
		\scalebox{0.85}{
			\begin{tikzpicture}[scale=1, nodestyle/.style={draw,circle, inner sep=1.5pt},baseline=(s)]
    
    \node [nodestyle] (s) at (0,0) {$s$};
    \node [nodestyle, rectangle, minimum height=12pt] (sa) [on grid, above right=1cm of s] {$s,\alpha$};
    \node [nodestyle, rectangle, minimum height=12pt] (sb) [on grid, below right=1cm of s] {$s,\beta$};
    \node [nodestyle] (sa1) at ($(sa) + (1,0.5)$) {};
    \node [nodestyle] (sa2) at ($(sa) + (1,-0.5)$) {};
    \node [nodestyle] (sb1) at ($(sb) + (1,0.5)$) {};
    \node [nodestyle] (sb2) at ($(sb) + (1,-0.5)$) {};
    
    \draw (s) edge[->] node[above, pos=0.25] {\scriptsize$\alpha$} (sa);
    \draw (s) edge[->] node[below, pos=0.25] {\scriptsize$\beta$} (sb);
    
    \draw (sa) edge[->, bend left=15] node[right] {\scriptsize} (sa1);
    \draw (sa) edge[->, bend right=15] node[right] {\scriptsize} (sa2);
    \draw (sb) edge[->, bend left=15] node[right] {\scriptsize} (sb1);
    \draw (sb) edge[->, bend right=15] node[right] {\scriptsize} (sb2);
    
    \draw (sa) edge[->, dashed, bend right=15] node[right] {\scriptsize} (sa1);
    \draw (sa) edge[->, dashed, bend left=15] node[right] {\scriptsize} (sa2);
    \draw (sb) edge[->, dashed, bend right=15] node[right] {\scriptsize} (sb1);
    \draw (sb) edge[->, dashed, bend left=15] node[right] {\scriptsize} (sb2);
    
\end{tikzpicture}
		}
		\label{fig:pmdp:S}
	}
	\subfigure[$\subst{\pmdp'}{R}^\sched$]{
		\scalebox{0.85}{
			\begin{tikzpicture}[scale=1, nodestyle/.style={draw,circle, inner sep=1.5pt},baseline=(s)]

\node [nodestyle] (s) at (0,0) {$s$};
\node [nodestyle, rectangle, minimum height=12pt] (sa) [on grid, above right=1cm of s] {$s,\alpha$};
\node [nodestyle, rectangle, minimum height=12pt] (sb) [on grid, below right=1cm of s] {$s,\beta$};
\node [nodestyle] (sa1) at ($(sa) + (1,0.5)$) {};
\node [nodestyle] (sa2) at ($(sa) + (1,-0.5)$) {};
\node [nodestyle] (sb1) at ($(sb) + (1,0.5)$) {};
\node [nodestyle] (sb2) at ($(sb) + (1,-0.5)$) {};

\draw (s) edge[->] node[above, pos=0.25] {\scriptsize$\alpha$} (sa);

\draw (sa) edge[->, bend left=15] node[right] {\scriptsize} (sa1);
\draw (sa) edge[->, bend right=15] node[right] {\scriptsize} (sa2);
\draw (sb) edge[->, bend left=15] node[right] {\scriptsize} (sb1);
\draw (sb) edge[->, bend right=15] node[right] {\scriptsize} (sb2);

\draw (sa) edge[->, dashed, bend right=15] node[right] {\scriptsize} (sa1);
\draw (sa) edge[->, dashed, bend left=15] node[right] {\scriptsize} (sa2);
\draw (sb) edge[->, dashed, bend right=15] node[right] {\scriptsize} (sb1);
\draw (sb) edge[->, dashed, bend left=15] node[right] {\scriptsize} (sb2);

\end{tikzpicture}
		}
		\label{fig:pmdp:Ssched}
	}
	\caption{Illustration of the substitution of a pMDP.}
	\label{fig:pmdp}
\end{figure*}

\begin{example}\label{ex:substmdp}
Consider the pMDP $\pmdp$ in Figure~\vref{fig:pmdp:M}, where the state $s$ has two enabled actions $\alpha$ and $\beta$.
The strategy $\sigma$ given by $\{s \mapsto \alpha\}$ applied to $\pmdp$ yields a pMC, which is subject to substitution, cf.\ Figure~\vref{fig:pmc:Msched}.
\end{example}
The parameter substitution of a pMDP (cf.~Figure~\vref{fig:pmdp:M}) yields an SG---as in Figure~\vref{fig:pmdp:S}. It represents, for all strategies of the pMDP, the parameter-substitution (as in Definition~\vref{def:approxPMC}) of each induced pMC.
To ensure that in the SG each state can be assigned to a unique player,  we split states in the pMDP which have both (parametric) probabilistic branching and non-determinism, such that states have either probabilistic branching or non-determinism, but not both.
The reformulation is done as follows:
After each choice of actions, auxiliary states are introduced, such that the outcome of the action becomes deterministic and the probabilistic choice is delayed to the auxiliary state.
This construction is similar to the conversion of Segala's probabilistic automata into Hansson's alternating model~\cite{SegalaT05}. More precisely, we
\begin{compactitem}
\item split each state $s \in S$ into $\{s\} \uplus \{\langle s, \act \rangle \mid \act \in \Act(s) \}$, 
\item add a transition with probability one for each $s \in S$ and $\act \in \Act(s)$. The transition  leads from $s$ to $\langle s, \act \rangle$, and
\item move the probabilistic choice at $s$ \wrt $\act$ to $\langle s, \alpha \rangle$.  	
\end{compactitem}
Applying this to the pMDP from Figure~\vref{fig:pmdp:M}, we obtain the pMDP $\pmdp'$ in Figure~\vref{fig:pmdp:Mp}, where the state $s$ has only \emph{nondeterministic} choices leading to states of the form $\langle s, \act \rangle$ with only \emph{probabilistic} choices.
The subsequent substitution on the probabilistic states yields the  SG  $\subst{\pmdp'}{R}$, where one player represents the nondeterminism of the original pMDP $\pmdp$, while the other player decides whether parameters should be set to their lower or upper bound in the region $R$. 
For the construction, we generalise $V_s$ to state-action pairs:
For a pMDP, a state $s$ and action $\alpha$, let 
\[\Var_{s,\act} = \{\, p \in \Var \mid p \text{ occurs in } \probmdp(s,\alpha,s') \text{ for some } s'\in S \, \}.\]
\begin{definition}[Substitution (pMDPs)]	\label{def:approxPMDP}
For pMDP $\pmdpInit$ and region $R$, let  SG \[\subst{\pmdp}{R} = (\spOne \uplus S_{\pTwo}, \sinit, \Act_{\substAbbrev}, \probsg_{\substAbbrev})\] with 
\begin{itemize}
\item $ \spOne = S$ 
\item $S_{\pTwo} = \{\langle s, \alpha \rangle \mid \act \in \Act(s) \}$, 
\item	$\Act_{\substAbbrev} = \Act \uplus \big(\biguplus_{\langle s, \act \rangle \in \spTwo} \Act_s^\alpha\big)$ where \[\Act_s^\alpha = \{ u \colon V_{s,\act} \rightarrow \R \mid u(p) \in B(p) \;\forall p \in V_{s,\act} \},\] and,
\item
	\[
	\probsg_{\substAbbrev}(t,\beta,t') = 
	\begin{cases}
	1                                        & \text{if } t \in \spOne, \beta \in \Act(t), t' {=} \langle t,\beta\rangle \in \spTwo,\\
	\probmdp(s,\alpha,t')[\beta]                & \text{if } t {=} \langle s,\alpha \rangle \in \spTwo, \beta \in \Act_s^\alpha, t' \in \spOne,\\
	0                                        & \text{otherwise.}
	\end{cases}
	\]
	\end{itemize}
	
	 be the \emph{(parameter-)substitution of $\pmdp$ and $R$}.
\end{definition}
%
We relate the SG $\subst{\pmdp}{R}$ under different strategies for player $\pOne$ with the substitution in the strategy-induced pMCs of $\pmdp$. 
We observe that the strategies for player $\pOne$ in $\subst{\pmdp}{R}$ coincide with strategies in $\pmdp$.
Consider the induced MDP $(\subst{\pmdp}{R})^\sched$ with a strategy $\sched$ for player $\pOne$.
The MDP $(\subst{\pmdp}{R})^\sched$ is obtained from $\subst{\pmdp}{R}$ by erasing transitions not agreeing with $\sched$.
In $(\subst{\pmdp}{R})^\sched$ player $\pOne$-state have a single enabled action, while player $\pTwo$-states have multiple available  enabled actions.
\begin{example}
Continuing Example~\vref{ex:substmdp}, applying strategy $\sigma$ to $\subst{\pmdp}{R}$ yields $(\subst{\pmdp}{R})^\sched$, see Figure~\vref{fig:pmdp:Ssched}. The MDP $(\subst{\pmdp}{R})^\sched$ matches the MDP $\subst{\pmdp^{\sched}}{R}$
apart from intermediate states of the form $\langle s, \alpha \rangle$:
 The outgoing transitions of $s$ in $\subst{\pmdp^\sigma}{R}$ coincide with the outgoing transitions of $\langle s, \alpha \rangle$ in $(\subst{\pmdp}{R})^\sched$, where $\langle s, \alpha \rangle$ is the unique successor of $s$.
\end{example}
The following corollary formalises that $(\subst{\pmdp}{R})^\sched$ and $\subst{\pmdp^{\sched}}{R}$ induce the same reachability probabilities.
\begin{corollary}
	\label{cor:mdpApprox}
	For pMDP $\pmdp$, graph-preserving region $R$, target states $T\subseteq S$, and strategies $\sched \in \SchedOne^{\subst{\pmdp}{R}}$ and $\altsched \in \Sched^{\subst{\pmdp^\sigma}{R}}$, it holds that
	\begin{align*}
		\reachPrT[(\subst{\pmdp^\sigma}{R})^\altsched] 
	=
	\reachPrT[\subst{\pmdp}{R}^{\sched, \widehat{\altsched}}]	
	\end{align*}
	with $\widehat{\altsched} \in \Sched_\pTwo^{\subst{\pmdp}{R}}$ 
	satisfies $\widehat{\altsched}(\langle s, \sched(s) \rangle) = \altsched(s)$.
\end{corollary}
Instead of performing the substitution on the pMC induced by $\pmdp$ and $\sched$, we can perform the substitution on $\pmdp$ directly and preserve the reachability probability.

Consequently, and analogously to the pMC case (cf. Theorem~\vref{th:dtmcApprox}), we can derive whether $ \pmdp, R \models_\clubsuit \varphi$ by analysing a stochastic game. 
For this, we consider various standard variants of model checking on stochastic games.
\begin{definition}[Model-relation on SGs]
	For an SG $\psg$, property~$\varphi$, and quantifiers $\mathcal{Q}_1, \mathcal{Q}_2$, we define $\psg \models^{\mathcal{Q}_1,\mathcal{Q}_2} \varphi$ as:
	\begin{align*}  \mathcal{Q}_1 \schedOne \in \SchedOne^{\subst{\pmdp}{R}}.~\mathcal{Q}_2 \sched_\pTwo\in \Sched_\pTwo^{\subst{\pmdp}{R}}\quad\psg^{\schedOne, \sched_\pTwo} \models \varphi \end{align*}
\end{definition}
The order of players, for these games, does not influence the outcome \cite{DBLP:conf/dimacs/Condon90,Sha53}.

\begin{theorem}
\label{th:mdpApprox}
Let $\pmdp$ be a pMDP, $R$ a region, and $\varphi$ a reachability property, subject to Assumption~\ref{assumptionpmc}\footnote{straightforwardly lifting locally monotone pMCs to locally monotone pMDPs}.  Then:
\begin{align*}
\subst{\pmdp}{R} \models^{\forall,\forall} \varphi \text{ implies }& \pmdp, R \models_d \varphi  \text{, and }\\
\subst{\pmdp}{R} \models^{\exists,\forall} \varphi \text{ implies }& \pmdp, R \models_a \varphi .	
\end{align*}
\end{theorem}
\begin{proof}
We only prove the second statement using $\varphi = \reachPropneglT$, other reachability properties are similar.
A proof for the (simpler) first statement can be derived in an analogous manner.
We have that 
$\pmdp, R \models_a \reachPropneglT$ iff for all $u \in R$ there is a strategy $\sched$ of ${\pmdp}$ for which the reachability probability in the MC $\pmdp^\sched[u]$ exceeds the threshold $\lambda$, \ie,
\[
\pmdp, R \models_a \reachPropneglT \iff \min_{u\in R} \max_{\sched \in \Sched^{\pmdp}}  \reachPrT[{\pmdp^\sched[u]}]  > \lambda.
\]
A lower bound for this probability is obtained as follows:
\begin{align*}
    & \min_{u\in R}\max_{\sched \in \Sched^{\pmdp}} \big( \reachPrT[{\pmdp^\sched[u]}] \big)
\\ 
\ge & \max_{\sched \in \Sched^{\pmdp}} \min_{u\in R}\big( \reachPrT[{\pmdp^\sched[u]}] \big) 
\\
\overset{\ast}{\ge} & \max_{\sched \in \Sched^{\pmdp}} \min_{\altsched \in \Sched^{\subst{\pmdp^\sigma}{R}}}\big( \reachPrT[(\subst{\pmdp^\sigma}{R})^\altsched] \big) 
 \\
 \overset{\ast\ast}{=}  &   \max_{\sched \in \SchedOne^{\subst{\pmdp}{R}}}  \min_{\altsched \in \Sched_\pTwo^{\subst{\pmdp}{R}}} \big(\reachPrT[\subst{\pmdp}{R}^{\sched, \altsched}] \big).
 \end{align*}
The inequality $\ast$ is due to Corollary~\vref{cor:dtmcApprox}.
The equality $\ast\ast$ holds by Corollary~\vref{cor:mdpApprox}.
Then:
\begin{align*}
&\subst{\pmdp}{R} \models^{\exists,\forall} \reachPropneglT  \\ 
\iff 
&
\exists \sched \in \SchedOne^{\subst{\pmdp}{R}}.~\forall \altsched \in \Sched_\pTwo^{\subst{\pmdp}{R}} \\ & \qquad \psg^{\sched, \altsched} \models \reachPropneglT  \\
\iff 
& \max_{\sched \in \SchedOne^{\psg}}  \Big( \min_{\altsched \in \Sched_\pTwo^{\psg}} \big(\reachPrT[\psg^{\sched, \altsched}] \big) \Big) > \lambda
\\  \implies & \min_{u\in R}\max_{\sched \in \Sched^{\pmdp}} \big( \reachPrT[{\pmdp^\sched[u]}] \big) > \lambda
 \\ \iff & \pmdp, R \models_a \reachPropneglT.
\end{align*}
\end{proof}

\section{Approximate Synthesis by Parameter Space Partitioning}
Parameter space partitioning is our iterative approach to the approximate synthesis problem. 
It builds on top of region verification, discussed above, and is, conceptually, independent of the methods used for verification discussed later.

\label{sec:psp}
Parameter space partitioning is best viewed as a counter-example guided abstraction refinement (CEGAR)-like~\cite{CGJLV00} approach to successively divide the parameter space into accepting and rejecting regions.
The main idea is to compute a sequence $\left( R^i_a \right)_i$ of simple accepting regions that successively extend each other.
Similarly, an increasing sequence $\left( R^i_r \right)_i$ of simple rejecting regions is computed.
At the $i$-th iteration, $R^i = R^i_a \cup R^i_r$ is the covered fragment of the parameter space.
The iterative approach halts when $R^i$ is at least $c$\% of the entire parameter space.
Termination is guaranteed: in the limit a solution to the exact synthesis problem is obtained as $\lim_{i \rightarrow \infty} R_a^i = R_a$ and $\lim_{i \rightarrow \infty} R_r^i = R_r$. 

\begin{figure}[t]
\centering
 \includegraphics[scale=0.5]{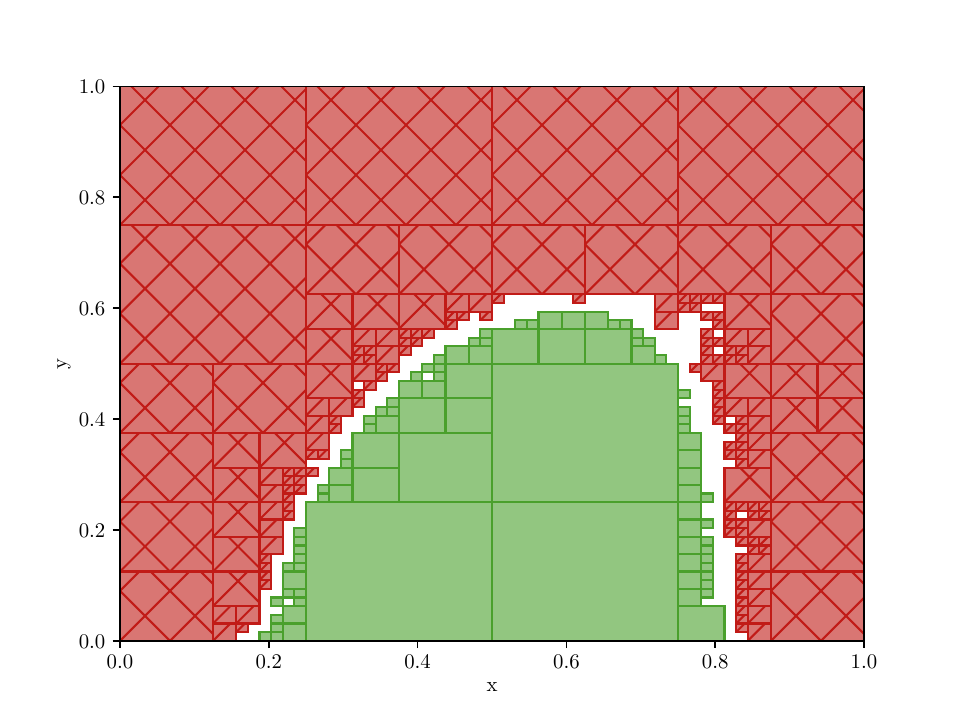}
 \caption{Parameter space partitioning into safe and unsafe regions.}
 \label{fig:regions_die_2}
\end{figure}

Let us describe the synthesis loop for the approximate synthesis as depicted in Figure~\vref{fig:problemrelation} in detail. 
In particular, we discuss how to generate \emph{candidate regions} that can be dispatched to the verifier along with a \emph{hypothesis} whether the candidate region is accepting or rejecting. 
We focus on \emph{rectangular regions} for several reasons: 
\begin{itemize}
\item the automated generation of rectangular regions is easier to generalise to multiple dimensions,
\item {earlier experiments~\cite{DJJ+15} revealed that rectangular regions lead to a more efficient SMT-based verification of regions (described in Section~\ref{sec:exact_mc}), and}
\item {model-checking based region verification (described in Section~\ref{sec:approx_mc}) requires rectangular regions.}
\end{itemize}
A downside of rectangular regions is that they are neither well-suited to approximate a region partitioning given by a diagonal, nor to cover well-defined regions that are not rectangular themselves.
\begin{remark}
{In the following, we assume that the parameter space is given by a rectangular well-defined region $R$. 
If the parameter space is not rectangular,  we over-approximate $R$ by a rectangular region $\hat{R} \supseteq R$. 
If the potential over-approximation of the parameter space $\hat{R}$ is not well-defined,  then we iteratively approximate $\hat{R}$ by a sequence of well-defined and ill-defined\footnote{A region $R$ is ill-defined if \emph{no} instantiation in $R$ is well-defined.} regions. }
The regions in the sequence of well-defined regions are then subject to the synthesis problem.
Constructing the sequence of regions is done analogously to the partitioning into accepting and rejecting regions.
\end{remark}
Before we present the procedure in full detail, we first outline a naive refinement procedure by means of an example.
\begin{example}[Naive refinement loop]
\label{ex:naiverefinement}
Consider the parametric die from Example~\vref{ex:psp}. 
Suppose we want to synthesise the partitioning as depicted in Figure~\vref{fig:regions_die}. 
We start by verifying the full parameter space $R$ against $\varphi$.
The verifier returns \texttt{false}, as $R$ is not accepting. 
Since $R$ (based on our knowledge at this point) might be rejecting, we invoke the verifier with $R$ and $\neg\varphi$, yielding \texttt{false} too.
Thus, the full parameter space $R$ is inconsistent.
We now split $R$ into four equally-sized regions, all of which are inconsistent. 
Only after splitting again, we find the first accepting and rejecting regions.
After various iterations, the procedure leads to the partitioning in Figure~\vref{fig:regions_die_2}.
\end{example}
Algorithm~\vref{alg:naiverefinement} describes this naive region partitioning procedure. 
It takes a pSG, a region $R$, a specification $\varphi$, and a (demonic or angelic) satisfaction relation as input.
It first initialises a (priority) queue $Q$ with $R$. 
In each iteration, a subregion $R'$ of $R$ is taken from the queue, the counter $i$ is incremented, and the sequence of accepted and rejected regions is updated.
There are three possibilities. 
Either $R'$ is accepting (or rejecting), and $R_a^{i}$ ($R_r^{i}$) extends $R_a^{i-1}$ ($R_r^{i-1}$) with $R'$, or $R'$ is inconsistent.
In the latter case, we split $R'$ into a finite set of subregions that are inserted into the queue $Q$.
Regions that are not extended are unchanged. 

\begin{algorithm}[t]
\caption{Naive refinement loop}
\begin{newalgorithm}{naive-refinement}{pSG $\psg$, rectangular region $R$, $\clubsuit \in \{ a, d \}$, specification\ $\varphi$}
	\State $i \colonequals 0$ \\
	\State $Q \colonequals \{ R \}$, $R_a^i \colonequals \emptyset$, $R_r^i \colonequals \emptyset$ \\
	\While{$Q\neq \emptyset$}
			\State $i \colonequals i + 1$\\
		\State $R' \colonequals  Q$.pop \\
		\If{$\psg, R' \models_\clubsuit \varphi$}
			\State $R^i_a \colonequals R^{i-1}_a \cup R'$, $R^i_r \colonequals R^{i-1}_r$ \\
		\ElseIf{$\psg, R' \models_\clubsuit \neg\varphi$ }
			\State $R^i_a \colonequals R^{i-1}_a$, $R^i_r \colonequals R^{i-1}_r \cup R'$ \\
		\Else
			\State $R^i_a \colonequals R^{i-1}_a$, $R^i_r \colonequals R^{i-1}_r$ \\
			\State $Q \colonequals Q \cup \text{\textbf{split}}(R')$ \\
		\EndIf
    \EndWhile
    \Return{Accepting region $R^i_a$, Rejecting region $R^i_r$}
  \end{newalgorithm}
  \label{alg:naiverefinement}
\end{algorithm}
The algorithm only terminates if $R_a$ and $R_r$ are a finite union of hyper-rectangles.
However, the algorithm can be terminated after any iteration yielding a sound approximation.
The algorithm ensures $\lim_{i \rightarrow \infty} R^i = R$, if we order $Q$ according to the size of the regions. 
We omit the technical proof here; the elementary property is that the regions are Lebesgue-measurable (and have a positive measure by construction).

The naive algorithm has a couple of structural weaknesses:
\begin{itemize}
\item 
It invokes the verification algorithm twice to determine that the full parameter space is inconsistent. 
\item 
It does not provide any (diagnostic) information from a verification invocation yielding \texttt{false}. 
\item It checks whether a region is accepting before it checks whether it is rejecting. 
    This order is suboptimal if the region is rejecting. 
\item If the region is inconsistent, it splits the region into $2^n$ equally large regions. 
   Instead, it might be beneficial to select a smaller number of regions (only split in one dimension).
\item Uninformed splitting yields many inconsistent subregions. 
Splitting in only one dimension even increases the number of verification calls yielding \texttt{false}.  
\end{itemize}

In the remainder of this section, we discuss ways to alleviate these weaknesses. 
The proposed improvements are based on empirical observations about the benchmarks
and are in line with the implementation in our tool \prophesy.
In particular, we tailor the heuristics to ``well-behaved'' models and specifications, which reflect the benchmarks from various domains. 
The notion of being well-behaved refers to 
\begin{itemize}
	\item a limited number of connected accepting and rejecting regions with smooth (albeit highly non-linear) borders between these regions.
	\item a limited number of accepting (rejecting) instantiations that are close to a rejecting (accepting) instantiations. We call instantiations that form a border between $R_a$ and $R_r$ \emph{border} instantiations.  
\end{itemize}
The parameter space depicted in Figure~\vref{fig:regions_die_2}
is well-behaved. 
It features only two connected regions, with a smooth border between them. 
Furthermore, the regions have a considerable interior, or equivalently, many instantiations are not too close to the border.
We remark that we do \emph{rely} on these assumptions to hold, but \prophesy will  be slow on models that are not well-behaved.

\subsection{Sampling} 
A simple but effective improvement is to verify an instantiated model $\psg[u]$ for some instantiation (a \emph{sample}) $u \in R$.
The verification result either reveals that the region is not accepting, if $\psg[u]\not\models_\clubsuit \varphi$, or not rejecting, if $\psg[u]\models_\clubsuit \varphi$.
Two samples within a region $R$ may suffice to conclude that $R$ is inconsistent.
In order to quickly find inconsistent regions by sampling, it is beneficial to seek for border instantiations. 
To this end, a good strategy is to start with a coarse (random) sampling to get a first indication of border instantiations. 
We then select additional instantiations by intra-/extrapolation between these samples.

\begin{figure*}
\centering
	\subfigure[(Uniform) Sampling]{
		\includegraphics[scale=0.2]{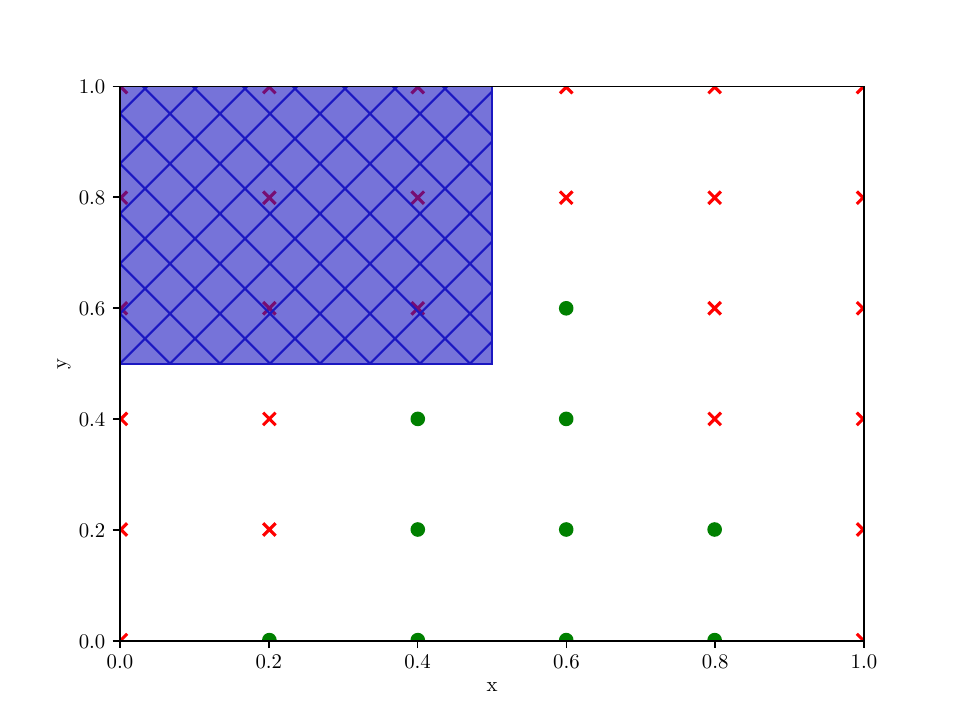}
		\label{fig:pspex:1}
	}
	\subfigure[Generating candidates]{
	\includegraphics[scale=0.2]{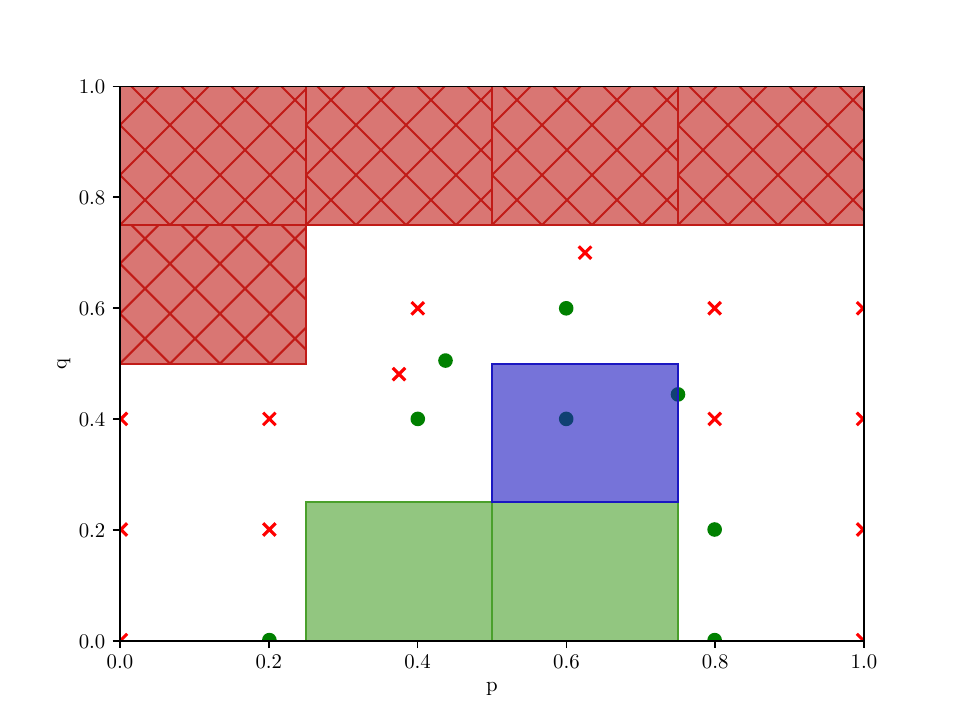}
	\label{fig:pspex:2}
	}
	\subfigure[Preliminary result]{
	\includegraphics[scale=0.2]{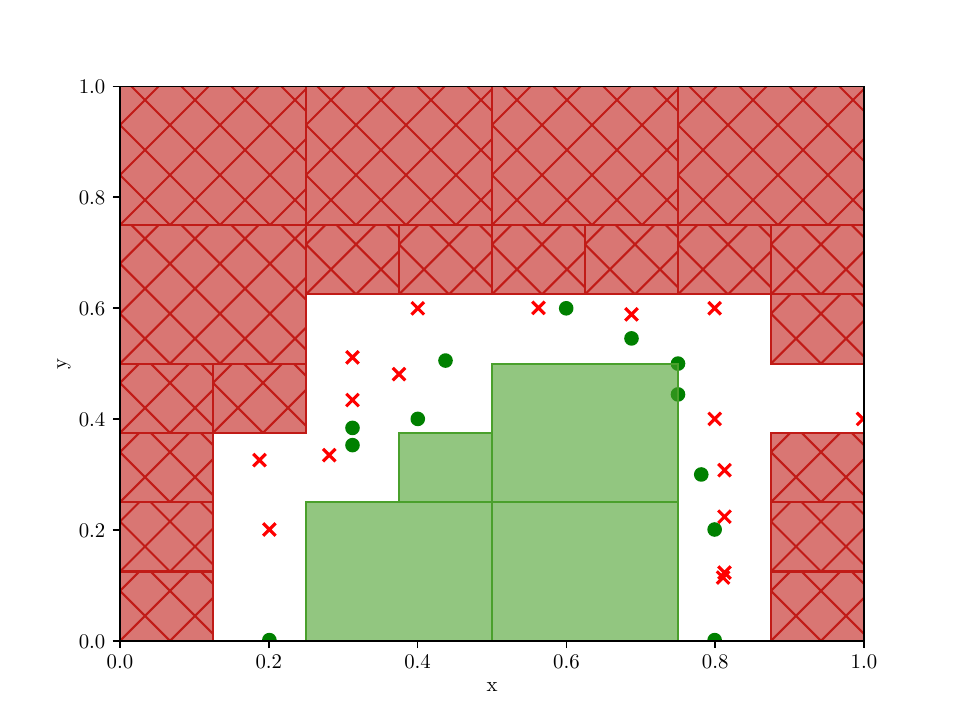}
	\label{fig:pspex:3}
	}
	\caption{Parameter space partitioning in progress: Images generated by \prophesy.}
\end{figure*}

\begin{example}
\label{ex:refine1}
We discuss how sampling may improve the naive refinement loop as discussed in Example~\vref{ex:naiverefinement}.
Figure~\vref{fig:pspex:1} shows a uniform sampling.
Red crosses indicate that the instantiated pMC satisfies $\neg\varphi$, while green dots indicate that the instantiation satisfies $\varphi$.
The blue rectangle is a candidate region (with the hypothesis $\neg\varphi$, indicated by the hatching), which is consistent with all samples.
\end{example}
\subsection{Finding region candidates}\label{sec:regionpartitioning:findingcandidates}We use the sampling results to steer the selection of a candidate region that may either be accepting or rejecting.
A simple strategy is to split regions that we found to be inconsistent via sampling.  
\begin{example}
Consider the parameter space with six samples depicted in Figure~\vref{fig:regioncandidates:samples}.	
	After verifying only six instantiated models, we conclude that the parameter space is inconsistent.
\end{example}
\begin{figure*}
\centering
	\subfigure[]{
		\begin{tikzpicture}[scale=2.5]
	\draw[very thick, ->] (0,0) -- (1,0);
	\draw[very thick, ->] (0,0) -- (0,1);
	\draw[dotted] (1,1) -- (1,0);
	\draw[dotted] (.5,1) -- (.5,0);
	
	\draw[dotted] (1,1) -- (0,1);
	\draw[dotted] (1,.5) -- (0,.5);
	
	\node at (0.5,-0.1) {\scriptsize $\nicefrac{1}{2}$ };
	\node at (1,-0.1) {\scriptsize $1$};
	\node at (-0.1,0.5) {\scriptsize $\nicefrac{1}{2}$};
	\node at (-0.1,1) {\scriptsize $1$};
	\node at (-0.2,1.1) {\scriptsize $p$};
	\node at (1.1,-0.2) {\scriptsize $q$};

	\node[draw,thick, circle, green,scale=0.4] at (0.2,0.3) {};
	\node[draw,thick, circle, green,scale=0.4] at (0.45,0.75) {};
	\node[draw,thick, circle, green,scale=0.4] at (0.68,0.4) {};
	
	\node[draw,thick, cross, red,scale=8] at (0.68,0.94) {};
	
	\node[draw,thick, cross, red,scale=8] at (0.78,0.65) {};
	
	\node[draw,thick, cross, red,scale=8] at (0.95,0.45) {};
\end{tikzpicture}
		\label{fig:regioncandidates:samples}
}
	\subfigure[]{
	\begin{tikzpicture}[scale=2.5]
	\draw[very thick, ->] (0,0) -- (1,0);
	\draw[very thick, ->] (0,0) -- (0,1);
	\draw[dotted] (1,1) -- (1,0);
	\draw[dotted] (.5,1) -- (.5,0);
	
	\draw[dotted] (1,1) -- (0,1);
	\draw[dotted] (1,.5) -- (0,.5);
	
	\node at (0.5,-0.1) {\scriptsize $\nicefrac{1}{2}$ };
	\node at (1,-0.1) {\scriptsize $1$};
	\node at (-0.1,0.5) {\scriptsize $\nicefrac{1}{2}$};
	\node at (-0.1,1) {\scriptsize $1$};
	\node at (-0.2,1.1) {\scriptsize $p$};
	\node at (1.1,-0.2) {\scriptsize $q$};

	\node[draw,thick, circle, green,scale=0.4] at (0.2,0.3) {};
	\node[draw,thick, circle, green,scale=0.4] at (0.45,0.75) {};
	\node[draw,thick, circle, green,scale=0.4] at (0.68,0.4) {};
	
	\node[draw,thick, cross, red,scale=8] at (0.68,0.94) {};
	
	\node[draw,thick, cross, red,scale=8] at (0.78,0.65) {};
	
	\node[draw,thick, cross, red,scale=8] at (0.95,0.45) {};
	
	\draw[color=blue, pattern=north west lines, pattern color=blue] (0,0) rectangle (0.5, 1);
	\draw[color=yellow!70!red, pattern=vertical lines, pattern color=yellow!70!red] (0.5,0.5) rectangle (1, 1);
\end{tikzpicture}
	\label{fig:regioncandidates:equal}
}

	\subfigure[]{
	\begin{tikzpicture}[scale=2.5]
	\draw[very thick, ->] (0,0) -- (1,0);
	\draw[very thick, ->] (0,0) -- (0,1);
	\draw[dotted] (1,1) -- (1,0);
	\draw[dotted] (.5,1) -- (.5,0);
	
	\draw[dotted] (1,1) -- (0,1);
	\draw[dotted] (1,.5) -- (0,.5);
	
	\node at (0.5,-0.1) {\scriptsize $\nicefrac{1}{2}$ };
	\node at (1,-0.1) {\scriptsize $1$};
	\node at (-0.1,0.5) {\scriptsize $\nicefrac{1}{2}$};
	\node at (-0.1,1) {\scriptsize $1$};
	\node at (-0.2,1.1) {\scriptsize $p$};
	\node at (1.1,-0.2) {\scriptsize $q$};

	\node[draw,thick, circle, green,scale=0.4] at (0.2,0.3) {};
	\node[draw,thick, circle, green,scale=0.4] at (0.45,0.75) {};
	\node[draw,thick, circle, green,scale=0.4] at (0.68,0.4) {};
	
	\node[draw,thick, cross, red,scale=8] at (0.68,0.94) {};
	
	\node[draw,thick, cross, red,scale=8] at (0.78,0.65) {};
	
	\node[draw,thick, cross, red,scale=8] at (0.95,0.45) {};

	\draw[color=yellow!70!red, pattern=vertical lines, pattern color=yellow!70!red] (0.47,0.42) rectangle (1, 1);
		\draw[color=blue, pattern=north west lines, pattern color=blue] (1,1) rectangle (0.78, 0.65);
\end{tikzpicture}
	\label{fig:regioncandidates:growing}
}
	\subfigure[]{
	\begin{tikzpicture}[scale=2.5]
	\draw[very thick, ->] (0,0) -- (1,0);
	\draw[very thick, ->] (0,0) -- (0,1);
	\draw[dotted] (1,1) -- (1,0);
	\draw[dotted] (.5,1) -- (.5,0);
	
	\draw[dotted] (1,1) -- (0,1);
	\draw[dotted] (1,.5) -- (0,.5);
	
	\node at (0.5,-0.1) {\scriptsize $\nicefrac{1}{2}$ };
	\node at (1,-0.1) {\scriptsize $1$};
	\node at (-0.1,0.5) {\scriptsize $\nicefrac{1}{2}$};
	\node at (-0.1,1) {\scriptsize $1$};
	\node at (-0.2,1.1) {\scriptsize $p$};
	\node at (1.1,-0.2) {\scriptsize $q$};

	\node[draw,thick, circle, green,scale=0.4] at (0.2,0.3) {};
	\node[draw,thick, circle, green,scale=0.4] at (0.68,0.78) {};
	\node[draw,thick, circle, green,scale=0.4] at (0.68,0.4) {};
	
	\node[draw,thick, cross, red,scale=8] at (0.68,0.94) {};
	
	\node[draw,thick, cross, red,scale=8] at (0.78,0.65) {};
	
	\node[draw,thick, cross, red,scale=8] at (0.95,0.45) {};

		\draw[color=blue, pattern=north west lines, pattern color=blue] (0,0) rectangle (0.68, 0.78);
\end{tikzpicture}
	\label{fig:regioncandidates:largest}	
}

\caption{Creating region candidates based on samples.}
\end{figure*}
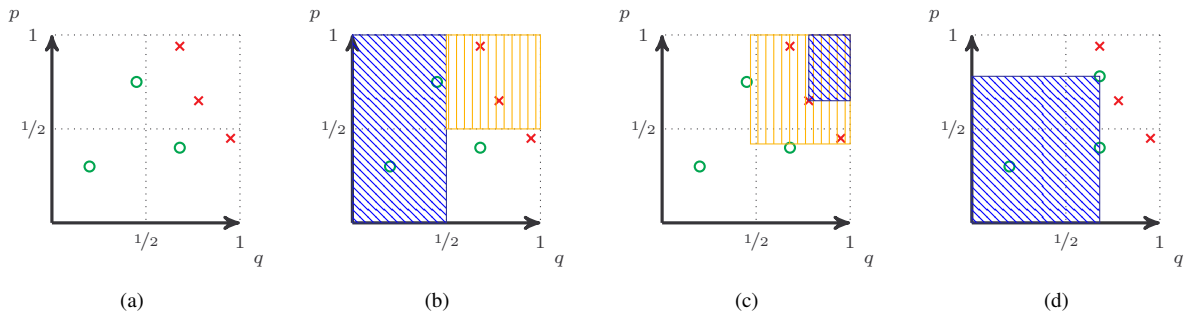
\begin{algorithm}[t]
\caption{Sampling-based refinement loop}
\begin{newalgorithm}{sampling-refinement}
    {pSG $\psg$, rectangular region $R$, $\clubsuit \in \{ a, d \}$, specification $\varphi$}
    \State $i\colonequals0$, $Q \colonequals \{ (R, \text{\textbf{sample}}(R)) \}$, $R_a^i \colonequals \emptyset$, $R_r^i \colonequals \emptyset$\\
	\While{$Q\neq \emptyset$}
			\State $i \colonequals i + 1$\\
		\State $(R', X') \colonequals  Q$.pop \\
		\If{$\psg, X' \models_\clubsuit \varphi$ \textbf{and} $\psg, R' \models_\clubsuit \varphi$}
			\State $R^i_a \colonequals R^{i-1}_a \cup R'$, $R^i_r \colonequals R^{i-1}_r$ \\
		\ElseIf{$\psg, X' \models_\clubsuit \neg\varphi$ \textbf{and} $\psg, R' \models_\clubsuit \neg\varphi$}
			\State $R^i_a \colonequals R^{i-1}_a,R^i_r \colonequals R^{i-1}_r \cup R'$ \\
		\Else
			\State $R^i_a \colonequals R^{i-1}_a$, $R^i_r \colonequals R^{i-1}_r$ \\
			\State $Q \colonequals Q \cup \text{\textbf{split}}(R', X')$ \\
		\EndIf
    \EndWhile
    \Return{Accepting region $R^i_a$, Rejecting region $R^i_r$}
  \end{newalgorithm}
  \label{alg:samplingrefinement}
\end{algorithm}
The use of samples allows to improve the naive refinement scheme as given in Algorithm~\vref{alg:naiverefinement}.
This improvement is given in Algorithm~\vref{alg:samplingrefinement}.
For each region $R$, we have a finite set $X$ of samples.
For each sample $u \in X$, it is known whether $\psg[u] \models_\clubsuit \varphi$.
The queue $Q$ now contains pairs $(R,X)$.

In each iteration, a pair $(R',X')$ where $R'$ is (as before) a subregion of $R$ is taken from the queue.
Then, we distinguish (again) three possibilities.
Only when all samples in $X'$ satisfy $\varphi$, it is verified whether $R'$ is accepting.
If $R'$ is accepting, we proceed as before: $R_a^{i}$ is extended by $R'$ while $R_r^{i}$ remains unchanged.
In the symmetric case that all samples in $X'$ refute $\varphi$, we proceed in a similar way by verifying whether $R'$ rejects $\varphi$.
Otherwise, $R'$ is split into a finite set of subregions with corresponding subsets of $X'$, and added to the queue $Q$.
In case the verification engine provides a counterexample, we can add this counterexample as a new sample.
We thus ensure that for all $(R', X') \in Q$, $u \in X'$ implies $u \in R'$.
The algorithm can be easily extended such that sampling is also done once a region without samples is obtained: rather than inserting $(R', \emptyset)$ into $Q$, we insert the entry $(R', \text{\textbf{sample}}(R'))$.

\begin{example}
	After several more iterations, the refinement loop started in Example~\vref{ex:refine1} has proceeded to the state in Figure~\vref{fig:pspex:2}.
	First, we see that the candidate region from Figure~\vref{fig:pspex:1} was not rejecting. The verification engine gave a counterexample in form of an accepting sample (around $p \mapsto 0.45, q\mapsto 0.52$).
	Further iterations with smaller regions had some successes, but some additional samples were generated as counterexamples. The current blue candidate is to be checked next. 
	In Figure~\vref{fig:pspex:3}, we see a further continuation, with even smaller regions being verified. Note the white box on the right border: It has been checked, but the verification timed out without a conclusive answer.
	 Therefore, we do not have a counterexample in this subregion.
\end{example}

It remains to discuss some methods to split a region, 
and how we may discard some of the constructed regions. 
We outline more details below.

\subsubsection{How to split}

Splitting of regions based on the available samples can be done using different strategies.
We outline two basic approaches. 
These approaches can be easily mixed and extended, and their performance heavily depends on the concrete example at hand.

\paragraph{Equal splitting.}
This approach splits regions in equally-sized regions; the main rationale is that this generates small regions with concise bounds (the bounds are typically powers of two).
Splitting in equally sized regions can be done recursively: 
One projects all samples down to a single dimension, and splits if both accepting and rejecting samples are in the region.  
The procedure halts if all samples in a region are either accepting or rejecting. 
The order in which parameters are considered plays a crucial role. 
Typically, it is a good idea to first split along the larger dimensions.  
\begin{example}
A split in equally-sized regions is depicted in Figure~\vref{fig:regioncandidates:equal}, where first the left region candidate is created. 
The remaining region can be split either horizontally or vertically to immediately generate another region candidate. 
A horizontal split in the remaining region yields a region without any samples.
\end{example}
The downside of equal splitting is that the position of the splits are not adapted based on the samples.
Therefore, the number of splits might be significantly larger than necessary, leading to an increased number of verification calls.

\paragraph{Growing rectangles.}
This approach attempts to gradually obtain a large region candidate\footnote{The approach shares its rationale with the approach formerly implemented in \prophesy~\cite{DJJ+15}, but is realised slightly differently to overcome challenges for n-dimensional hyper-rectangles.}.
{The underlying rationale is to quickly cover vast amounts of the parameter space. 
This is illustrated in Figure~\vref{fig:regioncandidates:largest} (notice that we adapted the samples for a consistent but concise description) where from an initial sampling a large rectangle is obtained as region candidate.}
\begin{example}
	Consider the shaded regions in Figure~\vref{fig:regioncandidates:growing}.
	 Starting from vertex $v=(1,1)$, the outer rectangle is maximised to not contain any accepting samples. 
	 Taking this outer rectangle as candidate region is very optimistic, it assumes that the accepting samples are on the border. 
	 A more pessimistic variant of growing rectangles is given by the inner shaded region. 
	 It takes a rejecting sample as vertex $v'$ such that the $v$ and $v'$ span the largest region.
\end{example}
The growing rectangles algorithm iterates over a subset of the hyper-rectangle's vertices: 
For each vertex (referred to as \emph{anchor}), among all possible sub-hyper-rectangles containing the anchor and only accepting or only rejecting samples, the largest is constructed.
\begin{example}
\label{ex:growing_rectangles_two}
The growing rectangles approach pessimistically takes anchor $(0,0)$ as anchor and yields the candidate region in Figure~\vref{fig:regioncandidates:largest}.
\end{example} 
The verification fails more often on large regions (either due to time-outs or due to the over-approximation).
Consequently, choosing large candidate regions comes at the risk of failed verification calls, and fragmentation of the parameter space in more subregions. 

Furthermore, growing rectangles requires a fall-back splitting strategy: To see why, consider  Figure~\vref{fig:regions_die_2}.
The accepting (green) region does not contain any anchors of the full parameter space, therefore the hypothesis for any created subregion is always rejection. 
Thus, no subregion containing a (known) accepting sample is ever considered as a region candidate.

\subsubsection{Neighbourhood analysis}
\label{sec:neighbourhoods}
Besides considering samples within a region, we would like to illustrate that analysis of a region $R$ can and should take information from outside of $R$ into account. 
First, take Figure~\vref{fig:regioncandidates:equal}, and assume that the left region is indeed accepting. 
The second generated region contains only rejecting samples, but it is only rejecting if all points, including all those on the border to the left region, are rejecting. 
In other words, the border between the accepting and rejecting regions needs to exactly follow the border between the generated region candidates.
The latter case does not occur often, so it is reasonable to shrink or split the second generated region.
 Secondly, a sensible hypothesis for candidate regions without samples inside is helpful, especially for small regions or in high dimensions.
Instead of spawning new samples, we take samples and decided regions outside of the candidate region into account to create a hypothesis.
Concretely, we infer the hypothesis for regions without samples via the  closest known region or sample.

\subsection{Requirements on verification back-ends}
In this section, we have described techniques for iteratively partitioning the parameter space into accepting and rejecting regions.
The algorithms rely on verifying  regions (and sets of samples) against the specification $\varphi$.
The way in which verification is used in the iterative parameter space partitioning scheme imposes the following requirements on the verification back-end:
\begin{enumerate}[i)]
\item 
The verification should work \emph{incrementally}.
That is to say, verification results from previous iterations should be re-used in successive iterations.
Verifying different regions share the same model (pMC or pMDP). 
A simple example of working incrementally is to reuse minimisation techniques for the model over several calls. 
If a subregion is checked, the problem is even incremental in a more narrow sense: 
any bounds etc. obtained for the super-region are also valid for the subregion.
\item 
If the verification procedure fails, i.e.\ if the verifier returns \texttt{false}, obtaining additional \emph{diagnostic information} in the form of a counterexample is beneficial. 
A counterexample here is a sample which refutes the verification problem at hand.
\end{enumerate}
This wish list is very similar to the typical requirements that theory solvers in lazy SMT frameworks should fulfil~\cite{DBLP:series/faia/2009-185}. 
Therefore, SMT-based verification approaches naturally match the wish-list.
Parameter-lifting can work incrementally: it reuses the graph-structure to avoid rebuilding the MDP, and it may use previous model checking results to improve the time until the model checker converges. 
Parameter-lifting, due to its approximative nature, does provide only limited diagnostic information: In particular, it provides information which parameters would be assigned with the upper or lower bounds based on the strategy that optimizes the MDP/SG.
  
\section{Implementation}
\label{sec:impl}
All the algorithms and constructions in this paper have been implemented, and are publicly available via \prophesy\footnote{\url{github.com/moves-rwth/prophesy}, archived at \url{doi.org/10.5281/zenodo.7697154}}.
In particular, \prophesy supports algorithms for:
\begin{itemize}	
\item \textbf{the exact synthesis problem:} 
via \emph{computing the solution function}, using either of the three variants of state elimination, discussed in Section~\ref{sec:solution_fct}.
\item \textbf{the verification problem:}
via an \emph{encoding to an SMT-solver} as in Section~\ref{sec:exact_mc} or by employing the \emph{parameter lifting} method as in Section~\ref{sec:approx_mc} and \ref{sec:nondet}.
\item \textbf{the approximate synthesis problem:}
via \emph{parameter space partitioning}, that iteratively generates verification calls as described in Section~\ref{sec:psp}.
\end{itemize}

\prophesy is implemented in \python, and designed as a flexible toolbox for developing and experimenting with parameter synthesis. \prophesy internally heavily relies on high-performance routines of the probabilistic model checker \storm~\cite{DBLP:conf/cav/DehnertJK017} and the SMT Z3. \prophesy is built in a modular way, such that it is easy to use different backend solvers.
The computation of the solution function and the parameter lifting presented in the experiments have been implemented in \storm.

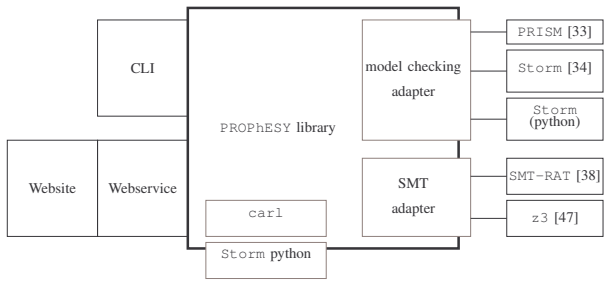
\begin{figure}
\centering
\scalebox{1}{
\begin{tikzpicture}
	\draw (0,2.2) rectangle (1.5,3.8);
	\node at (.75,3) {\scriptsize CLI};
	\draw (0,0.2) rectangle (1.5,1.8);
	\node at (.75,1) {\scriptsize Webservice};
	\draw (-1.5,0.2) rectangle (0,1.8);
	\node at (-.75,1) {\scriptsize Website};
	\draw[very thick] (1.5,0) rectangle (6,4);
	\node at (3,2) {\scriptsize \prophesy library};
	\draw[gray, fill=white] (4.4,1.8) rectangle (6.2,3.8);
	\node[align=center] at (5.25,2.8) {\scriptsize model checking\\\scriptsize adapter};
	\draw[gray, fill=white] (4.4,0.2) rectangle (6.2,1.5);
	\node[align=center] at (5.25,0.85) {\scriptsize SMT\\\scriptsize adapter};
	
	\draw[gray, fill=white] (1.8,-0.5) rectangle (3.8,0.1);
	\node[align=center] at (2.8,-0.1) {\scriptsize \storm python};
	\draw[gray, fill=white] (1.8,0.2) rectangle (3.8,0.8);
	\node[align=center] at (2.8,0.6) {\scriptsize \tool{carl}};
	
	\draw (6.8, 3.8) rectangle (8.4, 3.4);
	\node[align=center] at (7.6,3.6) {\scriptsize \prism~\cite{KNP11}};
	\draw (6.8, 3.3) rectangle (8.4, 2.6);
	\node[align=center] at (7.6,3.0) {\scriptsize \storm~\cite{DBLP:conf/cav/DehnertJK017}};
	
	\draw (6.8, 2.5) rectangle (8.4, 1.8);
	\node[align=center] at (7.6,2.3) {\scriptsize \storm};
	\node[align=center] at (7.6,2.1) {\scriptsize (python)};
	
		\draw (6.8, 0.2) rectangle (8.4, 0.8);
	\node[align=center] at (7.6,0.5) {\scriptsize \tool{z3}~\cite{dMB08}};
		\draw (6.8, 0.9) rectangle (8.4, 1.5);
	\node[align=center] at (7.6,1.2) {\scriptsize \tool{SMT-RAT}~\cite{DBLP:conf/sat/CorziliusKJSA15}};
	
	\draw (6.8, 2.15) -- (6.2,2.15);
	\draw (6.8, 3.6) -- (6.2,3.6);
	\draw (6.8, 0.5) -- (6.2,0.5);
	\draw (6.8, 1.2) -- (6.2,1.2);
	\draw (6.8, 2.95) -- (6.2,2.95);
	
\end{tikzpicture}
}
\caption{High-level architecture of \prophesy and its backends}
\label{fig:highlevelprophesy}
\end{figure}

\prophesy can be divided in three parts:
\begin{enumerate}[i)]
\item First and foremost, it presents a library consisting of:
\begin{inparaenum}[a)]
\item data structures for parameter spaces and instantiations, solution functions, specifications, etc., built around the \python bindings of the library  \carl\footnote{\url{https://moves-rwth.github.io/pycarl/}} (featuring computations with polynomials and rational functions),
\item  algorithms such as guided sampling, various candidate region generation procedures, decomposition of regions, etc., methods that require tight integration with the model are realised via the python bindings of \storm\footnote{\url{https://moves-rwth.github.io/stormpy/}},
\item abstract interfaces to backend tools, in particular probabilistic model checkers, and SMT-checkers, together with some concrete adapters for the different solvers, see Figure~\vref{fig:highlevelprophesy}.\end{inparaenum}
\item An extensive command-line interface which provides simple access to the different core functionalities of the library,  ranging from sampling to full parameter synthesis. 
\item A prototypical web-service running on top of the library, which allows users to interact with the parameter synthesis via a web-interface. 
\end{enumerate}

\prophesy is constructed in a modular fashion: besides the python bindings for \carl, all non-standard packages and tools (in particular model checkers and SMT solvers) are optional. 
Naturally, the full power of \prophesy can only be used if these packages are available.
Besides the methods presented in this paper, \prophesy contains two further mature parameter synthesis methods:
\begin{inparaenum}[i)]
\item particle-swarm optimisation inspired by~\cite{chen2013model}, and
\item convex optimisation from~\cite{DBLP:conf/atva/CubuktepeJJKT18}.
\end{inparaenum}

The information in the remainder details the implementation and the possibilities provided by \prophesy. 
The section contains some notions from probabilistic model checking~\cite{BK08,katoen2016probabilistic,DBLP:reference/mc/BaierAFK18}. We refrain from providing detailed descriptions of these notions, as it would go beyond the scope of this paper.

\subsection{Model construction and preprocessing \scriptsize\emph{(Realised in \storm)}}

The model checker \storm supports the creation of pMCs and pMDPs from both PRISM-language model descriptions~\cite{KNP11} and JANI-specifications~\cite{DBLP:conf/tacas/BuddeDHHJT17}. 
The latter can be used as intermediate format to support, e.g., digital-clock PTAs with parameters written in \modest~\cite{DBLP:journals/fmsd/HahnHHK13}, or to support expected time properties of generalised stochastic Petri nets~\cite{DBLP:journals/sigmetrics/MarsanBCDF98} with parametric rates and/or weights.
Parametric models can be built using the matrix-based, explicit representation, as well as the symbolic, decision diagram (dd)-based engine built on top of \sylvan~\cite{DBLP:journals/sttt/DijkP17}. 
Both engines support the computation of qualitative properties, an essential preprocessing step, and bisimulation minimisation on parametric models, as described in~\cite{param_sttt}.
We advocate the use of the \storm-\python API adapter: 
Its interactive nature avoids the repetition of expensive steps. In particular, it allows for the incremental usage of parameter lifting and sampling.

The support for rational functions is realised via the library \carl\footnote{\url{https://github.com/moves-rwth/carl-storm}}.
The rational function is stored as a tuple consisting of multivariate polynomials. 
These polynomials are by default stored in a partially factorised fashion, cf.~\cite{jansen-et-al-qest-2014}. 
Each factor (a polynomial) is stored as an ordered sparse sum of terms, each term consists of the coefficient and a sparse representation of variables with their non-zero exponents.
For manipulating the (rational) coefficients, we exploit \gmp\footnote{\url{https://gmplib.org/}} or  \cln\footnote{\url{https://www.ginac.de/CLN/}}. The former is thread-safe, while the latter performs slightly better with single-thread usage. 
Computation of GCDs in multivariate polynomials is done either via \ginac~\cite{DBLP:journals/jsc/BauerFK02} or \cocoa~\cite{CoCoALib}.

\subsection{Solution function computation \scriptsize\emph{(Realised in \storm)}}
The computation of solution functions for pMCs as discussed in Section~\ref{sec:solution_fct} is implemented for a variety of specifications:
\begin{itemize}
  \item reachability and reach-avoid probabilities,
  \item expected rewards, including expected time of continuous-time Markov chains,
  \item step-bounded reachability probabilities, and
  \item long-run average probabilities and rewards.	
\end{itemize}

The computation is realised either via state elimination, or via Gaussian elimination. An implementation of set-based transition elimination is available for symbolic representations of the pMC.

\subsubsection{State elimination}
\label{sec:impl:stateelim}

As the standard sparse matrix representation used by \storm is not suitable for fast removal and insertion of entries, a flexible sparse matrix with faster delete and insert operations is used. 

The order in which states are eliminated has a severe impact on the performance~\cite{DJJ+15}. 
 \storm supports a variety of static (pre-computed) and dynamic orderings for the elimination: 
 \begin{itemize}
    \item several static orders (forward (reversed), backward (reversed)) based on the order of state-generation by the model construction algorithms. This latter order is typically  determined by a depth-first search through the high-level model description\footnote{this order is destroyed during the computation of a bisimulation quotient},
    \item orders based on the topology of the pMC, e.g., based on the decomposition in strongly connected components,
    \item orders (Regex) which take into account the in-degree (the number of incoming transitions at a state), inspired by \cite{DBLP:conf/wia/Sakarovitch05,DBLP:journals/fuin/Han13},
    \item orders (SPen, DPen) which take into account the complexity of the rational function corresponding to the transition probability. The complexity is defined by the degree and number of terms of the occurring polynomials.
 \end{itemize}
The orders are computed as penalties for states, and the order prefers states with a low penalty.
For dynamic orderings (Regex, DPen), the penalties are recomputed as the in-degree of states and complexity of transition probabilities change during state elimination.

\subsubsection{Gaussian elimination}
\storm supports Eigen~\cite{eigenweb} as a linear equation system solver over the field of rational functions.
 It uses the ``supernodal'' (supernodes) LU factorisation. The matrix is permuted by the column approximate minimum degree permutation (COLAMD) algorithm to reorder the matrix.
 One advantage is that this solver is based on sparse model-checking algorithm for parameter-free models. The solver therefore, in addition to the properties supported by state elimination,  supports the construction in~\cite{DBLP:conf/tacas/BaierKKM14} for conditional probabilities \emph{and} rewards.

\subsubsection{Set-based transition elimination}
This elimination method is targeted for symbolic representations of the Markov chain.
Set-based transition elimination is implemented via matrix-matrix multiplications. 
In every multiplication, a copy of the dd-representation of a matrix over variables $(\vec{s},\vec{t})$ is made. 
The copy uses renamed  dd-variables $(\vec{t}, \vec{t'})$.
Then, a multiplication of the original matrix with the copy can be done on the dd level  yielding a matrix $(\vec{s}, \vec{t'})$. 
Renaming $\vec{t'}$ to $\vec{t}$ yields a matrix on the original dd-variables.

\subsection{Parameter lifting \scriptsize\emph{(Realised in \storm)}}
For parameter lifting (Section~\ref{sec:approx_mc} and~\ref{sec:nondet}), the major effort beyond calling standard model-checking procedures is the construction of the substituted (lifted) model.
As parameter lifting for different regions does not change the topology of the lifted model, it is beneficial to create a template of the lifted model once, and to substitute the values according to the region at hand. 
The substitution operation can be sped up by exploiting the following observation:
Typically, transition probability functions coincide for many transitions. 
Thus, we evaluate each occurring function once and substitute the outcome directly at all occurrences.
Moreover, for a growing number of regions to be checked, any one-time preprocessing of the lifted model eventually pays off.   In particular, we apply minimisation techniques before construction of the lifted model. We use both bisimulation minimisation as well as state elimination of parameter-free transitions.
These minimisations drastically reduce the run-time of checking a single region.
We use numerical methods first: for regions that we want to classify as accepting (or rejecting) we resort to the analysis of MDPs using policy iteration with rational numbers. 
For that, we initialise the policy iteration with a guess based on the earlier numerical results.

\subsection{SMT-based region verification \scriptsize\emph{(Realised in \prophesy)}}
This complete region checking procedure is realised by constructing SMT queries, as elaborated in Section~\ref{sec:exact_mc}.
When invoking the SMT solver, we use some features of the SMT-lib standard \cite{BarFT-SMTLIB}. 
First of all, when checking several regions, we use backtrack-points to only partly reset the solver:
More precisely, the problem description is given by a conjunction of subformulae, where the conjunction is represented by a stack. We first push the constraints for the problem to the stack, save a backtrack point, and then store the region. 
Once we have checked a particular region, we backtrack to the \emph{backtrack point}, that is, we remove the constraints for the particular region from the problem description. 
This way, we reuse simplifications and data structures the solver constructed for the problem description covering the model (and not the region).
To support both verifying the property and its negation, the problem description is slightly extended. 
We add two Boolean variables (\textsl{accepting} and \textsl{rejecting}). 
The following gives an example of the encoding together with checking whether a region $R_1$ is accepting, and a region $R_2$ is rejecting, using the notation of Section~\ref{sec:exact_mc}.
\begin{align*}
	& x = f_{\pdtmc,\varphi} 
	 \land \big( \textsl{accepting} \implies x \geq \lambda  \big)
	\land \big(  \textsl{rejecting} \implies x < \lambda \big) \\
	 & (\textsf{push}) \\
	  & \textsl{ accepting} \land \Upphi(R_1) \\
	  & (\textsf{pop}) \qquad 
	  (\textsf{push}) \\
	  & \textsl{ rejecting} \land \Upphi(R_2)
\end{align*}

\subsection{Sampling \scriptsize\emph{(Realised in \prophesy)}}
We accelerate the selection of regions by getting a rough picture through sampling, as discussed in Section~\ref{sec:psp}.
 We support two engines for computing the samples: Either via model checking, or by instantiating the solution function. Sampling on the solution function should always be done exactly, as the evaluation  of the typically highly-nonlinear solution functions is (again typically) numerically unstable.
In each iteration, based on the current set of samples, a new set of sampling candidates is computed.
The choice of the new samples can be modified in several ways. 
The standard used here is via linear interpolation between accepting and rejecting samples.

\subsection{Partitioning \scriptsize\emph{(Realised in \prophesy)}}
For the construction of region candidates, we split the initial regions according to our heuristic (quads or growing rectangles, cf.\ Section~\ref{sec:regionpartitioning:findingcandidates}) until none of the regions is inconsistent. 
We sort the candidate regions based on their size in descending order.
Furthermore, we prefer regions where we deem verification to be less costly: Candidate regions that are supposed to be accepting and are further away from samples or regions that are rejecting are preferred over those regions which have rejecting samples or regions in their neighbourhood.

\section{Experimental Evaluation}\label{sec:eval}
In this section, we review the scalability of the presented approaches based on a selection of benchmarks.

\subsection{Set-up}
\subsubsection{Benchmarks}
We consider five case studies from the literature. 
The selection represents various application domains. 

\paragraph{NAND multiplexing.}
With integrated circuits being built at ever smaller scale, they are more prone to defects and/or to exhibit transient failures~\cite{DBLP:journals/pieee/HaselmanH10}.
One way to overcome these deficiencies is the implementation of redundancy at gate-level. In particular, one aims to construct reliable devices from unreliable components.
\model{NAND} multiplexing is such a technique, originally due to von~Neumann~\cite{vN56}.
Automated analysis of NAND multiplexing via Markov chain model checking was considered first in \cite{NPKS05}.
They also studied the influence of gate failures in either of the stages of the multiplexing by sampling various values.
We use the pMC from~\cite{DJJ+15}, that replaced fixed probabilities in the original formulation with parameters.
We analyse the effect of changing failure probabilities of the gates on the reliability of the multiplexed NAND.

\paragraph{Herman's self-stabilising protocol.}
In distributed systems, tokens are used to grant privileges (e.g., access to shared memory) to processes.
Randomisation is an essential technique to break the symmetry among several processes~\cite{DBLP:conf/stoc/Angluin80}.
\model{Herman's} probabilistic algorithm~\cite{DBLP:journals/ipl/Herman90} is a token circulation algorithm for ring structures. 
In each step, every process possessing a token passes the token along with probability $p$ and keeps the token with probability $1{-}p$.
The algorithm is self-stabilising, i.e., started from any illegal configuration with more than one token the algorithm recovers to a legal configuration with a unique token.
The recovery time crucially depends on the probability of passing the token, and an optimal value for $p$ depends on the size of the system~\cite{DBLP:journals/fac/KwiatkowskaNP12}.
We investigate the expected recovery time by parameter synthesis, inspired by~\cite{DBLP:conf/srds/AflakiVBKS17}.

\paragraph{Mean-time-to-failure of a computer system.}
In reliability engineering, fault trees are a prominent model to describe how a system may fail based on faults of its various components~\cite{TriBob2017,RS15}.
Dynamic fault trees (DFTs,~\cite{DBB92}) extend these fault trees with a notion of a state, and allow to model spare management and temporal dependencies in the failure behaviour. 
State-of-the-art approaches for dynamic fault trees translate such fault trees into Markov chains~\cite{CSD00,DBLP:journals/tdsc/BoudaliCS10,DBLP:journals/tii/VolkJK18}; evaluation of the mean-time-to-failure boils down to the analysis of the underlying Markov chain.
Probabilities and rewards originate from the failure rate of the components in the described system. 
Such failure rates are often not known (precisely), especially during design time. Therefore, they may be represented by parameters. 
We take the \model{HECS} DFT~\cite{VS02} benchmark describing the failure of a computer system with an unknown failure rate for the software interface and the spare processor, as first described in \cite{DBLP:conf/safecomp/0001JK16}.
We analyse how this failure rate affects the expected time until the failure (mean-time-to-failure) of the complete computer system.

\paragraph{Network scheduling.}
This benchmark~\cite{DBLP:conf/globecom/YangMZ11} concerns the wireless downlink scheduling of traffic to different users, with hard deadlines and prioritised packets. 
The system is time-slotted: time is divided into periods and each period is divided into an equal number of slots. 
At the start of each time period, a new packet is generated for each user with a randomly assigned priority. 
The goal of scheduling is to, in each period, deliver the packets to each user before the period ends. 
Packets not delivered by the end of a period are dropped.
Scheduling is non-trivial, as successful transmissions are not stochastically independent, i.e., channels have a (hidden) internal state.
The system is described as a partially observable Markov decision process~\cite{DBLP:books/daglib/0023820}, a prominent formalism in the AI community. 
We take the \model{Network} model from~\cite{DBLP:journals/rts/Norman0Z17}, and consider the pMC that describes randomised finite memory controllers that solve this scheduling problem,
based on a translation from~\cite{DBLP:conf/uai/Junges0WQWK018}.
Concretely, the parameters represent how the finite memory controller randomises. 
We evaluate the effect of the randomisation in the scheduling on the expected packet loss.

\paragraph{Bounded retransmission protocol.}
The bounded retransmission protocol (\model{BRP},~\cite{HSV94,DBLP:conf/tacas/DArgenioKRT97}) is a variant of the alternating bit protocol. 
It can be used as part of an OSI data link layer, to implement  retransmitting corrupted file chunks between a sender and a receiver.
The system contains two channels; from sender to receiver and vice versa.
BRP is a famous benchmark in (non-parametric) probabilistic model checking,
based on a model in~\cite{DBLP:conf/papm/DArgenioJJL01}.
We consider the parametric version from~\cite{param_sttt}.
The parameters naturally reflect the channel qualities. The model contains non-determinism as the arrival of files on the link layer cannot be influenced.
This non-determinism hampers a manual analysis.
The combination of parametric probabilities and non-determinism naturally yields a pMDP.
We analyse the maximum probability that a sender eventually does not report a successful transmission.

\begin{remark}
Other benchmarks and a thorough performance evaluation have been presented before in \cite{DJJ+15} (for state elimination and parameter space partitioning) and \cite{QDJJK16} (for parameter lifting).
\end{remark}

\subsubsection{Benchmark statistics}
Table~\vref{tab:benchmarkinfo} summarises relevant information about the concrete instances that we took from the benchmarks.
\begin{table}[]
\centering
\caption{Detailed information for models in the benchmark set}
\label{tab:benchmarkinfo}
{\scriptsize\begin{tabular}{c|cc|c|rr|r}
\hline
\hcentl{id}         & \hcent{benchmark}              & \hcentl{instance}                         & \hcentl{$|V|$} & \hcent{states} & \hcentl{transitions} & \hcent{time} \\
\hline
\hline

\multirow{2}{*}{ 1} & \multirow{12}{*}{\model{BRP}}  & \multirow{2}{*}{MAX=2,N=16}               & \multirow{12}{*}{2}
                                                                                                             &    1439 &        1908 &   0.06\\
                    &                                &                                           &           &     664 &         928 &   0.22\\ \cline{5-7}
\multirow{2}{*}{ 2} &                                & \multirow{2}{*}{MAX=2,N=256}              &           &   20639 &       27348 &   0.57\\
                    &                                &                                           &           &   10264 &       14368 & 370.83\\ \cline{5-7}
\multirow{2}{*}{ 3} &                                & \multirow{2}{*}{MAX=2,N=512}              &           &   41119 &       54484 &   1.11\\
                    &                                &                                           &           &   20504 &       28704 & 197.69\\ \cline{5-7}
\multirow{2}{*}{ 4} &                                & \multirow{2}{*}{MAX=5,N=16}               &           &    2801 &        3783 &   0.10\\
                    &                                &                                           &           &    1354 &        1912 &   1.23\\ \cline{5-7}
\multirow{2}{*}{ 5} &                                & \multirow{2}{*}{MAX=5,N=256}              &           &   40721 &       55143 &   1.15\\
                    &                                &                                           &           &   21034 &       29752 &3305.07\\ \cline{5-7}
\multirow{2}{*}{ 6} &                                & \multirow{2}{*}{MAX=5,N=512}              &           &   81169 &      109927 &   2.25\\
                    &                                &                                           &           &   42026 &       59448 & 345.21\\
\hline

\multirow{2}{*}{ 7} & \multirow{4}{*}{\model{HECS}}  & \multirow{2}{*}{m=1,k=1,i=1}        & \multirow{4}{*}{2}
                                                                                                             &    129 &         489 &   0.02\\
                    &                                &                                           &           &     25 &          71 &   0.00\\ \cline{5-7}
\multirow{2}{*}{ 8} &                                & \multirow{2}{*}{m=1,k=1,i=2}        &           &    145 &         589 &   0.02\\
                    &                                &                                           &           &     49 &         173 &   0.00\\
\hline

\multirow{2}{*}{ 9} & \multirow{8}{*}{\model{Herman}}& \multirow{2}{*}{N=3}                      & \multirow{8}{*}{1}
                                                                                                             &      9 &          36 &   0.02\\
                    &                                &                                           &           &      3 &           5 &   0.00\\ \cline{5-7}
\multirow{2}{*}{10} &                                & \multirow{2}{*}{N=5}                      &           &     33 &         276 &   0.03\\
                    &                                &                                           &           &      5 &          15 &   0.00\\ \cline{5-7}
\multirow{2}{*}{11} &                                & \multirow{2}{*}{N=7}                      &           &    129 &        2316 &   0.11\\
                    &                                &                                           &           &     16 &         137 &   0.02\\ \cline{5-7}
\multirow{2}{*}{12} &                                & \multirow{2}{*}{N=9}                      &           &    513 &       20196 &   0.92\\
                    &                                &                                           &           &    347 &       15009 &   0.12\\
\hline

\multirow{2}{*}{13} & \multirow{12}{*}{\model{NAND}} & \multirow{2}{*}{K=2,N=2}                  & \multirow{12}{*}{2}
                                                                                                             &    178 &         243 &   0.03\\
                    &                                &                                           &           &    125 &         167 &   0.00\\ \cline{5-7}
\multirow{2}{*}{14} &                                & \multirow{2}{*}{K=2,N=20}                 &           & 154942 &      239832 &   2.81\\
                    &                                &                                           &           & 102012 &      154722 &   0.91\\ \cline{5-7}
\multirow{2}{*}{15} &                                & \multirow{2}{*}{K=2,N=30}                 &           & 681362 &     1065797 &  12.56\\
                    &                                &                                           &           & 474847 &      732768 &   4.65\\ \cline{5-7}
\multirow{2}{*}{16} &                                & \multirow{2}{*}{K=5,N=10}                 &           &  35112 &       52647 &   0.63\\
                    &                                &                                           &           &  23603 &       34093 &   0.21\\ \cline{5-7}
\multirow{2}{*}{17} &                                & \multirow{2}{*}{K=5,N=20}                 &           & 384772 &      594792 &   7.04\\
                    &                                &                                           &           & 288102 &      436332 &   3.17\\ \cline{5-7}
\multirow{2}{*}{18} &                                & \multirow{2}{*}{K=5,N=30}                 &           &1697732 &     2653937 &  31.45\\
                    &                                &                                           &           &1345507 &     2074758 &  18.49\\
\hline

\multirow{2}{*}{19} & \multirow{12}{*}{\model{Network}}& \multirow{2}{*}{c=2,K=2,T=2}              & \multirow{2}{*}{8}
                                                                                                             &     52 &         133 &   0.00\\
                    &                                &                                           &           &     52 &         133 &   0.00\\ \cline{5-7}
\multirow{2}{*}{20} &                                & \multirow{2}{*}{c=2,K=2,T=3}              & \multirow{2}{*}{16}
                                                                                                             &    106 &         269 &   0.01\\
                    &                                &                                           &           &    106 &         269 &   0.00\\ \cline{5-7}
\multirow{2}{*}{21} &                                & \multirow{2}{*}{c=2,K=2,T=4}              & \multirow{2}{*}{24}
                                                                                                             &    164 &         411 &   0.01\\
                    &                                &                                           &           &    164 &         411 &   0.00\\ \cline{5-7}
\multirow{2}{*}{22} &                                & \multirow{2}{*}{c=2,K=4,T=2}              & \multirow{2}{*}{20}
                                                                                                             &    136 &         365 &   0.01\\
                    &                                &                                           &           &    136 &         365 &   0.00\\ \cline{5-7}
\multirow{2}{*}{23} &                                & \multirow{2}{*}{c=2,K=4,T=3}              & \multirow{2}{*}{36}
                                                                                                             &    262 &         691 &   0.01\\
                    &                                &                                           &           &    262 &         691 &   0.00\\ \cline{5-7}
\multirow{2}{*}{24} &                                & \multirow{2}{*}{c=2,K=4,T=4}              & \multirow{2}{*}{52}
                                                                                                             &    392 &        1023 &   0.01\\
                    &                                &                                           &           &    392 &        1023 &   0.00\\
\hline
\end{tabular}
}
\end{table}
The \emph{id} is used for reference.
The \emph{benchmark} refers to the name of the benchmark-set, while the \emph{instance} describes the particular instance from this benchmark set.
We give the total number of \emph{parameters $|V|$} both in the transition matrix as well as in the reward structure whenever applicable.
For the remainder of the columns, we give two numbers per benchmark instance:
The upper row describes the original model, the latter describes the (strong) bisimulation quotient.
The columns give the number of \emph{states} and \emph{transitions}.
The last row gives the \emph{time} (in seconds) required for constructing the model (top) and constructing the bisimulation quotient (bottom). We remark that all benchmarks have a limited number of parameters: Systems with many parameters are beyond the reach of the methods discussed here, but can be analysed with respect to simpler synthesis questions (such as finding one suitable instantiation). We refer to the related work for a discussion of such methods.

\subsubsection{Evaluation}

We conducted the empirical evaluation on an HP BL685C G7 with Debian 9.6.
Each evaluation run could use 8 cores with 2.1GHz each.
However, unless specified otherwise, algorithms use a single core.
We set the timeout to 1 hour and the memory limit to 16GB.
We used \prophesy version 2.0, together with the \storm-python bindings version 1.3.1, z3 version 4.8.4.
All benchmark files are made available via \prophesy\footnote{Benchmarks are in the subfolder \texttt{benchmark\_files}}.

\subsection{Exact synthesis via the solution function}
To evaluate the exact synthesis approach, we use 
state elimination with 7 different heuristics, 
set-based transition elimination, 
and Gaussian elimination.
All configurations are evaluated with and without strong bisimulation.

First, we show the sizes of the solution function:
The results are summarised in Table~\vref{tab:solution_function}.
\begin{table}[]
\centering
\caption{Empirical performance of computing the solution function}
\label{tab:solution_function}
{\scriptsize\begin{tabular}{c|rrrr|r|rr}
\hline
\hcentl{id}         & \hcent{degree}      & \hcent{degree}        & \hcent{\# terms}      & \hcentl{\# terms}   & \hcentl{success}    & \hcent{time} & \hcent{time}  \\
                    & \hcent{num}         & \hcent{denom}         & \hcent{num}           & \hcentl{denom}      &                     & \hcent{mc}   & \hcent{total} \\
\hline
\hline

\multirow{2}{*}{ 7} &\multirow{2}{*}{23}  & \multirow{2}{*}{24}   & \multirow{2}{*}{234}  & \multirow{2}{*}{247}& \multirow{2}{*}{16} &    2.00 &    2.09 \\
                    &                     &                       &                       &                     &                     &    0.64 &    0.72 \\ \cline{2-8}
\multirow{2}{*}{ 8} &\multirow{2}{*}{31}  & \multirow{2}{*}{32}   & \multirow{2}{*}{408}  & \multirow{2}{*}{425}& \multirow{2}{*}{16} &    9.12 &    9.21 \\
                    &                     &                       &                       &                     &                     &    3.00 &    3.08 \\ \cline{2-8}
\hline

\multirow{2}{*}{ 9} &\multirow{2}{*}{0}   & \multirow{2}{*}{2}    & \multirow{2}{*}{1}    & \multirow{2}{*}{2}  & \multirow{2}{*}{18} &    0.00 &    0.09 \\
                    &                     &                       &                       &                     &                     &    0.00 &    0.08 \\ \cline{2-8}
\multirow{2}{*}{10} &\multirow{2}{*}{4}   & \multirow{2}{*}{6}    & \multirow{2}{*}{5}    & \multirow{2}{*}{6}  & \multirow{2}{*}{17} &    0.04 &    1.55 \\
                    &                     &                       &                       &                     &                     &    0.00 &    0.10 \\ \cline{2-8}
\multirow{2}{*}{11} &\multirow{2}{*}{28}  & \multirow{2}{*}{30}   & \multirow{2}{*}{29}   & \multirow{2}{*}{30} & \multirow{2}{*}{11} &    0.62 &    0.82 \\
                    &                     &                       &                       &                     &                     &    0.37 &    0.56 \\ \cline{2-8}
\multirow{2}{*}{12} &\multirow{2}{*}{150} & \multirow{2}{*}{152}  & \multirow{2}{*}{151}  & \multirow{2}{*}{152}& \multirow{2}{*}{ 8} &  247.00 &  248.14 \\
                    &                     &                       &                       &                     &                     &  114.49 &  115.64 \\ \cline{2-8}
\hline

\multirow{2}{*}{13} &\multirow{2}{*}{10}  & \multirow{2}{*}{0}    & \multirow{2}{*}{32}   & \multirow{2}{*}{1}  & \multirow{2}{*}{18} &    0.00 &    0.11 \\
                    &                     &                       &                       &                     &                     &    0.00 &    0.09 \\ \cline{2-8}
\multirow{2}{*}{14} &\multirow{2}{*}{100} & \multirow{2}{*}{0}    & \multirow{2}{*}{2106} & \multirow{2}{*}{1}  & \multirow{2}{*}{15} &   43.05 &   46.88 \\
                    &                     &                       &                       &                     &                     &   15.46 &   19.35 \\ \cline{2-8}
\multirow{2}{*}{15} &\multirow{2}{*}{150} & \multirow{2}{*}{0}    & \multirow{2}{*}{4653} & \multirow{2}{*}{1}  & \multirow{2}{*}{13} &  469.29 &  486.74 \\
                    &                     &                       &                       &                     &                     &  110.54 &  128.48 \\ \cline{2-8}
\multirow{2}{*}{16} &\multirow{2}{*}{110} & \multirow{2}{*}{0}    & \multirow{2}{*}{1220} & \multirow{2}{*}{1}  & \multirow{2}{*}{15} &    6.30 &    7.24 \\
                    &                     &                       &                       &                     &                     &    3.30 &    4.25 \\ \cline{2-8}
\multirow{2}{*}{17} &\multirow{2}{*}{200} & \multirow{2}{*}{0}    & \multirow{2}{*}{4640} & \multirow{2}{*}{1}  & \multirow{2}{*}{13} &  245.47 &  256.05 \\
                    &                     &                       &                       &                     &                     &   88.18 &   98.71 \\ \cline{2-8}
\multirow{2}{*}{18} &\multirow{2}{*}{330} & \multirow{2}{*}{0}    & \multirow{2}{*}{10260}& \multirow{2}{*}{1}  & \multirow{2}{*}{ 1} & 3031.34 & 3083.88 \\
                    &                     &                       &                       &                     &                     & 3031.34 & 3083.88 \\ \cline{2-8}
\hline

\multirow{2}{*}{19} &\multirow{2}{*}{1}   & \multirow{2}{*}{0}    & \multirow{2}{*}{23}   & \multirow{2}{*}{1}  & \multirow{2}{*}{16} &    0.00 &    0.07 \\
                    &                     &                       &                       &                     &                     &    0.00 &    0.06 \\ \cline{2-8}
\multirow{2}{*}{20} &\multirow{2}{*}{1}   & \multirow{2}{*}{0}    & \multirow{2}{*}{111}  & \multirow{2}{*}{1}  & \multirow{2}{*}{16} &    0.01 &    0.08 \\
                    &                     &                       &                       &                     &                     &    0.01 &    0.07 \\ \cline{2-8}
\multirow{2}{*}{21} &\multirow{2}{*}{1}   & \multirow{2}{*}{0}    & \multirow{2}{*}{519}  & \multirow{2}{*}{1}  & \multirow{2}{*}{16} &    0.04 &    0.11 \\
                    &                     &                       &                       &                     &                     &    0.03 &    0.09 \\ \cline{2-8}
\multirow{2}{*}{22} &\multirow{2}{*}{1}   & \multirow{2}{*}{0}    & \multirow{2}{*}{65}   & \multirow{2}{*}{1}  & \multirow{2}{*}{16} &    0.01 &    0.08 \\
                    &                     &                       &                       &                     &                     &    0.01 &    0.08 \\ \cline{2-8}
\multirow{2}{*}{23} &\multirow{2}{*}{1}   & \multirow{2}{*}{0}    & \multirow{2}{*}{289}  & \multirow{2}{*}{1}  & \multirow{2}{*}{16} &    0.07 &    0.15 \\
                    &                     &                       &                       &                     &                     &    0.03 &    0.10 \\ \cline{2-8}
\multirow{2}{*}{24} &\multirow{2}{*}{1}   & \multirow{2}{*}{0}    & \multirow{2}{*}{1377} & \multirow{2}{*}{1}  & \multirow{2}{*}{16} &    0.40 &    0.48 \\
                    &                     &                       &                       &                     &                     &    0.12 &    0.20 \\ \cline{2-8}
\hline

\end{tabular}
}
\end{table}
The \emph{id} references the corresponding benchmark instance in Table~\vref{tab:benchmarkinfo}.
The \model{BRP} pMDP is not included. The set of all strategies prevents the computation of the solution function for all induced pMCs.
The next four columns display properties of the resulting rational function.
We give the degree of both the numerator (\emph{degree num}) and denominator (\emph{degree denom}), as well as the number of terms in both polynomials (\emph{\# terms num}, \emph{\# terms denom}).
The next column gives the number of configurations (out of the 18) which \emph{successfully} finished within the time limit.
The last two columns indicate timings.
We give the times (in seconds) to compute the solution function (\emph{time mc}) and the \emph{total time} including model building, (optional) bisimulation minimisation and computing the solution function.
For these timings we give two numbers per benchmark instance:
The upper row describes the median value over all successful configurations and the lower row describes the best result obtained.
Thus, \emph{while functions often grow prohibitively large, medium-sized functions can still be computed.}
Contrary to model checking for parameter-free models, model building is typically \emph{not} the bottleneck.

\begin{figure}
\centering
\includegraphics{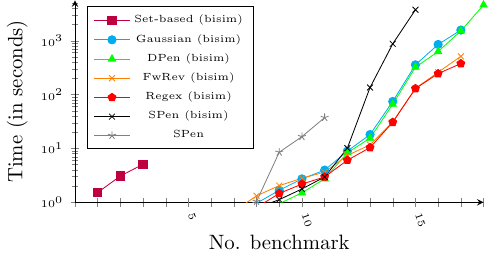}
\caption{Cumulative solving times for solution function computation}
\label{fig:results_sf_comparison}
\end{figure}

Furthermore, we see that the selected heuristic is indeed crucial. Consider instance 11: 11 heuristics successfully compute the solution function (and most of them within a second). However, 7 others yield a timeout.
That leads us to compare some heuristics in Figure~\vref{fig:results_sf_comparison}.
The plot depicts the cumulative solving times for selected configurations over all 18 benchmark instances (excluding \model{BRP}).
Gaussian and set-based refer to these approaches, respectively, all other configurations are variants of state elimination, cf.\ Section~\ref{sec:impl:stateelim}, (bisim) denotes that bisimulation minimisation is used.
The x-axis represents the number of solved instances and the (logarithmic) y-axis represents the time in seconds.
A point $(x,y)$ in the plot represents the $x$ fastest instances which could be solved within a total time of $y$ seconds.
For 15 instances, one of the depicted configurations was the fastest overall. 
\textsl{Regex} based configurations were the fastest eight times, \textsl{DPen} based ones four times and three times configurations based on \textsl{FwRev} were fastest.
From these numbers, we conclude that \emph{the selection of the heuristic is essential, and depending on the model to be analysed.} 
From the graph, we further observe that although using a Gaussian elimination yields good performance, state-elimination approaches can (significantly) outperform the Gaussian elimination on some benchmarks. 
The \textsl{DPen} solves all instances (the only configuration to do so), but \textsl{Regex} is overall (slightly) faster.
The uninformed \textsl{FwRev} with bisimulation works surprisingly well for these benchmarks (but that is mostly coincidence).
The set-based elimination is clearly inferior on the benchmarks considered here, but allows to analyse some models with a very regular structure and a gigantic state space, e.g., a parametric Markov chain for the analysis of the \model{bluetooth} protocol~\cite{DBLP:journals/sttt/DuflotKNP06}.
\subsection{Three types of region verification}
\begin{figure}
\centering
\includegraphics{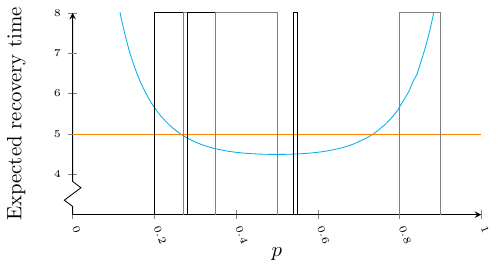}
\caption{Plot for \model{Herman} model with seven processes and parameter $p$ (Benchmark Id: 11)}
\label{fig:herman7}
\end{figure}

We evaluate region verification using two SMT-based approaches (SF: based on first computing the Solution Function, or ETR: encoding the equations into Existential Theory of the Reals), and PLA.
In particular, we present some results for the \model{Herman} benchmark: it features a single parameter, and therefore is well-suited for the illustration of some concepts.
We visualised the results for instance 11 in Figure~\vref{fig:herman7}.
The x-axis represents the probability $p$ and the y-axis the expected recovery time.
We indicate the solution function in blue.
The threshold in the following is set to $\lambda=5$ and indicated by the orange horizontal line. The black columns depict six different regions\footnote{Strictly speaking, regions are given by the intervals for the parameter, we depict the columns for better visibility.} that are evaluated with region checking.
For each region we want to verify whether the expected recovery time is at least $5$. 
The results are summarised in (the upper part of) Table~\vref{tab:regions}.
\begin{table}[]
\centering
\caption{Empirical performance of region verification algorithms.}
\label{tab:regions}
{\scriptsize\setlength{\tabcolsep}{5pt}
\begin{tabular}{c|c|c|c|c|r}
\hcentl{id}          & \hcentl{$\lambda$}    & \hcentl{region}                                     &\hcentl{techn.} & \hcent{result} & \hcent{time} \\
\hline
\hline
\multirow{18}{*}{11} & \multirow{18}{*}{5}   & \multirow{3}{*}{[0.20, 0.27]}                       & ETR & inconsistent  &  12,11  \\
                     &                       &                                                     & PLA & unknown &   0.01  \\
                     &                       &                                                     & SF  & unknown &\timeout \\ \cline{3-6}
                     &                       & \multirow{3}{*}{[0.27, 0.28]}                       & ETR & reject  &  20.68  \\
                     &                       &                                                     & PLA & reject  &   0.01  \\
                     &                       &                                                     & SF  & unknown &\timeout \\ \cline{3-6}
                     &                       & \multirow{3}{*}{[0.28, 0.35]}                       & ETR & reject  &  53.47  \\
                     &                       &                                                     & PLA & unknown &   0.01  \\
                     &                       &                                                     & SF  & unknown &\timeout \\ \cline{3-6}
                     &                       & \multirow{3}{*}{[0.35, 0.50]}                       & ETR & reject  &  23.41  \\
                     &                       &                                                     & PLA & reject  &   0.00  \\
                     &                       &                                                     & SF  & unknown &\timeout \\ \cline{3-6}
                     &                       & \multirow{3}{*}{[0.54, 0.55]}                       & ETR & reject  &  22.35  \\
                     &                       &                                                     & PLA & reject  &   0.01  \\
                     &                       &                                                     & SF  & unknown &\timeout \\ \cline{3-6}
                     &                       & \multirow{3}{*}{[0.80, 0.90]}                       & ETR & unknown &\timeout \\
                     &                       &                                                     & PLA & accept  &   0.01  \\
                     &                       &                                                     & SF  & unknown &\timeout \\
\hline

\multirow{15}{*}{13} & \multirow{15}{*}{0.3} & \multirow{3}{*}{[0.01, 0.99] $\times$ [0.70, 0.90]} & ETR & accept  & 16.20 \\
                     &                       &                                                     & PLA & unknown &  0.01 \\
                     &                       &                                                     & SF  & accept  &  0.16 \\ \cline{3-6}
                     &                       & \multirow{3}{*}{[0.01, 0.99] $\times$ [0.90, 0.99]} & ETR & inconsistent & 19.41 \\
                     &                       &                                                     & PLA & unknown &  0.01 \\
                     &                       &                                                     & SF  & inconsistent &  0.04 \\ \cline{3-6}
                     &                       & \multirow{3}{*}{[0.01, 0.50] $\times$ [0.65, 0.70]} & ETR & accept  & 45.61 \\
                     &                       &                                                     & PLA & unknown &  0.01 \\
                     &                       &                                                     & SF  & accept  &  0.13 \\ \cline{3-6}
                     &                       & \multirow{3}{*}{[0.01, 0.50] $\times$ [0.75, 0.90]} & ETR & accept  &  4.58 \\
                     &                       &                                                     & PLA & accept  &  0.01 \\
                     &                       &                                                     & SF  & accept  &  0.12 \\ \cline{3-6}
                     &                       & \multirow{3}{*}{[0.01, 0.99] $\times$ [0.40, 0.50]} & ETR & reject  & 19.82 \\
                     &                       &                                                     & PLA & reject  &  0.00 \\
                     &                       &                                                     & SF  & reject  &  0.08 \\
\hline
\end{tabular}
}
\end{table}
The first column \emph{id} references the benchmark instance and the second column gives the threshold \emph{$\lambda$}.
The next columns indicate the considered \emph{region} and the \emph{technique}.
The last columns give the \emph{result} of the region verification and the \emph{time} (in seconds) needed for the computation.
The timeout (\timeout) was set to 120 seconds.

For benchmark instance 11,
Parameter lifting (PLA) computes a result within milliseconds and the computation time is independent of the considered region.
The SMT-based techniques take longer and the SF technique in particular does not terminate within two minutes.
However, the ETR technique could yield a result for region $[0.28, 0.35]$ whereas PLA could not give a conclusive answer due to its inherent over-approximation.

\begin{figure}
\centering
\includegraphics{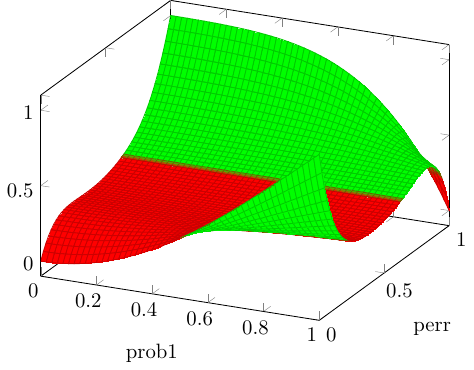}
\caption{Plotting the solution function for \model{NAND} $K=2,N=2$ (Benchmark Id: 13) and parameters prob1 and perr.}
\label{fig:nand}
\end{figure}

We now consider the region verification on the \model{NAND} model with two parameters.
We visualised the solution function for instance 13 in Figure~\vref{fig:nand}.
The considered threshold is $\lambda=0.3$.
Green coloured parts indicate parameter instantiations leading to probabilities above $\lambda$ and red parts lie below $\lambda$.
The results of the verification for different regions are given in (the lower part of) Table~\vref{tab:regions}.
PLA is again the fastest technique, but for larger regions close to the threshold PLA can often not provide a conclusive answer.
Contrary to before, SF is superior to ETR.

The performance of the SMT-based techniques (again) greatly depends on the considered region.
It is only natural that \emph{the size of the region, and the difference to the threshold have a significant influence on the performance of region verification}.
These observations are general and do \emph{hold on all other benchmarks}.
Furthermore, parameter lifting seems broadly applicable, and in the setting evaluated here, clearly faster than SMT-based approaches. 
Parameter lifting over-approximates and therefore might only give a decisive result in a refinement loop such as parameter space partitioning.
The SMT-based approaches are a valuable fallback. When relying on the SMT techniques, it is heavily model-dependent which performs better.
Table~\vref{tab:partitioning} at the end of the next section gives some additional results, indicating the performance of the different verification techniques.

\subsection{Approximative synthesis via parameter space partitioning}
We now evaluate the parameter space partitioning.
We use the implementation in \prophesy with the three verification procedures evaluated above.
Therefore, we focus here on the actual parameter space partitioning.

\begin{figure}
\centering
\includegraphics{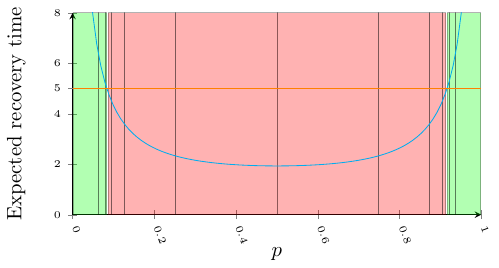}
\caption{Parameter space partitioning for \model{Herman} $N=5$ (Benchmark Id: 10) with parameter $p$}
\label{fig:herman5}
\end{figure}

First, consider again \model{Herman} for illustration purposes.
Region verification is not applicable for instance 10 (with threshold 5), as neither all instantiations accept nor all reject the specification.
Instead, parameter space partitioning delivers which of these instantiations accept, and which reject the specification.
The resulting parameter space partitioning is visualised in Figure~\vref{fig:herman5}.

\begin{figure*}[tbh]
    \centering
    \subfigure[\model{Herman}, $N=5$, with $\lambda=5$]{
        \scalebox{.66}{
            \includegraphics{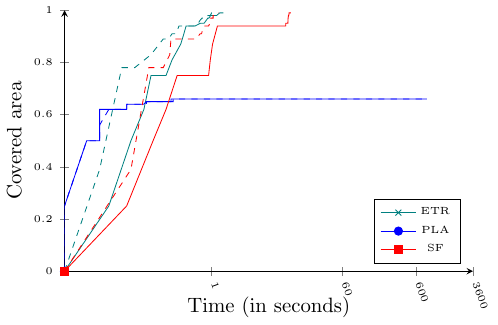}
        }
        \label{fig:results_covered_area_herman}
    }
    \subfigure[\model{NAND} $K=2,N=2$ with  $\lambda=0.1$]{
        \scalebox{.66}{
            \includegraphics{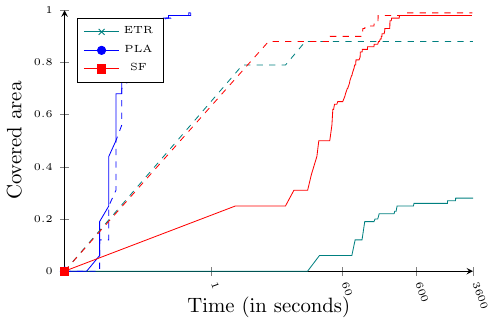}
        }
        \label{fig:results_covered_area_nand_01}
    }
    \subfigure[\model{NAND} $K=2,N=2$ with $\lambda=0.3$]{
        \scalebox{.66}{
            \includegraphics{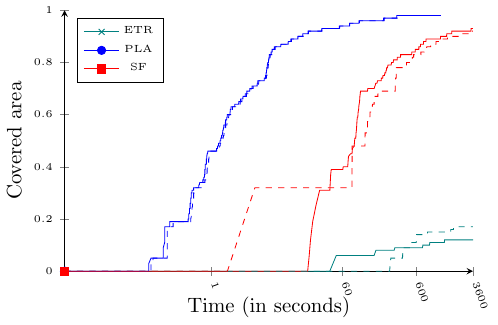}
        }
        \label{fig:results_covered_area_nand_03}
    }
    \caption{Covered areas for parameter space partitioning on different models and thresholds}
    \label{fig:results_covered_areas}
\end{figure*}

Next, we compare the three verification techniques---each with two different methods for selecting candidate regions--in Figure~\vref{fig:results_covered_areas}.
Figure~\vref{fig:results_covered_area_herman} depicts the computation on the \model{Herman} model with 5 processes and threshold $\lambda=5$.
The plot depicts the covered area for all three techniques with both quads (straight lines) and rectangles (dashed lines) as regions.
The x-axis represents the computation time (in seconds) on a logarithmic scale and the y-axis represents the percentage of covered area.
A point $(x,y)$ in the plot represents $y$ percent of the parameter space which could be covered within $x$ seconds.

For \model{Herman}, SMT-based techniques perform better than PLA.
PLA was able to cover 64\% of the parameter space within milliseconds.
However, in the remaining hour only 2\% more space was covered.
The SMT-based techniques were able to cover at least 99\% of the parameter space within 15 seconds.
Moreover, the rectangles cover the parameter space faster than quads.
We also perform the parameter space partitioning on the \model{NAND} model with two different thresholds:
We compare the parameter space partitioning techniques for threshold $\lambda=0.1$ in Figure~\vref{fig:results_covered_area_nand_01}, and for threshold $\lambda=0.3$ in Figure~\vref{fig:results_covered_area_nand_03}.
For \model{NAND}, the PLA technique performs better than the SMT-based techniques.
For threshold $\lambda=0.1$, PLA could cover at least 99\% of the parameter space within 1 second. 
The main reason is that the border is in a corner of the parameter space.
Additionally, the SMT-based techniques with rectangles are significantly faster than the quads for this threshold.
For threshold $\lambda=0.3$, more region verification steps were necessary.
PLA still outperforms ETR and SF.
However, the use of rectangles over quads does not lead to a better performance for this threshold.
At any point in time, there can be very significant differences between the heuristics for candidate generation, especially in settings where single region verification calls become expensive.

Finally, we summarise an overview of the performance in  Table~\vref{tab:partitioning}. For brevity, we pruned some rows, especially if the present approaches already struggle with smaller instances.
\begin{table}[]
\centering
\caption{Empirical performance of parameter space partitioning variations.}
\label{tab:partitioning}
{\scriptsize\setlength{\tabcolsep}{5pt}
\begin{tabular}{c|c|rrr|rr|rr}
\hcentl{id}          &\hcentl{techn.} & \hcent{time} & \hcent{time} & \hcentl{time} & \hcent{area} & \hcentl{area} & \hcent{percent} & \hcent{percent}     \\
                     &                & \hcent{50\%} & \hcent{90\%} & \hcentl{98\%} & \hcent{cov}  & \hcentl{safe} & \hcent{reg gen} & \hcent{analysis} \\
\hline
\hline

\multirow{3}{*}{ 1} & ETR & ---   & ---   & ---    & 0.20 & 0.20 & 0.00 \%  & 99.82 \% \\
                    & PLA & 0.04  & 0.19  & 3.09   & 0.99 & 0.83 & 31.65 \% & 12.36 \% \\
                    & SF  & ---   & ---   & ---    & 0.00 & 0.00 & ---      & ---      \\ \cline{2-9}
\multirow{3}{*}{ 2} & ETR & ---   & ---   & ---    & 0.00 & 0.00 & 0.00 \%  & 89.50 \% \\
                    & PLA & 0.33  & 0.34  & ---    & 0.97 & 0.97 & 0.00 \%  & 81.18 \% \\
                    & SF  & ---   & ---   & ---    & 0.00 & 0.00 & ---      & ---      \\ \cline{2-9}
\multirow{3}{*}{ 4} & ETR & ---   & ---   & ---    & 0.03 & 0.03 & 0.00 \%  & 99.61 \% \\
                    & PLA & 0.06  & 0.30  & 7.86   & 0.99 & 0.63 & 38.45 \% & 10.83 \% \\
                    & SF  & ---   & ---   & ---    & 0.00 & 0.00 & ---      & ---      \\ \cline{2-9}
\multirow{3}{*}{ 5} & ETR & ---   & ---   & ---    & 0.00 & 0.00 & 0.00 \%  & 9.73 \%  \\
                    & PLA & 0.70  & ---   & ---    & 0.87 & 0.87 & 0.00 \%  &7.72 \%   \\
                    & SF  & ---   & ---   & ---    & 0.00 & 0.00 & ---      & ---      \\ \cline{2-9}
\hline
\multirow{3}{*}{ 7} & ETR & ---   & ---   & ---    & 0.47 & 0.00 & 0.00 \%  & 99.64 \% \\
                    & PLA & 0.00  & 0.00  & 0.00   & 1.00 & 0.00 & 0.00 \%  & 0.00 \%  \\
                    & SF  & ---   & ---   & ---    & 0.00 & 0.00 & ---      & ---      \\ \cline{2-9}
\multirow{3}{*}{ 8} & ETR & ---   & ---   & ---    & 0.00 & 0.00 & 0.00 \%  & 99.77 \% \\
                    & PLA & 0.01  & 0.01  & 0.01   & 1.00 & 1.00 & 0.00 \%  & 0.91 \%  \\
                    & SF  & ---   & ---   & ---    & 0.00 & 0.00 & ---      & ---      \\
\hline
\multirow{3}{*}{ 9} & ETR & 0.02  & 30.19 & 70.41  & 0.99 & 0.05 & 0.00 \%  & 98.89 \% \\
                    & PLA & 0.08  & ---   & ---    & 0.55 & 0.06 & 0.15 \%  & 73.89 \% \\
                    & SF  & 0.02  & 0.09  & 0.23   & 0.99 & 0.05 & 0.00 \%  & 18.60 \% \\ \cline{2-9}
\multirow{3}{*}{10} & ETR & 0.12  & 0.45  & 1.29   & 0.99 & 0.16 & 0.00 \%  & 57.09 \% \\
                    & PLA & 0.03  & ---   & ---    & 0.66 & 0.17 & 0.15 \%  & 74.84 \% \\
                    & SF  & 0.24  & 1.20  & 11.30  & 0.99 & 0.16 & 0.00 \%  & 90.63 \% \\ \cline{2-9}
\multirow{3}{*}{12} & ETR & ---   & ---   & ---    & 0.00 & 0.00 & 0.00 \%  & 99.66 \% \\
                    & PLA & 1.75  & ---   & ---    & 0.56 & 0.43 & 0.15 \%  & 75.39 \% \\
                    & SF  & ---   & ---   & ---    & 0.00 & 0.00 & 0.00 \%  & 96.20 \% \\
\hline
\multirow{3}{*}{13} & ETR & ---   & ---   & ---    & 0.28 & 0.28 & 0.00 \%  & 99.80 \% \\
                    & PLA & 0.05  & 0.09  & 0.49   & 0.99 & 0.98 & 8.22 \%  & 15.53 \% \\
                    & SF  & 28.70 & 202.98& 357.90 & 0.98 & 0.98 & 0.00 \%  & 96.26 \% \\ \cline{2-9}
\multirow{3}{*}{14} & ETR & ---   & ---   & ---    & 0.00 & 0.00 & 0.00 \%  & 85.15 \% \\
                    & PLA & 3.08  & 16.08 & 152.36 & 0.99 & 0.15 & 32.01 \% & 47.66 \% \\
                    & SF  & ---   & ---   & ---    & 0.00 & 0.00 & 0.00 \%  & 98.68 \% \\ \cline{2-9}
\multirow{3}{*}{15} & ETR & ---   & ---   & ---    & 0.00 & 0.00 & ---      & ---      \\
                    & PLA & 20.27 & 91.18 & 854.48 & 0.99 & 0.14 & 30.27 \% & 61.95 \% \\
                    & SF  & ---   & ---   & ---    & 0.00 & 0.00 & 0.00 \%  & 92.56 \% \\ \cline{2-9}
\multirow{3}{*}{16} & ETR & ---   & ---   & ---    & 0.00 & 0.00 & 0.00 \%  & 98.95 \% \\
                    & PLA & 0.55  & 4.65  & 55.99  & 0.99 & 0.19 & 33.04 \% & 25.99 \% \\
                    & SF  & ---   & ---   & ---    & 0.00 & 0.00 & 0.00 \%  & 99.42 \% \\ \cline{2-9}
\multirow{3}{*}{17} & ETR & ---   & ---   & ---    & 0.00 & 0.00 & 0.00 \%  & 18.75 \% \\
                    & PLA & 8.79  & 40.99 & 326.12 & 0.99 & 0.16 & 33.23 \% & 54.62 \% \\
                    & SF  & ---   & ---   & ---    & 0.00 & 0.00 & 0.00 \%  & 94.39 \% \\ \cline{2-9}
\multirow{3}{*}{18} & ETR & ---   & ---   & ---    & 0.00 & 0.00 & ---      & ---      \\
                    & PLA & 53.69 &254.13 &1861.31 & 0.99 & 0.16 & 33.21 \% & 60.37 \% \\
                    & SF  & ---   & ---   & ---    & 0.00 & 0.00 & ---	    & ---      \\
\hline
\multirow{3}{*}{19} & ETR & ---   & ---   & ---    & 0.00 & 0.00 & 0.00 \%  & 99.28 \% \\
                    & PLA & ---   & ---   & ---    & 0.12 & 0.12 & 0.00 \%  & 99.54 \% \\
                    & SF  & ---   & ---   & ---    & 0.32 & 0.32 & 0.00 \%  & 98.22 \% \\ \cline{2-9}
\hline
\end{tabular}
}
\end{table}
The \emph{id} is a reference to the benchmark instance.
The \emph{technique} is given in the next column.
In the next three columns we give for each technique the \emph{time} (in seconds) needed to cover at least 50\%, 90\% and 98\% of the complete parameter space.
The next two columns give the complete \emph{covered area}---i.e.\ the sum of the sizes of all accepting or rejecting regions--- when terminating the parameter space partitioning after 1h, together with the \emph{safe area}, i.e.\ the sum of the sizes of all accepting regions.
The last two columns indicate the percentage of the total time spent in generating the regions (\emph{time reg gen}) and verifying the regions (\emph{time analysis}).
\emph{PLA is almost always superior, but not on all benchmarks (and not on all (sub)regions.}
Depending on the model, SF or ETR are the best SMT-based technique.
There might be room for improvement by portfolios and machine-learned algorithm selection schemes.

\section{Related Work and discussion}
\label{sec:discussion}
\noindent We discuss related work with respect to various relevant topics.

\paragraph{Complexity.}
For graph-preserving pMCs, many complexity results are collected in~\cite{DBLP:journals/jcss/JungesK0W21}, including results from~\cite{DBLP:journals/corr/Chonev17}. 
In particular, the complement of the verification problem, i.e., the question whether there exists an instantiation in a region that satisfies a reachability property, is ETR-complete for both pMDPs and pMCs\footnote{It holds that P $\subset$ ETR $\subseteq$ PSPACE. A prominent ETR-complete problem is whether a multivariate polynomial has a real-valued root.}. For any fixed number of parameters, the problem can be solved in polynomial time~\cite{gandalf-journal}. This paper also considers a richer fragment of the logic PCTL.

\paragraph{Computing a solution function.}
This approach was pioneered by \cite{Daws04} and significantly improved by~\cite{param_sttt}.
Both \prism~\cite{KNP11} and \param~\cite{PARAM10} support the computation of a solution function based on the latter method. 
It has been adapted in \cite{jansen-et-al-qest-2014} to an elimination of SCCs and a more clever representation of rational functions. This representation  has been adapted by \storm~\cite{DBLP:conf/cav/DehnertJK017}.
In~\cite{DBLP:journals/tse/FilieriTG16}, computing a solution function via a computer algebra system was considered. That method targets small, randomly generated pMCs with many parameters.
Recently,~\cite{gandalf-journal} explored the use of one-step fraction-free Gaussian elimination to reduce the number of GCD computations.
For pMDPs,~\cite{DBLP:conf/nfm/HahnHZ11} experimented with the introduction of discrete parameters to reflect strategy choices---this method, however, scales poorly.
In~\cite{DBLP:journals/ai/DelgadoSB11} and~\cite{DBLP:journals/ai/DelgadoBDS16}, variants of value iteration with a dd-based representation of the solution function are presented. Fast sampling on (concise representations of) the solution function is considered in~\cite{DBLP:conf/atva/GainerHS18,DBLP:conf/cav/HoltzenJVMSB20}.

\paragraph{Equation system formulation.}
Regarding pMDPs, instead of introducing a Boolean structure, one can lift the linear program formulation for MDPs to a nonlinear program (NLP). 
This lifting has been explored in \cite{bartocci2011model}, and shown to be not feasible in general.
\nils{A string of results rely on convex programming approaches.
For instance,} although the general NLP does not lie in the class of convex problems, a variety of verification related problems can be expressed by a sequence of geometric programs, 
which is exploited in \cite{DBLP:conf/tacas/Cubuktepe0JKPPT17}.
Alternatively, finding satisfying parameter instantiations in pMDPs under demonic non-determinism and with affine transition probabilities can be approached by iteratively solving a convex-concave program that approximates the original NLP~\cite{DBLP:conf/atva/CubuktepeJJKT18}.
\nils{A comprehensive overview of exploiting convex programming is presented in~\cite{DBLP:journals/tac/CubuktepeJJKT22}.}
Alternatively, more efficient solvers can be used~\cite{DBLP:journals/iandc/Chatzieleftheriou18} for subclasses of pMDPs. 
An alternative parametric model with a finite set of parameter instantiations, but without the assumption that these instantiations are graph preserving is considered in \cite{shepherding}.

\paragraph{Model repair.}
The problem of model repair is related to parameter synthesis. In particular, for a Markov model and a refuted specification the problem is to transform the model such that the specification is satisfied. 
In the special case where repair amounts to changing transition probabilities, the underlying model is parametric as in this paper: the parameters are addive factors to be added to the original transition probabilities.
The problem was first defined and solved either by a nonlinear program or parameter synthesis in~\cite{bartocci2011model}.
A greedy approach was given in~\cite{DBLP:conf/nfm/PathakAJTK15} and efficient simulation-based methods are presented in~\cite{chen2013model}.
In addition, parametric models are used to rank patches in the repair of software~\cite{DBLP:conf/popl/LongR16}. 

\paragraph{Interval Markov chains.}
Instead of parametric transitions, interval MCs or MDPs feature intervals at their \nils{transitions~\cite{DBLP:conf/lics/JonssonL91,DBLP:journals/ai/GivanLD00,wiesemann2013robust,DBLP:conf/birthday/0001DLM21}}. 
These models do not allow for parameter dependencies, but verification is necessarily ``robust'' against all probabilities within the intervals, see for instance~\cite{seshia_et_al_cav_13}, where convex optimization is utilised, \nils{and~\cite{DBLP:conf/qest/HahnHHLT17,DBLP:journals/tomacs/HahnHHLT19}}, where efficient verification of multiple-objectives is introduced.
In~\cite{DBLP:conf/time/AndreD16,DBLP:journals/tcs/BartDFLMT18}, these models are extended to so-called parametric interval MCs, where interval bounds themselves are parametric.
\nils{Extensions to richer models such as partially observable MDPs are considered in~\cite{DBLP:conf/ijcai/Suilen0CT20,DBLP:conf/aaai/Cubuktepe0JMST21}.}

\paragraph{Derivatives and monotonicity.}
Many systems behave monotonically in some of their system parameters. For example, most network protocols become more reliable if the communication channel reliability increases. If the solution function is monotonic, then parameter space partitioning can be accelerated~\cite{DBLP:conf/atva/SpelJK19}. Assessing monotonicity can be tightly integrated in a loop that uses parameter lifting~\cite{DBLP:conf/tacas/SpelJK21}. Finally, the derivative of the solution function can be used for gradient descent whenever the goal is to find a counterexample for region verification~\cite{DBLP:conf/vmcai/HeckSJMK22}.

\paragraph{Sensitivity analysis.}
Besides analysing in which regions the system behaves correctly \wrt the specification, it is often desirable to perform a sensitivity analysis~\cite{chen-et-al-concur-2014-pertubation,DBLP:journals/tse/SuFCR16}, \ie, to determine in which regions of the parameter space a small perturbation of the system leads to a relatively large change in the considered measure. 
In our setting, such an analysis can be conducted with little additional effort. Given a rational function for a measure of interest, its derivations \wrt all parameters can be easily computed. Passing the derivations with user-specified thresholds to the SMT solver then allows for finding parameter regions in which the system behaves robustly. Adding the safety constraints described earlier, the SMT solver can find regions that are both safe \emph{and} robust.

\paragraph{Parameters with distributions.}
Rather than a model in which the parameter values are chosen from a set, they can be equipped with a distribution. 
The verification outcome consists then of confidence intervals rather than absolute guarantees. 
In \cite{DBLP:journals/sosym/MeedeniyaMAG14}, simulation based methods are used, whereas \cite{DBLP:journals/tr/CalinescuGJPRT16,DBLP:conf/tacas/CalinescuJP16} use statistical methods on a solution function.
pMDPs with a distribution over the parameters are considered in \cite{DBLP:conf/qest/ArmingBCKS18}.
\nils{Sampling-based methods that rely on the so-called scenario-approach~\cite{campi2011sampling,campi2008exact} are presented in~\cite{DBLP:journals/sttt/BadingsCJJKT22,DBLP:conf/tacas/Cubuktepe0JKT20}}.

\paragraph{Ensuring graph preservation.}
Checking graph-preservation is closely related to checking whether a well-defined point instantiation exists, which has an exponential runtime in the number of parameters~\cite{lanotte}.
For parametric interval Markov chains, the question whether there exists a well-defined instantiation is referred to as \emph{consistency} and received attention in \cite{DBLP:conf/time/AndreD16,DBLP:conf/forte/PetrucciP18}.

\paragraph{Robust strategies.}
Robust strategies for pMDPs, as mentioned in Remark~\vref{rem:robust}, are considered in, among others, \cite{wiesemann2013robust,DBLP:conf/icml/MannorMX12}. 
These and other variants of synthesis problems on pMDPs were compared in \cite{DBLP:journals/corr/ArmingBS17}.
A variant where parameters are not non-deterministically chosen, but governed by a prior over these parameters, has recently been considered~\cite{DBLP:conf/qest/ArmingBCKS18}.
\nils{In \cite{DBLP:conf/qest/PolgreenWHA16}, data-driven bounds on parameter ranges are obtained, and properties are validated using parameter synthesis techniques.}

\paragraph{Continuous time.}
Parametric CTMCs were first considered by \cite{DBLP:conf/rtss/HanKM08}.
A method using relaxations similarly to parameter lifting has been proposed in \cite{DBLP:conf/cav/BrimCDS13}. 
The method was improved in \cite{DBLP:conf/cmsb/CeskaDKP14} and implemented in \prismpsy \cite{DBLP:conf/tacas/CeskaPPBK16}.
A combination with sampling-based algorithms to find good parameter instantiations is explored in \cite{DBLP:journals/jss/CalinescuCGKP18}.
Parameter synthesis with statistical guarantees has been explored in \cite{DBLP:journals/iandc/BortolussiMS16,DBLP:conf/tacas/BortolussiS18}.
\nils{Moreover, a sampling-based approach for so-called uncertain parametric CTMCs that have a distribution over the parameter values obtains statistical guarantees on reachability probabilities~\cite{DBLP:conf/cav/BadingsJJSV22}.}
Finally, in \cite{gouberman2019markov}, finding good parameter instantiations is considered by identifying subsets of parameters that have a strictly positive or negative influence on the property at hand.

\paragraph{Connection to other models.}
Furthermore,~\cite{DBLP:conf/uai/Junges0WQWK018} establishes connections to the computation of strategies in partially observable MDPs~\cite{DBLP:books/daglib/0023820}, a prominent model in AI. 
In \cite{DBLP:conf/concur/WinklerJPK19}, the connection to concurrent stochastic games is shown. 
pMCs can be used to accelerate solving hierarchical Markov models~\cite{DBLP:conf/aips/NearyVCT22,DBLP:conf/cav/JungesS22} and for parameter synthesis in Bayesian networks~\cite{DBLP:conf/ecsqaru/SalmaniK21}.
Finally, in~\cite{hidden-parameter-mdp-lacerda}, a method that maintains a belief over parameter values is introduced in a robotics context. 

\sj{Should we mention this new paper on robotics}

\section{Conclusion and Future Work}
\label{sec:conclusion}
This paper gives an extensive account of parameter synthesis for discrete-time Markov chain models.
In particular, we considered three different variants of parameter synthesis questions. 
For each problem variant, we give an account of the available algorithms from the literature, together with several extensions from our side. 
All algorithms are available in the open-source tool \prophesy.

\paragraph{Future work}
Future work in various directions is possible.
Many of the results here can be ported to the more general setting of weighted automata over the adequate semiring~\cite{droste2009handbook}, which can be interesting from a theoretical perspective. 
Algorithmically, we would like to develop methods which identify and exploit structural properties \nils{that are common to standard benchmarks for} Markov chains and Markov decision processes. 
First steps in this direction have been taken, e.g., by exploiting monotonicity~\cite{DBLP:conf/atva/SpelJK19}.
While graph-preservation is common in many applications, this restriction is not always natural. 
The decomposition presented in this paper yields an exponential blow-up in the number of parameters that we would like to avoid whenever possible. 
However, algorithms that do not rely on graph-preservation have not yet been integrated.
The techniques to cover the parameter space by sets of smaller and easy-to-verify regions are still rather naive: This is true both for region verification, where we split due to the approximation, and for parameter space partitioning. 
The above mentioned monotonicity is one possibility to accelerate the way we split.
\nils{In general, we plan to exploit parametric models in a data-driven context, where the structure provided by parameter dependencies can be exploited to accelerate learning of probabilistic models~\cite{DBLP:journals/corr/abs-2205-15827,DBLP:conf/fm/TapplerA0EL19}.}

\begin{acknowledgements}
The authors would like to thank Harold~Bruintjes and Florian~Corzilius for their contributions to \prophesy 1.0, Tom~Janson and Lutz~Klinkenberg for their help in developing \prophesy 2.0, Gereon Kremer as a long-term maintainer of \carl and the anonymous reviewers for their thorough feedback.
\end{acknowledgements}

\bibliographystyle{spbasic}      
\bibliography{literature}   

\end{document}